\documentclass[a4paper,11pt]{article}
\usepackage{jheppub}

\usepackage[utf8]{inputenc}
\usepackage{graphicx,cancel,float}
\usepackage{amssymb}
\usepackage{textcomp}
\usepackage{amsmath}
\usepackage{tabularx}
\usepackage{bm}
\usepackage{times}
\usepackage{color}
\usepackage{slashed}
\usepackage{multirow}
\usepackage{verbatim}
\usepackage{epsfig,epstopdf}
\usepackage{subfigure}
\allowdisplaybreaks[4]

\def\lsim{\mathrel{\raise.3ex\hbox{$<$\kern-.75em\lower1ex\hbox{$\sim$}}}}
\def\gsim{\mathrel{\raise.3ex\hbox{$>$\kern-.75em\lower1ex\hbox{$\sim$}}}}

\definecolor{orange}{rgb}{1,0.5,0}

\newcommand{\minigraph}[5][0.25in]{\begin{minipage}{#2}\begin{center}\includegraphics[width=#2]{#5}\\\vspace{#3}\hspace{#1}{\footnotesize #4}\end{center}\end{minipage}}

\preprint{CPPC-2024-03}
\subheader{\it \today}

\title{\Large{\bf The quark flavor-violating ALPs in light of B mesons and hadron colliders}}

\author[a]{Tong Li}
\emailAdd{litong@nankai.edu.cn}
\author[b]{Zhuoni Qian}
\emailAdd{zhuoniqian@hznu.edu.cn}
\author[c]{Michael A. Schmidt}
\emailAdd{m.schmidt@unsw.edu.au}
\author[a]{Man Yuan}
\emailAdd{yuanman@mail.nankai.edu.cn}

\affiliation[a]{School of Physics, Nankai University, Tianjin 300071, China}
\affiliation[b]{School of Physics, Hangzhou Normal University, Hangzhou, Zhejiang 311121, China}
\affiliation[c]{
Sydney Consortium for Particle Physics and Cosmology,\\
School of Physics, The University of New South Wales, Sydney, New South Wales 2052, Australia
}

\abstract{
The axion-like particle (ALP) may induce flavor-changing neutral currents (FCNCs) when the fermions' Peccei-Quinn charges are not generation universal. The search for flavor-violating ALP couplings with a bottom quark so far focused on FCNC processes of $B$ mesons at low energies. The recent measurements of $B\to K +X$ rare decays place stringent bounds on the quark flavor violations of a light ALP in different decay modes. In this work we propose a novel direct search for bottom flavor-violating interaction of a heavy ALP at the LHC and its upgrades, namely QCD production of an ALP associated with one $b$ jet and one light jet $p~p\to b~j~a$. We consider the decay of the ALP to photons, muons and invisible ALP decays.
The Boosted Decision Tree (BDT) algorithm is used to analyze the events and we train the BDT classifier by feeding in the kinematic observables of signal and backgrounds. Finally, we show the complementarity between the search prospects of hadron colliders and the low-energy $B$ meson constraints from $B$ meson mixing and $B$ meson decays to a light ALP.
}

\begin{document}

\maketitle
\setcounter{page}{2}

\newpage

\section{Introduction}

Axion-like particles (ALPs) are CP-odd pseudo-Nambu-Goldstone bosons as a result of the spontaneous symmetry breaking of a global Peccei-Quinn (PQ) U(1) symmetry. An ALP can appear in many theoretical constructions. One well-studied example is the QCD axion~\cite{Peccei:1977hh,Peccei:1977ur,Weinberg:1977ma,Wilczek:1977pj} (see Ref.~\cite{DiLuzio:2020wdo} for a recent review).
In general, the mass ($m_a$) and the symmetry breaking scale (often called decay constant $f_a$) associated with an ALP can be drastically different~\cite{Dimopoulos:1979pp,Tye:1981zy,Zhitnitsky:1980tq,Dine:1981rt,Holdom:1982ex,Kaplan:1985dv,Srednicki:1985xd,Flynn:1987rs,Kamionkowski:1992mf,Berezhiani:2000gh,Hsu:2004mf,Hook:2014cda,Alonso-Alvarez:2018irt,Hook:2019qoh}. Its mass range spans from sub-micro-eV~\cite{Kim:1979if,Shifman:1979if,Dine:1981rt,Zhitnitsky:1980tq,Turner:1989vc} to the TeV scale and even beyond~\cite{Rubakov:1997vp,Fukuda:2015ana,Gherghetta:2016fhp,Dimopoulos:2016lvn,Chiang:2016eav,Gaillard:2018xgk,Gherghetta:2020ofz}. Thus, the search for the ALPs requires rather different experimental strategies and facilities. Any experimental observation for such a particle would significantly renew our knowledge of physics beyond the Standard Model (SM)~\cite{Kuster:2008zz,DiLuzio:2020wdo,Choi:2020rgn}.

The flavor-changing neutral currents (FCNCs) of ALP are interesting in both theory and phenomenology.~The ALP couplings to the SM fermions carrying PQ charges are rather model-dependent in the Dine-Fischler-Srednicki-Zhitnitsky (DFSZ) type models~\cite{Zhitnitsky:1980tq,Dine:1981rt}. The couplings are determined by the generation textures of PQ charge matrices in flavor basis~\cite{Davidson:1981zd,Wilczek:1982rv,Davidson:1984ik,Peccei:1986pn,Krauss:1986wx,Geng:1988nc,Celis:2014iua}. If one loosens the assumption of the universality of the PQ current and allows non-universal PQ charges in a flavor symmetry, the flavor-violating ALP couplings to SM quarks or leptons arise at tree-level~\cite{Ema:2016ops,Calibbi:2016hwq,Arias-Aragon:2017eww,Bjorkeroth:2018ipq,delaVega:2021ugs,DiLuzio:2023ndz}. The mass mixing between
the heavy vectorlike quark in the Kim-Shifman-Vainshtein-Zakharov (KSVZ) type models~\cite{Kim:1979if,Shifman:1979if} and the SM quarks can also lead to flavor-violating interactions~\cite{Alonso-Alvarez:2023wig}. Or, the flavor-violating ALP couplings can be induced radiatively even if the tree-level flavor structure is absent~\cite{Gavela:2019wzg,Bauer:2020jbp,Bauer:2021mvw,Chakraborty:2021wda,Bertholet:2021hjl}. See Refs.~\cite{MartinCamalich:2020dfe,Bauer:2021mvw,Carmona:2021seb} and references therein for relevant phenomenological probes of the flavor-violating couplings of ALP.

The most promising search for the quark flavor-violating couplings of ALP is through the FCNC processes of mesons at low energies, such as the meson decays or the mixing
of neutral mesons. Recently, the first measurement of the rare decay $B^+\to K^+\nu\bar{\nu}$ at Belle II~\cite{Belle-II:2023esi} draws much attention in the community. The combination
of the inclusive and hadronic tagging results gives the branching ratio as
\begin{eqnarray}
{\rm BR}(B^+\to K^+\nu\bar{\nu})_{\rm Belle~II}=(2.3\pm 0.7)\times 10^{-5}\;.
\end{eqnarray}
This result is 2.7 standard deviations above the SM expectation~\cite{Parrott:2022zte,Becirevic:2023aov}.
It provides an excellent opportunity to confine the flavor-violating couplings to bottom quark as well as the invisible decay mode of ALP, together with the previous searches~\cite{Belle:2013tnz,Belle:2017oht}. See Ref.~\cite{Altmannshofer:2023hkn} for a recent ALP interpretation of Belle II measurement and Refs.~\cite{Berezhiani:1989fs,Berezhiani:1990jj,Berezhiani:1990wn,Ferber:2022rsf} for earlier considerations of $B$ meson decays to an ALP. Moreover, the LHCb, Belle II and BaBar also searched for the rare decays $B\to K^{(\ast)}\mu^+ \mu^-$~\cite{LHCb:2015nkv,LHCb:2016awg,Belle-II:2023ueh} as well as $B^+\to K^+ \gamma\gamma$~\cite{BaBar:2021ich}. They can also place experimental bounds on the visible decay modes of ALP and its $b$ quark flavor violation.
These bounds are considerable for $m_a\lesssim m_B$. The suppression of heavier ALP mediator with $m_a\gtrsim m_B$ would however forbid the on-shell two-body decay and weaken these low-energy constraints.

High-energy colliders have been the primary tool for the direct probe of new physics (NP) beyond the SM in the past decades.
The collider experiments would efficiently reveal the mass and interactions of the new particles at the energy frontier. After the CERN Large Hadron Collider (LHC), the luminosity upgrade of the LHC (HL-LHC) will take the lead in searching for new physics beyond the SM~\cite{Apollinari:2015wtw}. There also have been considerations to construct the next generation of hadron colliders with 100 TeV center-of-mass (c.m.) energy (FCC-hh)~\cite{FCC:2018byv,CEPC-SPPCStudyGroup:2015csa}. They provide an ideal environment to search for heavy ALP with quark flavor-violating interactions. The very recent studies of ALP quark FCNC at colliders focus on the flavor interactions of an ALP with a top quark~\cite{Carmona:2022jid,Esser:2023fdo,Rygaard:2023dlx,Blasi:2023hvb,Phan:2023dqw}. In this paper, we will investigate the potential probe of ALP flavor-violating couplings to the bottom quark at hadron colliders and the complementarity with the low-energy $B$ meson constraints.
We simulate the QCD production of ALP with one $b$ jet and one light jet $j$
\begin{eqnarray}
p~p \to b~j~a\;,
\end{eqnarray}
through either the FCNC $b$-$d$-$a$ coupling or $b$-$s$-$a$ coupling. The ALP is then considered to be short-lived and decay invisibly or decay into dimuon and diphoton final states.~The popular Boosted Decision Tree (BDT) algorithm is applied to analyze the events and we train the BDT classifier by feeding in the kinematic observables of signal and SM background events. Finally, we compare the sensitivity of LHC, HL-LHC and FCC-hh to the $b$ quark flavor-violating couplings of ALP with the constraints from $B$ meson measurements.

The rest of this paper is organized as follows. In Sec.~\ref{sec:ALPs}, we first describe the theoretical framework for the ALPs interactions with the SM fermions and emphasize the quark flavor-violating couplings.
We also present the low-energy constraints on the flavor-violating couplings to $b$ quark from leptonic and semi-leptonic $B$ meson decays, and $B$ meson oscillations.
In Sec.~\ref{sec:inv}, we analyze the search for $b$ quark FCNC of ALP associated with invisible decay products at hadron colliders. The exclusion limits at colliders are shown in comparision with the low-energy $B$ meson constraints. The numerical analyses for the dimuon and diphoton decay modes of ALP are given in Sec.~\ref{sec:dilepton} and Sec.~\ref{sec:diphoton}, respectively.
We summarize our results in Sec.~\ref{sec:Sum}.

\section{General Fermionic Interactions for ALPs}
\label{sec:ALPs}

\subsection{Theoretical formulation}
\label{sec:theory}

We introduce a generic massive CP-odd scalar $a$, presumably an ALP associated with a global U(1) symmetry spontaneously broken above the electroweak scale.
Besides the kinetic term for ALP, the most general effective Lagrangian for fermionic ALP interactions is given by~\cite{Brivio:2017ije,Calibbi:2020jvd,Bauer:2021mvw}
\begin{eqnarray}
\mathcal{L}_{af_if_j}
&=& {\partial_\mu a\over 2f_a} \bar{f}_i \gamma^\mu (c^V_{ij}+c^A_{ij}\gamma_5) f_j\;,
\end{eqnarray}
with complex $(c^{V,A}_{ij})^\ast=c^{V,A}_{ji}$ for $i\neq j$. For on-shell fermions, after applying the equations of motion, the above Lagrangian becomes
\begin{eqnarray}
\mathcal{L}_{af_if_j}
&=&-{i\over 2f_a}a \bar{f}_i [c^V_{ij}(m_i-m_j)+c^A_{ij}(m_i+m_j)\gamma_5] f_j\;.
\end{eqnarray}
Thus, the ALP fermion couplings are proportional to the linear combinations of fermion masses. The flavor-conserving currents are only induced by pseudo-scalar bilinear and $c_{ii}^A$ coefficient. The ALP generally has flavor-violating couplings from either scalar bilinear or pseudo-scalar bilinear. If they are not present at tree level, they are induced via electroweak interactions at the one-loop level. This in particular results in flavor-violating ALP decays into lighter fermions $a\to f_i \bar{f}_j+\bar{f}_i f_j$ for $m_a>m_i+m_j$ is
\begin{equation}\label{eq:2body-decay}
    \begin{aligned}
\Gamma_{a\to f_if_j}=&2\Gamma(a\to f_i \bar{f}_j) \\
=&{N_c\over 16\pi m_a^3 f_a^2} \Big\{|c^V_{ij}|^2\Big[m_a^2(m_i-m_j)^2-(m_i^2-m_j^2)^2\Big]
\\&\hspace{2.5cm} +|c^A_{ij}|^2\Big[m_a^2(m_i+m_j)^2-(m_i^2-m_j^2)^2\Big] \Big\}
\times
\lambda^{1/2}(m_a^2,m_i^2,m_j^2)\\
=& {N_c m_a m_i^2\over 16\pi f_a^2} \Big(|c^V_{ij}|^2+|c^A_{ij}|^2\Big) \Big(1-{m_i^2\over m_a^2}\Big)^2 ~~{\rm when}~m_j=0\;,
\end{aligned}
\end{equation}
where $\lambda(x,y,z)=x^2+y^2+z^2-2xy-2xz-2yz$.

The ALP lifetime sensitively depends on its mass $m_a$ and decay constant $f_a$. When $m_a\gg m_b$ and assuming only down-type quark couplings exist, the above partial width leads to the ALP lifetime as
\begin{eqnarray}
c\tau_a \lesssim 2\times 10^{-13}~{\rm mm} \Big( {10~{\rm GeV}\over m_a} \Big) \Big( {f_a\over 10^3~{\rm GeV}} \Big)^2  {1\over |c_{bq}^V|^2 + |c_{bq}^A|^2}\;.
\end{eqnarray}
The collider searches for heavy ALP with $b$ jets will be in the regime of prompt ALP decay.
For light ALP with $m_a<m_b$, we take the decay to muons as an example with
\begin{align}
     c\tau_a \lesssim 4.5~{\rm mm} \left(\frac{1 \,\mathrm{GeV}}{m_a}\right)\left(\frac{f_a}{10^5\, \mathrm{GeV}}\right)^2 \frac{1}{|c^A_{\mu\mu}|^2}\;.
 \end{align}
The ALPs in $B$ meson decays are usually long-lived due to the smaller ALP mass and the stronger limit on the decay constant.
The inclusion of additional decay modes would make the ALP more short-lived.

In summary, for $m_a\gg m_b$ in the LHC collider search region, the current bound on $f_a$ still allows for large parameter space for a prompt decayed ALP, and thus our focus for the prompt decay case in the collider analysis later. For $m_a < m_b$, bounds on $f_a$ from B physics pushes such light ALPs, if decaying mostly back to the SM particles, to be long-lived.

There are several constraints on ALPs from meson decays and oscillations. We follow the discussion in Refs.~\cite{MartinCamalich:2020dfe,Bauer:2021mvw} and focus on $B$ physics which constrains the ALP couplings $c_{bq}^{V,A}$. The $B$ physics provides complementary constraints to collider searches with $b$ jets.

The flavor-violating effective operators in the weak effective theory (WET) can be generated through the exchange of $W$ boson at loop-level~\cite{Chetyrkin:1996vx,Hiller:2014yaa}. They are thus suppressed by the Fermi constant $G_F$ as well as CKM matrix, for instance $G_F V_{ts}^\ast V_{tb} (\bar{s}_L \gamma_\mu b_L)\sum_q(\bar{q}\gamma^\mu q)$. If the ALP flavor-conserving coupling is present and the ALP radiates from one of the quark legs, the $b$-$s$-$a$ flavor-violating coupling can be generated by closing the loop of the $q$ quark. This kind of ALP flavor-violating coupling is constructed by vertices of two effective field theories which are dimension-six four-fermion operators in WET and dimension-five ALP flavor-conserving coupling.
Refs.~\cite{Bauer:2021wjo,Cornella:2023kjq} embedded the ALP chiral theory into the above weak effective Lagrangian for the calculation of $K\to \pi a$ decay, through the ALP-meson mixing, ALP emission of a meson or the ALP coupling to the four-fermion vertices (the first five Feynman diagrams of Fig.~1 in \cite{Bauer:2021wjo}). They suffer from a further suppression by $f_\pi^2 G_F$ compared to the off-diagonal ALP couplings to quarks at tree-level (the sixth diagram in their Fig.~1).

As a phenomenological analysis, we remain agnostic about the origin of the ALP flavor couplings to SM quarks. The ALP flavor-violating couplings were simply taken as independent parameters in an effective framework. We refer the reader to existing UV models which realize large flavor-violating couplings.
One example is the astrophobic axion proposed in Ref.~\cite{DiLuzio:2017ogq}.
The assignment of family dependent PQ charges in DFSZ-like models suppresses the diagonal couplings of the first family SM fermions and implies flavor violating axion couplings.

Other examples are given by models identifying the Peccei-Quinn symmetry $U(1)_{\rm PQ}$ with a global Froggatt-Nielsen $U(1)$ flavor symmetry in order to solve the flavor hierarchy of SM fermions, see e.g.~Refs.~\cite{Ema:2016ops,Bonnefoy:2019lsn,delaVega:2021ugs,Calibbi:2016hwq}.
This flavor symmetry enforces a Yukawa matrix structure where the diagonal elements $(1,1)$ and $(2,2)$ are suppressed compared to the $(3,3)$ and off-diagonal elements. For example Ref.~\cite{delaVega:2021ugs} realizes at the nearest-neighbour texture where only off-diagonal elements and the $(3,3)$ element are generated when truncating at dimension-7. The other Yukawa matrix elements are further suppressed.
The families of SM quarks gain non-universal PQ charges and sizable off-diagonal ALP couplings.
The ``axiflavon'' couplings to the SM fermions are proportional to the sum of flavor-dependent charges of $SU(2)_L$ doublet and singlet, see e.g.~Eq. (8) in \cite{Calibbi:2016hwq}.
The couplings are in general not diagonal in the fermion mass eigenstate basis but have flavor changing neutral currents. The $U(2)$ axiflavon in Ref.~\cite{Linster:2018avp} gains flavor violating couplings in vector currents which are more dominant than the diagonal elements.

\mathversion{bold}
\subsection{Leptonic $B$ meson decays}
\mathversion{normal}
Leptonic $B$ meson decays are sensitive probes for ALPs due to their chiral suppression in the SM~\cite{Bauer:2021mvw}. The branching ratio can be expressed in terms of the SM branching ratio~\cite{Bauer:2021mvw}
\begin{equation}
\begin{aligned}
\mathrm{BR}(B_q\to f\bar f) &= \mathrm{BR}(B_q\to f\bar f)_{\rm SM} \left|1+\frac{1}{1-x_a} \frac{c_{bq}^{A*} c_{ff}^A}{G_F f_a^2 \sqrt{2}\alpha \lambda_t C_{10}^{\rm SM}}\right|^2\;,
\\
\mathrm{BR}(B_q\to f\bar f)_{\rm SM} & = \frac{G_F^2 \tau_{B_q} m_{B_q}^3 f_{B_q}^2 |\lambda_t|^2 |C_{10}^{\rm SM}|^2}{16\pi^3} x_f(1-4x_f)^{1/2} \;,
\end{aligned}
\end{equation}
where $\lambda_t = V_{ts}^* V_{tb}$, $C_{10}^{\rm SM}\simeq -4.2$~\cite{Beneke:2017vpq}, $x_{a,f}=m_{a,f}^2/m_{B_q}^2$, $\alpha$ is the fine structure constant, and $f_{B_q}$ is the meson decay constant.
In particular, the decays to muons are constrained to be~\cite{ATLAS:2018cur,CMS:2019bbr,LHCb:2017rmj,LHCb:2020zud}
\begin{align}
    \mathrm{BR}(B_d\to \mu^+\mu^-) &=(0.6\pm0.7)\times 10^{-10}  (<1.6\times 10^{-10}\,\, \text{at 90\% CL})\;,\\
    \mathrm{BR}(B_s\to \mu^+\mu^-) &=(2.69^{+0.37}_{-0.35})\times 10^{-9}\;.
\end{align}
These limits apply to the product of coefficients $c_{bq}^{A\ast} c_{\mu\mu}^A/f_a^2$ if $m_a>m_{B_q}$ or if the ALP decays promptly and thus do not constrain the same ALP coupling combination as the collider searches.

As the ALP couplings to fermions are proportional to the fermion mass, its contribution to the pseudoscalar meson decay to neutrinos is suppressed. But, any additional invisible decay width of the ALP will be constrained by invisible $B_q$ decay.
For instance, Refs.~\cite{Bharucha:2022lty,Ghosh:2023tyz} proposed the ALP portal to freeze-in dark matter. BaBar obtained an upper limit on invisible $B_d$ decays~\cite{BaBar:2012yut}
\begin{equation}
    \mathrm{BR}(B_d\to \mathrm{inv}) \leq 2.4\times 10^{-5}
\end{equation}
at 90\% CL and recently, the authors of Ref.~\cite{Alonso-Alvarez:2023mgc} derived the first upper limit on the invisible $B_s$ decays using LEP data
\begin{align}
    \mathrm{BR}(B_s\to \mathrm{inv}) & \leq 5.4\times 10^{-4}
\end{align}
at 90\% CL.
For a 2-body decay into an invisible fermion-antifermion pair with mass $m_{f}$ and thus $x_{f}=m_{f}^2/m_{B_q}^2$, the invisible $B_q$ decay width can be expressed in terms of the ALP decay width $\Gamma(a\to f\bar f)$ as
\begin{align}
    \Gamma(B_q\to \mathrm{inv}) &
     =\frac{|c_{bq}^A|^2 f_{B_q}^2  }{4f_a^2 (1-x_a)^2} \frac{x_a^{3/2}\lambda^{1/2}(1,x_f,x_f)}{\lambda^{1/2}(x_a,x_f,x_f)}\, \Gamma(a\to f\bar f) \;.
\end{align}
The constraint on the invisible $B_q$ branching ratio can be translated in a lower bound on the axion decay constant
\begin{align}
    \frac{f_a^4}{|c_{bq}^A c_{ff}^A|^2}  > \frac{\tau_{B_q}\,f_{B_q}^2 m_{B_q}^3 }{16\pi \mathrm{BR}(B_q\to \mathrm{inv}) }\frac{x_ax_f\lambda^{1/2}(1,x_f,x_f)}{(1-x_a)^2}\;,
\end{align}
which results in relatively weak constraints on the axion decay constant $f_a/\sqrt{|c_{bq}^{A}c_{ff}^A|}\gtrsim 10^3$ GeV.
%

\mathversion{bold}
\subsection{Other $B$ meson decays with an ALP in the final state}
\mathversion{normal}
 We are considering three main scenarios depending on the dominant ALP decay mode~\cite{Bauer:2021mvw} (i)
 $B\to Ma (\to \mathrm{invisible})$, (ii) $B\to M a (\to \ell^+\ell^-)$, and (iii) $B\to Ma (\to\gamma\gamma)$ with $M=K^{(*)},\pi,\rho$. The calculation for each of the scenarios depends on the ALP mass. For a heavy ALP with $m_a\gtrsim m_B$, the ALP cannot be produced on-shell and its contribution to the semi-leptonic $B$ meson decay can be described within effective field theory (EFT). For lighter ALPs with $m_a<m_B-m_{M}$, the ALP can be produced on-shell and the decay rates are
\begin{eqnarray}
\Gamma(B \to P a) &=& \frac{m_B^3}{64\pi}{|c_{bq}^V|^2\over f_a^2}\Big(1-\frac{m_P^2}{m_B^2}\Big)^2 f_0^2(m_a^2)\lambda^{1/2}\left(1,{m_P^2\over m_B^2},{m_a^2\over m_B^2}\right)\;,\\
\Gamma(B \to V a) &=&
\frac{m_B^3}{64\pi}{|c_{bq}^A|^2\over f_a^2}A_0^2(m_a^2)\lambda^{3/2}\Big(1,{m_V^2\over m_B^2},{m_a^2\over m_B^2}\Big)\;,
\end{eqnarray}
where $q=s$ for $P=K, V=K^\ast$, and $q=d$ for $P=\pi, V=\rho$. For $\pi^0, \rho^0\sim (u\bar{u}-d\bar{d})/\sqrt{2}$, there is an additional overall factor of $1/2$.

\subsubsection{Invisible decays}
\label{sec:BKinv}
The branching ratio for semi-invisible decays receives two contributions
\begin{align}
    {\rm BR}(B\to P/V + {\rm inv.})_{\rm NP}={\rm BR}(B\to P/V + a) \left[ {\rm BR}(a\to {\rm inv.}) + e^{-r_{\rm det}/\beta\gamma c \tau_a} \mathrm{BR}(a\to f\bar f,\gamma\gamma)\right]\;,
\end{align}
where $r_{\rm det}$ the size of the detector and BR($a\to f\bar f,\gamma\gamma$) denotes the branching ratio for decays to fermion-antifermion pairs and two photons.
We focus on the scenario where the ALP decays dominantly invisibly and thus the first term dominates the branching ratio and it is approximately given by the product
\begin{equation}
    {\rm BR}(B\to P/V + {\rm inv.})_{\rm NP}={\rm BR}(B\to P/V + a)  {\rm BR}(a\to {\rm inv.}) \;.
\end{equation}
\begin{table}[tb!]
\centering
\begin{tabular}{c|c}
\hline
\hline
    $B$ decays & measurements \\
    \hline
    \hline
    ${\rm BR}(B^+\to K^+ + {\rm inv.})_{\rm NP}$ & $=(1.9\pm 0.7)\times 10^{-5}$~\cite{Belle-II:2023esi,Becirevic:2023aov} \\
    ${\rm BR}(B^+\to K^{\ast +} + {\rm inv.})_{\rm NP}$ & $<3.1\times 10^{-5}$~\cite{Belle:2013tnz,He:2022ljo}  \\
    ${\rm BR}(B^0\to K^{0} + {\rm inv.})_{\rm NP}$ & $<2.3\times 10^{-5}$~\cite{Belle:2017oht,He:2022ljo} \\
    ${\rm BR}(B^0\to K^{\ast 0} + {\rm inv.})_{\rm NP}$ & $<1.0\times 10^{-5}$~\cite{Belle:2017oht,He:2022ljo} \\
    \hline
    ${\rm BR}(B^+\to \pi^{+} + {\rm inv.})_{\rm NP}$ & $<1.4\times 10^{-5}$~\cite{Belle:2017oht,He:2022ljo} \\
    ${\rm BR}(B^+\to \rho^{+} + {\rm inv.})_{\rm NP}$ & $<3.0\times 10^{-5}$~\cite{Belle:2017oht,He:2022ljo} \\
    ${\rm BR}(B^0\to \pi^{0} + {\rm inv.})_{\rm NP}$ & $<8.9\times 10^{-6}$~\cite{Belle:2017oht,He:2022ljo} \\
    ${\rm BR}(B^0\to \rho^{0} + {\rm inv.})_{\rm NP}$ & $<4.0\times 10^{-5}$~\cite{Belle:2017oht,He:2022ljo} \\
    \hline
    \hline
\end{tabular}
\caption{The invisible decay modes of $B$ mesons and the corresponding measurement or bounds on NP contribution.
The upper bounds have been obtained by subtracting the lower bound of the SM prediction from the experimental branching ratio following the procedure in Refs.~\cite{He:2022ljo,He:2023bnk}.}
\label{tab:semi-invisible_BR}
\end{table}
Apart from the recent Belle II measurement of $B^+\to K^+ +\mathrm{inv}$~\cite{Belle-II:2023esi}, there are currently only upper limits. Subtracting the SM contributions, Ref.~\cite{He:2022ljo} derived upper limits on non-interfering new physics contributions, which are reproduced in Table~\ref{tab:semi-invisible_BR}.
We find the constraint on the quantity $c^{V(A)}_{bq}/f_a \sqrt{{\rm BR}(a\to {\rm inv.})}\lesssim 10^{-8}~{\rm GeV}^{-1}$ for $m_a\lesssim m_B$. In Sec.~\ref{sec:inv}, we will show the preferred region or upper limits on $c^{V(A)}_{bs}/f_a \sqrt{{\rm BR}(a\to {\rm inv.})}$ by $B\to K/K^\ast+{\rm inv.}$ and $c^{V(A)}_{bd}/f_a \sqrt{{\rm BR}(a\to {\rm inv.})}$ by $B\to \pi/\rho+{\rm inv.}$ in details.

\subsubsection{Decays to dileptons}
\label{sec:BKmumu}
LHCb and Belle II searched for displaced vertices in the decay $B\to K^{(*)} a (\to \mu^+\mu^-)$~\cite{LHCb:2015nkv,LHCb:2016awg,Belle-II:2023ueh}. The Belle-II analysis also places constraints on $B\to K^{(*)}e^+e^-$ which is however not relevant for the collider study in Sec.~\ref{sec:dilepton}. See also the recent phenomenological study of the Belle II sensitivity~\cite{Ferber:2022rsf}.

\begin{figure}[tb!]
\centering
\includegraphics[width=0.48\textwidth]{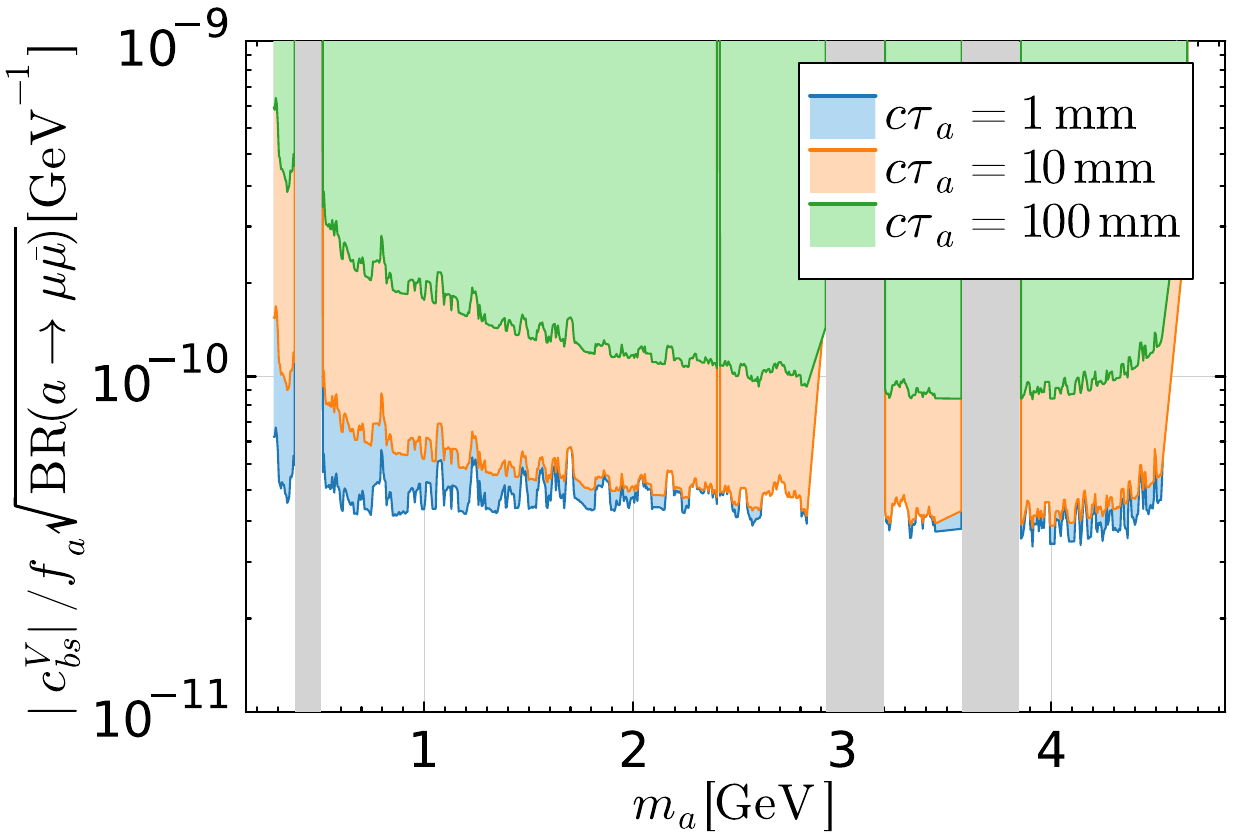}\hfill
\includegraphics[width=0.48\textwidth]{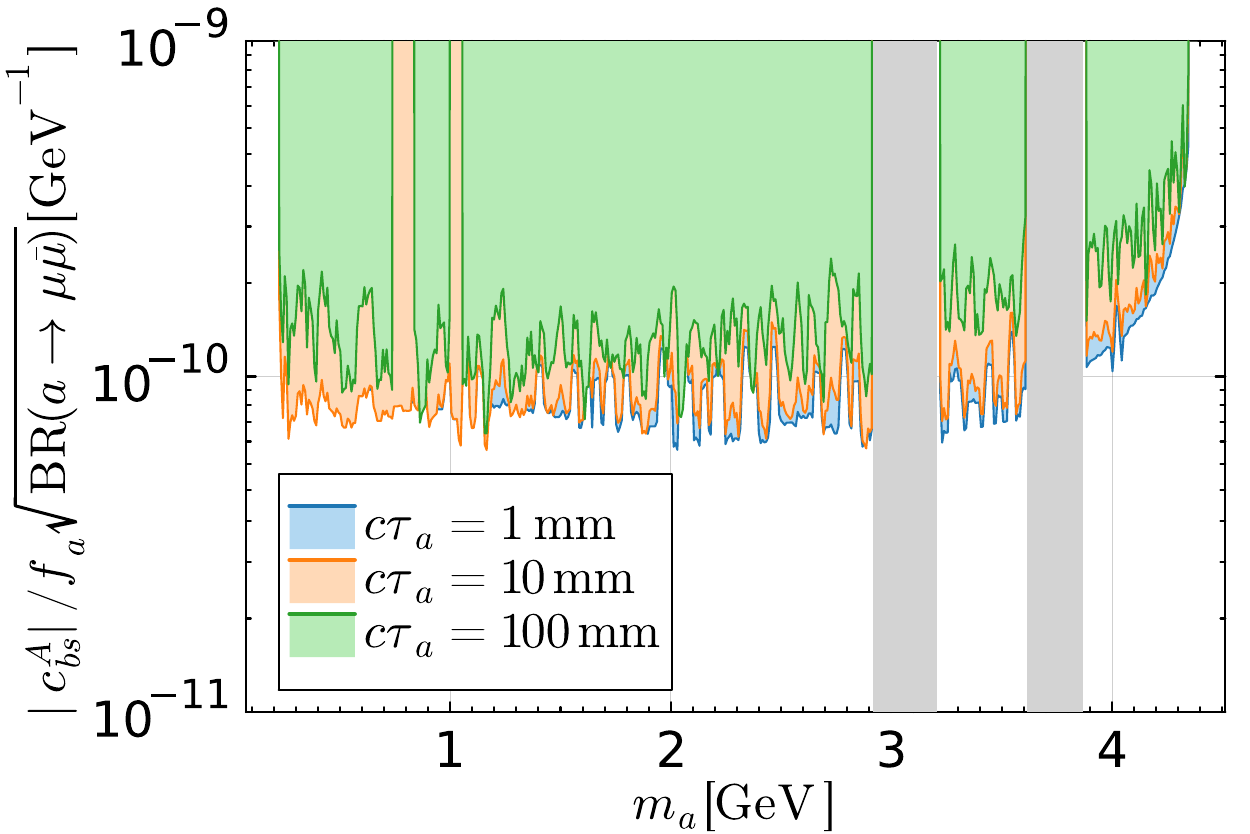}
\caption{Upper limits from displaced vertex searches of LHCb: $B^+\to K^+ a(\to \mu^+\mu^-)$ (left) and $B^0 \to K^{*0} a(\to \mu^+\mu^-)$ (right). The coloured regions are excluded. The LHCb $B^+\to K^+ a(\to\mu^+\mu^-)$ analysis is not sensitive for masses close to the $K_S^0$, $J/\psi$, $\psi(2S)$ and $\psi(3770)$ meson masses, which is indicated by gray shaded regions. We do not colour the partially vetoed mass range close to the $\phi$ and $\psi(4160)$ resonances. Similarly the $B^0\to K^{*0} a(\to \mu^+\mu^-)$ search looses sensitivity close to the $J/\psi$, $\psi(2S)$ and $\psi(3770)$ meson resonances. }
\label{fig:BKmumu}
\end{figure}

The most stringent constraint is placed by the LHCb search for $B^+\to K^+a(\to\mu^+\mu^-)$ with a displaced vertex~\cite{LHCb:2016awg} which presents constraints on the branching ratio $\mathrm{BR}(B^+\to K^+ a) \mathrm{BR}(a\to \mu^+\mu^-)$ as a function of the mass $m_a$ and lifetime $\tau_a$. Assuming short enough decay lengths of ALP, we recast the experimental limit on a bound on the combination $|c^{V(A)}_{bs}|/f_a \sqrt{\mathrm{BR}(a\to \mu^+\mu^-)}$ as a function of the axion mass $m_a$. The colored regions in Fig.~\ref{fig:BKmumu} are excluded at 95\% CL. The blank regions correspond to the vetoed $K_S^0$, $J/\psi$, $\phi$ and $\psi$ meson resonances. We find the constraint on the ALP coupling $|c^{V(A)}_{bs}|/f_a \sqrt{{\rm BR}(a\to \mu^+\mu^-)}\lesssim 10^{-10}~{\rm GeV}^{-1}$ for $m_a\lesssim m_B$. This is stronger by two orders of magnitude than the above constraint from invisible decay. Note the search looses its sensitivity for several masses due to hadronic resonances.
Similarly, the LHCb search for $B^0\to K^{*0}a (\to \mu^+\mu^-)$~\cite{LHCb:2015nkv} provides a constraint on the axial-vector coupling $c^A_{bs}$.

\subsubsection{Decays to photons}

\begin{figure}[tb!]
\centering
\includegraphics[width=0.6\textwidth]{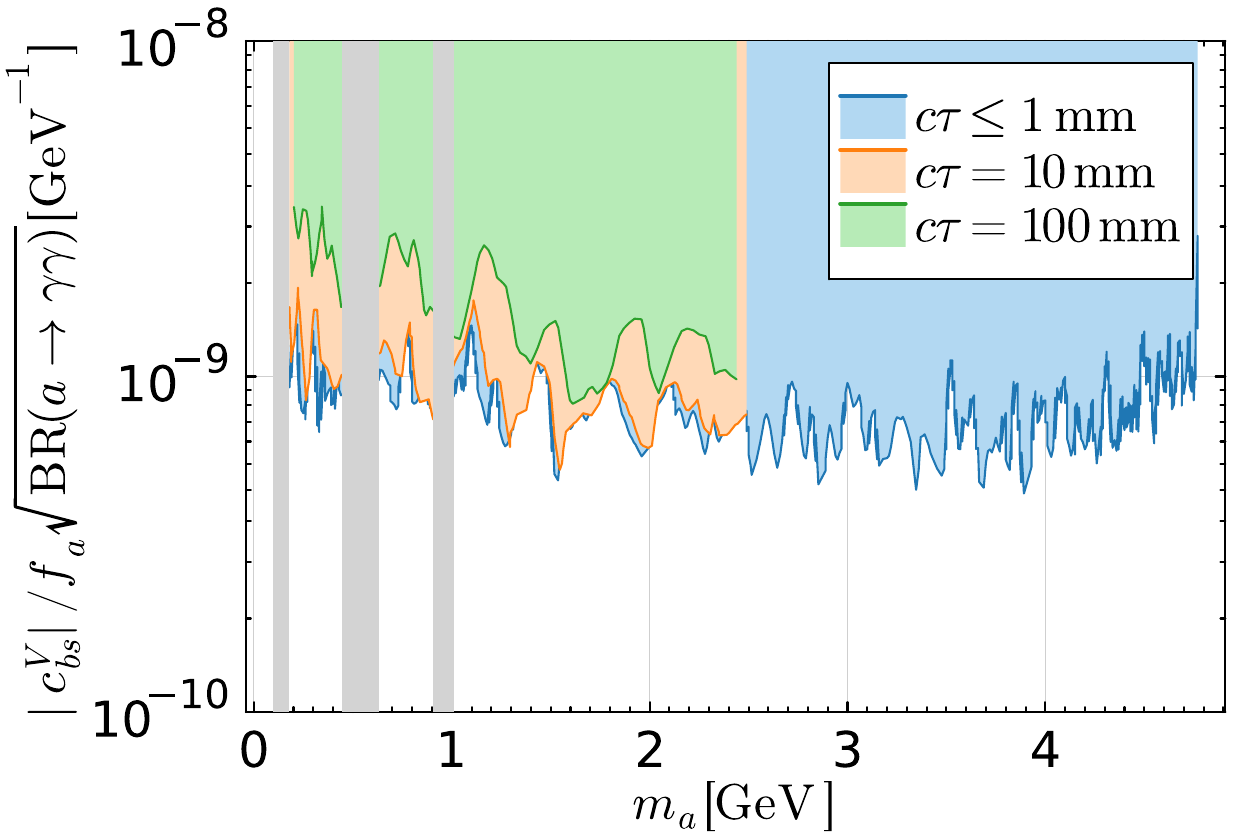}%
\caption{Upper limits from BaBar search for $B^+\to K^+ a(\to \gamma\gamma)$. The coloured regions are excluded and the gray shaded regions have been excluded in the analysis due to the vicinity to the $\pi^0$, $\eta$, and $\eta^\prime$ meson resonances.
}
\label{fig:BKgamgam}
\end{figure}

BaBar searched for ALPs decaying to a pair of photons. We recast the constraints in figures 3 and 4 of Ref.~\cite{BaBar:2021ich} as a constraint on the product of the ALP coupling $|c^V_{bs}|/f_a$ and the square root of the branching ratio to two photons and present the constraints in Fig.~\ref{fig:BKgamgam}. While the BaBar analysis~\cite{BaBar:2021ich} considered the mass range $0.175~\mathrm{GeV} < m_a<m_{B^+}-m_{K^+}$ for the prompt decay analysis, the search for long-lived ALPs was restricted to $m_a<2.5$ GeV. The mass ranges close to the pion, $\eta$ and $\eta^\prime$ masses have been excluded due to large peaking backgrounds. The 90\% CL limit on the coupling product $|c_{bs}^V|/f_a \sqrt{{\rm BR}(a\to \gamma\gamma)}$ turns out to be $5\times 10^{-10}\sim 10^{-9}~{\rm GeV}^{-1}$ for promptly decaying ALPs.

\mathversion{bold}
\subsection{$B$ meson oscillations}
\mathversion{normal}

At leading order in the heavy $b$ quark mass, $B$ meson mixing is described in terms of dimension-6 operators in heavy quark effective theory. The effective Hamiltonian describing $B$ meson mixing is~\cite{Ciuchini:1998ix}
\begin{align}
    \mathcal{H}_{\rm eff}^{\Delta B=2} = \sum_{i=1}^5 C_i(\mu_b) \mathcal{O}_i + \sum_{i=1}^3 \tilde C_i(\mu_b) \tilde{\mathcal{O}}_i
\end{align}
with the dimension-6 operators
\begin{equation}
\begin{aligned}
    \mathcal{O}_1 & = (\bar q^i \gamma^\mu b^i_L) (\bar q^j \gamma_\mu b^j_L)\;,
    &
    \mathcal{O}_2 & = (\bar q^i b^i_L) (\bar q^j  b^j_L)\;,
    &
    \mathcal{O}_3 & = (\bar q^i b^j_L) (\bar q^j  b^i_L)\;,
    \\
    \mathcal{O}_4 & = (\bar q^i b^i_L) (\bar q^j  b^j_R)\;,
    &
    \mathcal{O}_5 & = (\bar q^i b^j_L) (\bar q^j  b^i_R)\;,
\end{aligned}
\end{equation}
together with the operators $\tilde{\mathcal{O}}_{1,2,3}$ which are obtained by flipping the chirality $L\leftrightarrow R$ in $\mathcal{O}_{1,2,3}$. In the SM only the Wilson coefficient $C_1$ is induced. The ALP induces in addition the
three Wilson coefficients~\cite{Bauer:2021mvw}
\begin{equation}
\begin{aligned}
   C_2(\mu_a) &  = \frac{m_b^2(\mu_a)}{8m_a^2 f_a^2} \left(c^V_{bq}+c^A_{bq}\right)^2\;,
   \\
   \tilde C_2(\mu_a) &  = \frac{m_b^2(\mu_a)}{8m_a^2 f_a^2} \left(c^V_{bq}-c^A_{bq}\right)^2\;,
   \\
   C_4(\mu_a) &  = \frac{m_b^2(\mu_a)}{4m_a^2 f_a^2} \left(c^V_{bq} +c^A_{bq}\right)\left(c^V_{bq} -c^A_{bq}\right)
\end{aligned}
\end{equation}
 at tree level at the ALP mass scale $\mu_a\simeq m_a$. They are related to the Wilson coefficients at the hadronic scale $\mu_b=m_b(m_b)$ via renormalization group running~\cite{Bagger:1997gg}
\begin{equation}
\begin{aligned}
    C_2(\mu_b) &= (0.983 \eta^{-2.42} + 0.017\eta^{2.75}) C_2(\mu_a)\;,
    \\
    C_3(\mu_b) &=(-0.064 \eta^{-2.42}+0.064 \eta^{2.75}) C_2(\mu_a)\;,
    \\
    C_4(\mu_b) &= \eta^{-4} C_4(\mu_a) \;.
\end{aligned}
\end{equation}
The parameter $\eta$ which describes the running of $C_1$ is given by
\begin{align}
    \eta&=
    \left(\frac{\alpha_s(m_t)}{\alpha_s(\mu_b)}\right)^{6/23}
    \left(\frac{\alpha_s(\mu_a)}{\alpha_s(m_t)}\right)^{6/21}
\end{align}
assuming a heavy ALP with $m_a > m_t$.
The Wilson coefficients $C_3$ and $\tilde C_3$ are induced by renormalization group running, but suppressed compared to $C_2$ and $\tilde C_2$, respectively. They are included in the numerical analysis, but are not included in the analytic expression below.

The mass difference can be expressed in terms of the Wilson coefficients at the hadronic scale $\mu_b$~\cite{Bauer:2021mvw}
\begin{equation}
\begin{aligned}
\Delta M_q=
\Bigg| &-\frac{(\lambda_t^q)^2}{|\lambda_t^q|^2} \Delta M_q^{\rm SM}
\\ &
+ f_{B_q}^2 m_{B_q}\left[  \left(C_2(\mu_b)+\tilde C_2(\mu_b) \right)\eta_2(\mu_b) B_{B_q}^{(2)}(\mu_b)
+ C_4(\mu_b) \eta_4(\mu_b) B_{B_q}^{(4)}(\mu_b)\right]\Bigg|
\;,
\label{eq:DelMq}
\end{aligned}
\end{equation}
where the first term denotes the SM contribution with $\lambda_t^q\equiv V_{tq}^* V_{tb}$~\footnote{Compared with Ref.~\cite{Bauer:2021mvw}, we additionally include the dependence on the CKM matrix elements through $\lambda_t^q$ to recover the phase information in the off-diagonal element of the mass matrix. In addition, instead of their $k_d$ and $k_D$ couplings, we used a different basis with $[k_d]_{ij}={c^V_{ij}+c^A_{ij}\over 2}$ and $[k_D]_{ij}={c^V_{ij}-c^A_{ij}\over 2}$.}, $C_i$ and $\tilde C_i$ describe the ALP contributions to the different Wilson coefficients, $f_{B_q}$ is the decay constant of $B_q$, $m_{B_q}$ the $B_q$ meson mass, $B_{B_q}^{(i)}$ are hadronic parameters taken from \cite{DiLuzio:2019jyq} and reproduced in Table~\ref{tab:Mixing} (bottom) and the $\eta_i^q(\mu_b)$ normalization factors are defined as~\cite{Bagger:1997gg}
\begin{equation}
\begin{aligned}
    \eta_2(\mu_b)&=-\frac{5}{12} \frac{m_{B_q}^2}{m_b^2(\mu_b)}\;,
 &
    \eta_3(\mu_b)&=\frac{1}{12} \frac{m_{B_q}^2}{m_b^2(\mu_b)}\;, &
    \eta_4(\mu_b)&=\frac{1}{2} \frac{m_{B_q}^2}{m_b^2(\mu_b)} + \frac{1}{12}\;.
\end{aligned}
\end{equation}
The pre-factor of coefficient $C_4$ is positive and twice the negative of the pre-factor of $C_2+\tilde{C}_2$ and thus we find it to dominate the meson mass difference.
The SM prediction and the experimental measurements of the mass differences for $B_d$ and $B_s$ meson mixing are presented in Table~\ref{tab:Mixing} (top). Note that both observables have sizable theoretical errors, larger than the experimental errors. In fact the experimental errors are negligible compared to the theoretical errors when combining them in quadrature. Moreover, there are discrepancies between the SM predictions of different groups. The SM predictions reported by FLAG~\cite{FermilabLattice:2016ipl,FlavourLatticeAveragingGroupFLAG:2021npn} are larger and thus deviate further from the experimental measurements. The larger deviation could be explained by an ALP. See~\cite{Bauer:2021mvw} for a discussion how it affects the results. In this work, we take a conservative approach and do not attempt to explain any deviation, because the main focus of the work is on the sensitivity of current and future colliders to flavor-violating ALP scenarios. We thus only show the results for the SM prediction obtained in~\cite{DiLuzio:2019jyq}.

\begin{table}\centering

    \begin{tabular}{lcc}
    \hline
    & SM prediction & experiment  \\
    \hline
    $\Delta M_d [\mathrm{ps}^{-1}]$ & $0.533^{+0.022}_{-0.036}$ & $0.5064\pm0.0019$
    \\
    $\Delta M_s [\mathrm{ps}^{-1}]$ &  $18.4^{+0.7}_{-1.2}$ & $17.7656\pm0.0057$ \\
    \hline
    \end{tabular}

    \vspace{5mm}

    \begin{tabular}{cccc}
    \hline
    $i$ &  2& 3 & 4  \\
    \hline
        $f_{B_d}^2B_{B_d}^{(i)} [\mathrm{GeV}^2]$  & 
        $0.0288\pm0.0013$ & $0.0281\pm0.0020$ & $0.0387\pm0.0015$ \\ 
        $f_{B_s}^2B_{B_s}^{(i)} [\mathrm{GeV}^2]$ & 
        $0.0441\pm0.0017$ & $0.0454\pm0.0027$ & $0.0544\pm0.0019$ \\
    \hline    
    \end{tabular}

\caption{Top: SM prediction (based on weighted average)~\cite{DiLuzio:2019jyq} and experimental measurements~\cite{HFLAV:2016hnz,LHCb:2021moh} for the mass differences.
Bottom: Hadronic parameters $f_{B_q}^2B_{B_q}^{(i)}$ for the relevant operators reproduced from \cite{DiLuzio:2019jyq}.
}
\label{tab:Mixing}
\end{table}
For $B_s-\bar B_s$ mixing, the SM contribution in the first term of Eq.~(\ref{eq:DelMq}) is real and negative, since the imaginary part of $\lambda_t^s$ is small. As the central value of SM prediction $\Delta M_s^{\rm SM}$ is larger than the experimental measurement, a positive ALP contribution is preferred. For $B_d-\bar B_d$ mixing, $\lambda_t^d$ and thus the SM contribution in the first term of Eq.~(\ref{eq:DelMq}) is complex with similar magnitudes of the real and imaginary parts. In terms of the ALP couplings $c^{V,A}_{bq}/f_a$, the mass differences are
\begin{align}\label{eq:delMd}
\frac{\Delta M_d}{\mathrm{ps}} & = \left| 0.533 + \left(0.685\, (c_{bd}^V)^2 - 1.70\, (c_{bd}^A)^2 \right)e^{0.751 i\pi }\left(\frac{10~\mathrm{TeV}}{f_a}\right)^2 \right|
\\\label{eq:delMs}
    \frac{\Delta M_s}{\mathrm{ps}} & = \left| 18.4 + \left(0.938\,(c_{bs}^V)^2 - 2.58\, (c_{bs}^A)^2\right)e^{-0.988 i\pi }\left(\frac{10~ \mathrm{TeV}}{f_a}\right)^2 \right|
\end{align}
for $m_a=100$ GeV.
As mass differences are proportional to the square of the ALP couplings, the result is independent of the sign of the ALP couplings. Note, the coefficients of the axial-vector couplings are roughly twice as large as the ones for the vector couplings and enter with the opposite sign. This results in a cancellation of the ALP contribution to the meson mass splitting if the vector and axial-vector ALP couplings satisfy
\begin{align}\label{eq:delM_cancellation}
    (c_{bd}^V)^2 &\simeq 2.48 \,(c_{bd}^A)^2\;, &
    (c_{bs}^V)^2 & \simeq 2.75 \,(c_{bs}^A)^2
    \;.
\end{align}

\begin{figure}[tb!]
\centering

\includegraphics[width=0.47\textwidth]{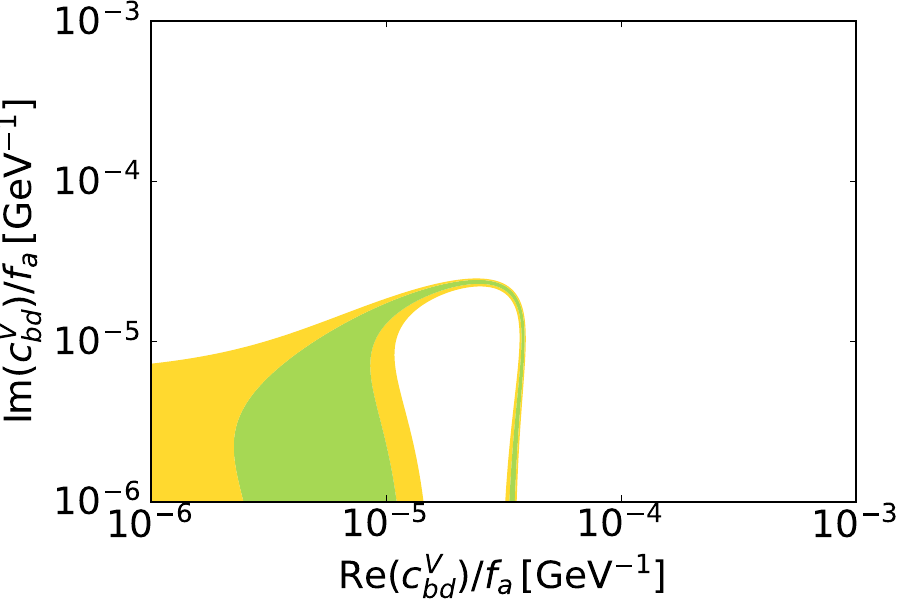}\hfill
\includegraphics[width=0.47\textwidth]{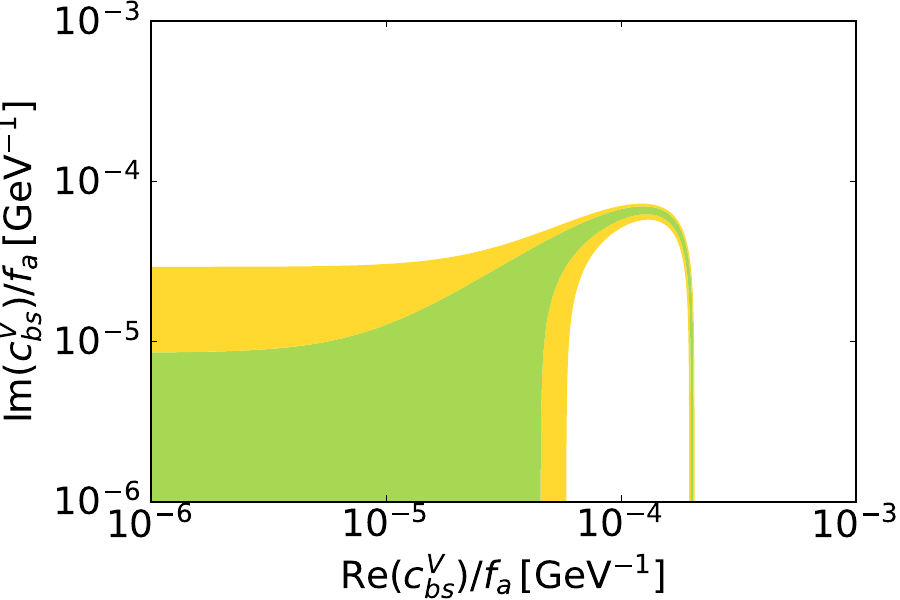}%

\vspace{5mm}

\includegraphics[width=0.47\textwidth]{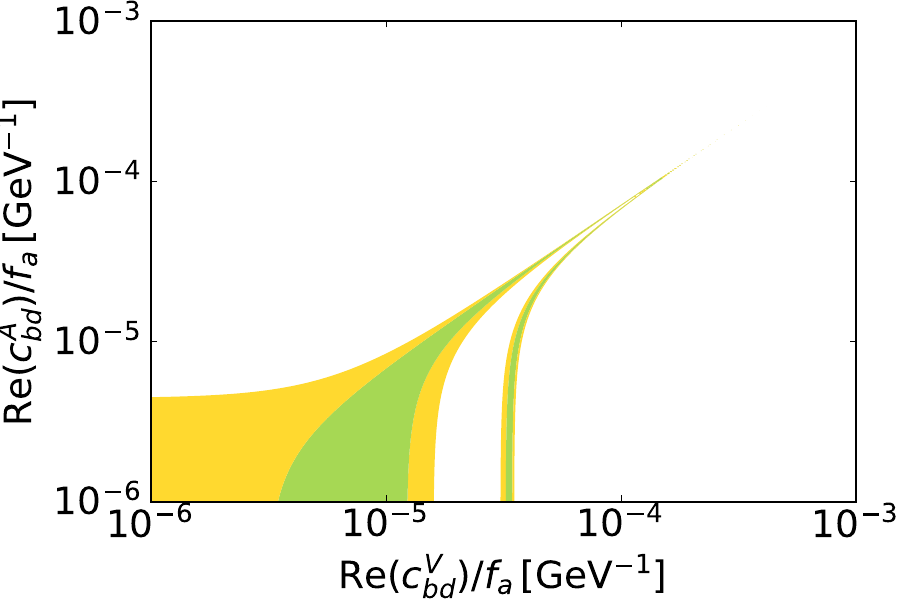}\hfill
\includegraphics[width=0.47\textwidth]{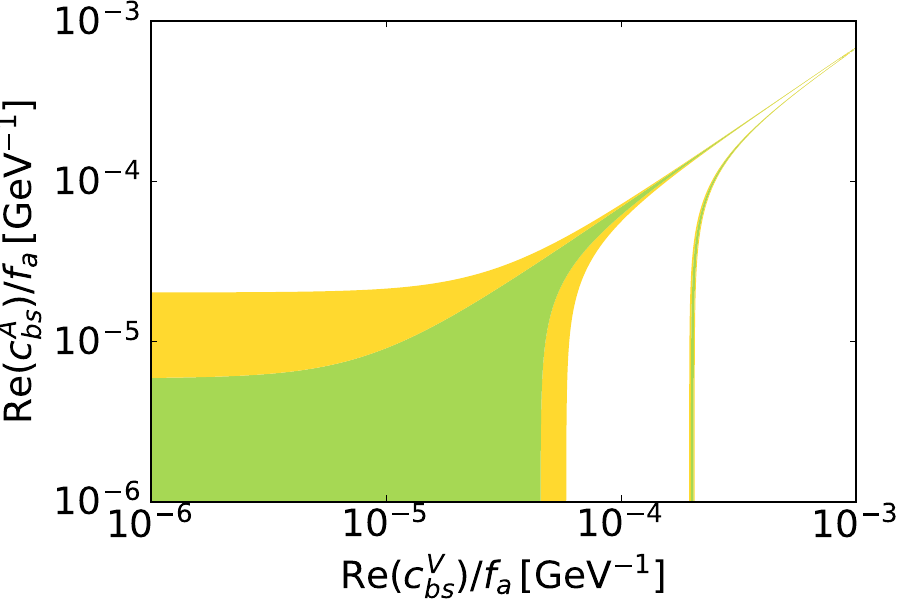}%

\vspace{5mm}

\includegraphics[width=0.47\textwidth]{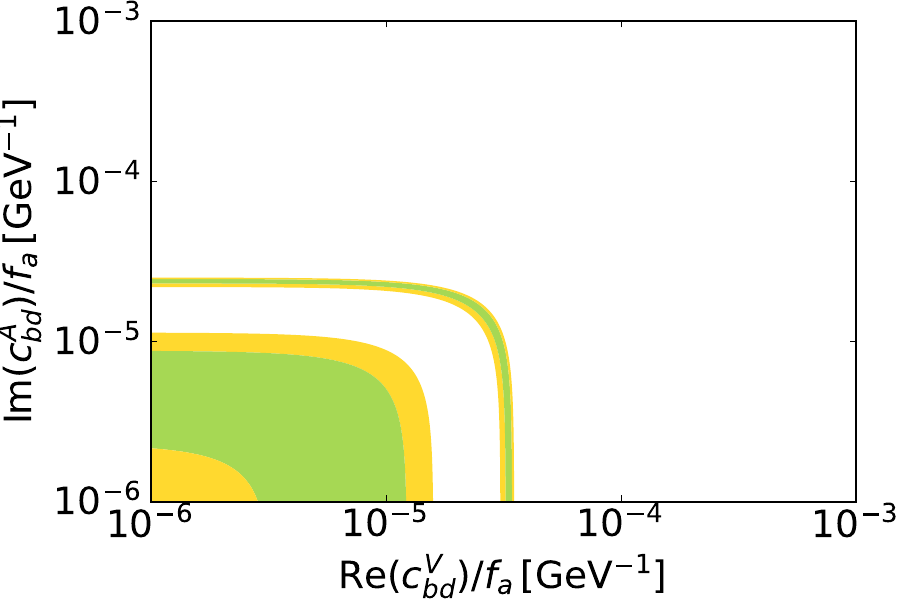}\hfill
\includegraphics[width=0.47\textwidth]{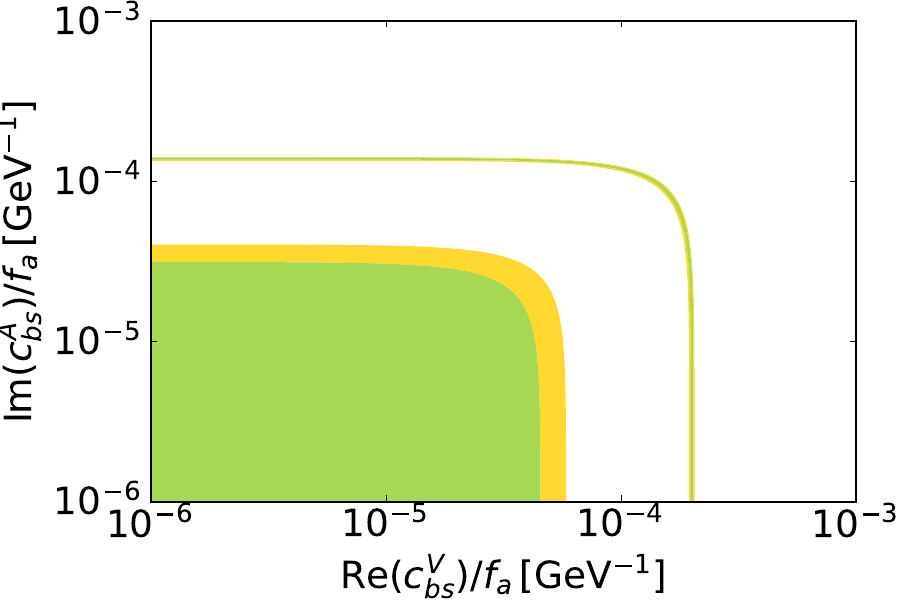}%

\caption{$1\sigma$ (green) and $2\sigma$ (yellow) contours of $c_{bq}^V/f_a$ (top), ${\rm Re}(c_{bq}^A)/f_a$ vs. ${\rm Re}(c_{bq}^V)/f_a$ (middle) and ${\rm Im}(c_{bq}^A)/f_a$ vs. ${\rm Re}(c_{bq}^V)/f_a$ (bottom) preferred by $B_d-\bar{B}_d$ (left) and $B_s-\bar{B}_s$ (right) mixing for fixed ALP mass $m_a=100$ GeV.
}
\label{fig:B-BbarC}
\end{figure}

In Fig.~\ref{fig:B-BbarC} we present the $1\sigma$ (green) and $2\sigma$ (yellow) preferred regions as a function of the positive real and imaginary parts of the ALP couplings $c_{bq}^{V,A}/f_a$ for an ALP mass $m_a=100$ GeV. The ALP couplings which are not explicitly shown have been set to zero. Negating one of the ALP couplings does not change the constraints. Although we only show the results for a fixed ALP mass $m_a=100$ GeV, it is straightforward to rescale the results to other ALP masses neglecting renormalization group effects, since the Wilson coefficients $C_2$, $\tilde C_2$ and $C_4$ are proportional to $(m_a f_a)^{-2}$.

Fig.~\ref{fig:B-BbarC} (top row) illustrates the dependence of the meson mass differences $\Delta M_q$ on the vector ALP couplings $c^V_{bq}/f_a$. For fixed real part, the imaginary part of the ALP coupling is bounded from above by
$\mathrm{Im}(c^V_{bd(bs)}) \lesssim 8 (22) \times 10^{-5} f_a/\mathrm{GeV} $.
For fixed small imaginary part, there are two disconnected preferred regions due to the destructive interference between the SM and NP contributions. This relaxes the upper bounds to
$\mathrm{Re}(c^V_{bd(bs)}) \lesssim 1 (6) \times 10^{-4} f_a/\mathrm{GeV}$.
The result for the axial-vector couplings is similar. Due to the relative minus sign in the expressions for the meson mass differences, the real and imaginary axis are swapped and the larger prefactor results in constraints which are more stringent by a factor
$\sim 1.57 (1.66)$ for $B_d-\bar B_d$ ($B_s-\bar B_s$) mixing.

The middle and lower panels of Fig.~\ref{fig:B-BbarC} show the strong correlation between the real and imaginary ALP couplings in the planes of ${\rm Re}(c_{bq}^A)/f_a$ vs.~${\rm Re}(c_{bq}^V)/f_a$ (middle) and ${\rm Im}(c_{bq}^A)/f_a$ vs.~${\rm Re}(c_{bq}^V)/f_a$ (bottom). The middle panel illustrates the possible cancellation between the real parts of the vector and axial-vector couplings for
$\mathrm{Re}(c_{bd}^V) \simeq 1.57\, \mathrm{Re}(c_{bd}^A)$ and $\mathrm{Re}(c_{bs}^V)\simeq 1.66\,\mathrm{Re}(c_{bs}^A)$.
While real axial-vector ALP couplings constructively interfere with the SM contribution, there is destructive interference between the real vector ALP couplings and the SM. The parameter space for imaginary ALP couplings has similar features as argued above.
The lower panel shows the region of parameter space, where vector and axial-vector couplings constructively interfere with each other.

While we agree with the order of magnitude for the constraints, the results in~\cite{Bauer:2021mvw} differ due to the missing phase information for the SM contribution in~\cite{Bauer:2021mvw} and an additional $\sim 1.4$ suppression of the NP contribution in the numerical expression for $\Delta M_q$ using $m_a=10$ GeV.\footnote{The numerical pre-factors can be straightforwardly calculated from the renormalization group equations, the normalization constants $\eta_i^q(\mu_b)$ and the running hadronic parameters $f_{B_q}^2 B_{B_q}^{(i)}(\mu_b) m_{B_q}$. Renormalization group corrections enhance the pre-factors of $C_2$ ($C_4$) by $1.153~(1.274)$ for $m_a=10$ GeV and the hadronic parameters are given in Table~\ref{tab:Mixing}.}

\section{FCNC search for invisibly decaying ALP at colliders}
\label{sec:inv}

\begin{figure}[thp!]
\begin{center}
      \minigraph{3.5cm}{-0.02in}{(1)}{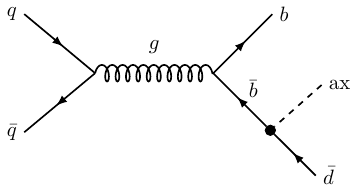}
      \minigraph{3.5cm}{-0.02in}{(2)}{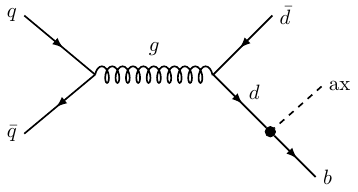}
      \minigraph{3.5cm}{-0.08in}{(3)}{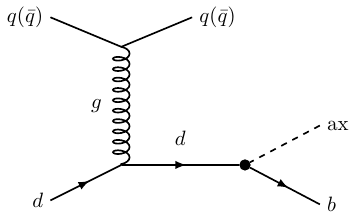}
      \minigraph{2.4cm}{-0.15in}{(4)}{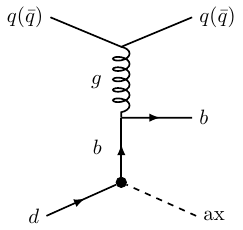}\\
      \vspace{0.5cm}
      \minigraph{3.5cm}{0.02in}{(5)}{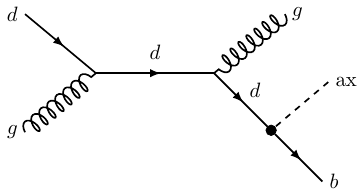}
      \minigraph{3.5cm}{0.02in}{(6)}{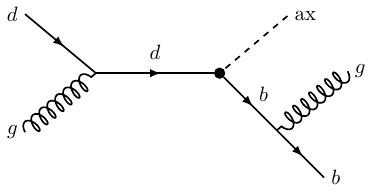}
      \minigraph{2.4cm}{-0.15in}{(7)}{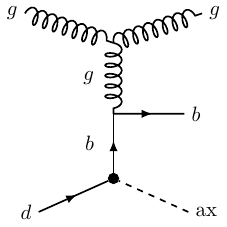}
      \minigraph{2.4cm}{-0.15in}{(8)}{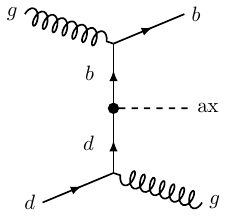}\\
      \vspace{0.5cm}
      \minigraph{3.5cm}{0.02in}{(9)}{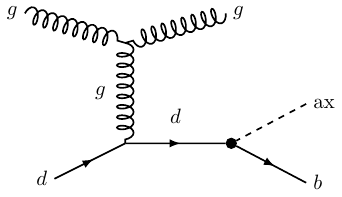}
      \minigraph{3.5cm}{0.02in}{(10)}{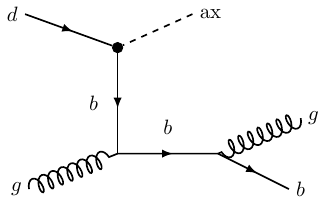}
      \minigraph{3.5cm}{0.02in}{(11)}{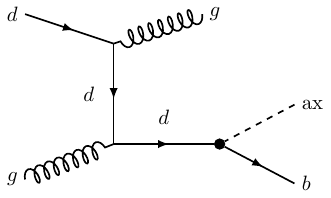}
      \minigraph{2.4cm}{-0.14in}{(12)}{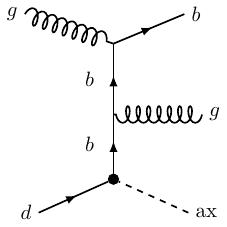}\\
      \vspace{0.5cm}
      \minigraph{3.2cm}{0.02in}{(13)}{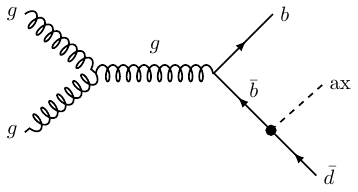}
      \minigraph{3.2cm}{0.02in}{(14)}{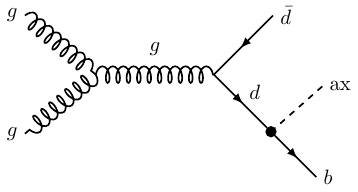}
      \minigraph{3.0cm}{0.02in}{(15)}{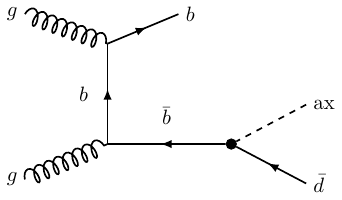}
      \minigraph{3.0cm}{0.02in}{(16)}{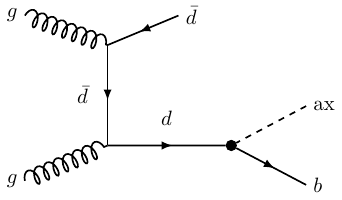}
      \minigraph{2.2cm}{-0.15in}{(17)}{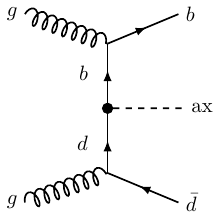}\\
\end{center}
\caption{The representative Feynman diagrams of the process $p~p \to b ~j~a$ induced by $b-d-a$ couplings at the LHC. The label ``ax'' denotes ALP $a$ and $q~(\Bar{q})= u,c,d,s~(\Bar{u},\Bar{c},\Bar{d},\Bar{s})$. }
\label{fig-FD}
\end{figure}

\begin{figure}[thp!]
\begin{center}
\minigraph{10cm}{-0.05in}{}{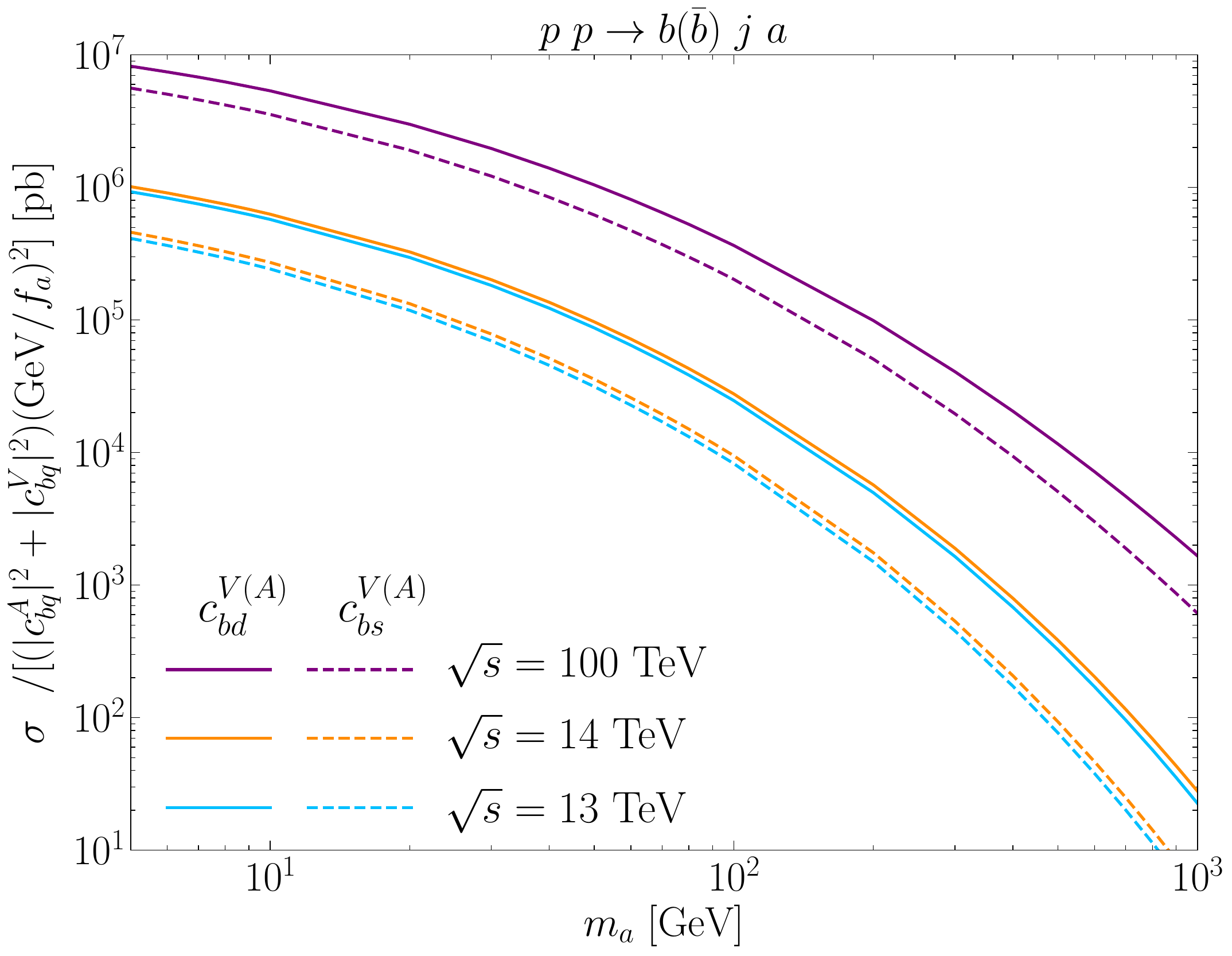}
\end{center}
\caption{
The parameter-independent cross section of $p~p \to b(\Bar{b}) ~j~ a$ as a function of $m_{a}$ with $\sqrt{s}=13$ (blue), 14 (orange) and 100 (purple) TeV, respectively. For comparison, we show two cases of FCNC with the ALP only coupled with $b$ quark and $d$ quark (solid line, $c_{bd}^{V(A)}$) and only coupled with $b$ quark and $s$ quark (dash line, $c_{bs}^{V(A)}$), respectively. The renormalization scale and factorization scale are taken as $\mu_R=\mu_F=\sqrt{\hat{s}}/2$ with $\sqrt{\hat{s}}$ being the partonic c.m. energy.
}
\label{fig-xsection}
\end{figure}

We first consider the production of ALP associated with one $b$ jet and one light jet $j$
\begin{equation}
p~p \to b (\Bar{b})~j~a\;.
\label{eqn-bjax}
\end{equation}
The representative Feynman diagrams of the process $p~p \to b ~j~a$ are shown in Fig.~\ref{fig-FD}. Here we only assume $b-d-a$ coupling for illustration and the charge conjugate diagrams are not displayed for simplicity. We define a ``bare'' cross section independent of the $b$-$q$-$a$ FCNC couplings as follows
\begin{eqnarray}
\sigma_0(bja)={\sigma(p~p \to b (\Bar{b}) ~j~a) \over (|c_{bq}^V|^2+|c_{bq}^A|^2) ({\rm GeV}/f_a)^2}\;.
\end{eqnarray}
The parameter-independent cross section $\sigma_0(bja)$ as a function of $m_a$ with $\sqrt{s}=13$ TeV (blue), 14 TeV (green) or 100 TeV (purple) is shown in Fig.~\ref{fig-xsection}, after applying the following parton-level cuts
\begin{eqnarray}
p_T(j,b)>20~{\rm GeV}\;,~~|\eta(j,b)|<3\;,~~\Delta R_{bj}>0.4\;.
\end{eqnarray}
We compare the results of both $b-d-a$ couplings $c_{bd}^{V(A)}$ (solid lines) and $b-s-a$ couplings $c_{bs}^{V(A)}$ (dashed lines). The small difference is due to the different parton distribution functions (PDFs) for $d$ and $s$ partons in Fig.~\ref{fig-FD} (3) - (12). Suppose the flavor-conserving couplings $\partial_\mu a\bar{q}\gamma^\mu\gamma_5q$ or $\partial_\mu a\bar{b}\gamma^\mu\gamma_5b$ exist, there would appear fake events containing $b~j~a$ by mis-identifying the quark jets. Here, we assume that the production processes are only induced by bottom quark flavor-violating couplings of ALP.

Suppose the tree-level ALP flavor-conserving and flavor-violating couplings are of the same order of magnitude, as discussed in Sec.~\ref{sec:theory}, the flavor-violating coupling generated via the four-fermion SM operators is more suppressed. Moreover, we expect that the production $q\bar{q}\to b s a$ via the effective operators has PDF suppression compared to our gluon fusion processes. Thus, the production via the SM effective operators and ALP flavor-conserving couplings can be ignored in our analysis.

We assume an invisible decay mode of the ALP.
The invisibly decaying ALP induces a large missing transverse energy $\cancel{E}_T$ (MET). We propose the search for invisible ALP associated with viable particles. The major SM backgrounds are thus
\begin{eqnarray}
p~p\to Z~j~j\;, Z~b~\bar{b}\;,
\end{eqnarray}
with the $Z$ boson's decay into neutrinos and the mis-identification of one jet $j\to b$ or $b\to j$. We also include reducible backgrounds $jjW^\pm$, $b\bar{b}W^\pm$ and $t\bar{t}$ with vetoed charged lepton from $W$ boson's leptonic decay.
Their K factors are 1.6 (2.5) for $b\bar{b}Z$ ($b\bar{b}W^\pm$)~\cite{FebresCordero:2009xzo}, 1.3 (2.3) for $jjZ$ ($jjW^\pm$)~\cite{Campbell:2003hd} and 1.8 for $t\bar{t}$~\cite{Ahrens:2011px}. The model file of ALP with FCNC couplings is produced by FeynRules~\cite{Alloul:2013bka} and is interfaced with MadGraph5\_aMC@NLO~\cite{Alwall:2014hca} to generate signal events. Both the signal and background events are then passed to Pythia 8~\cite{Sjostrand:2014zea} and Delphes 3~\cite{deFavereau:2013fsa} for parton shower and detector simulation, respectively.
The default Delphes cards for ATLAS, HL-LHC or FCC-hh are used for $b$-tagging efficiency. We pre-select the events with at least two jets and one of them tagged as $b$ jet satisfying
\begin{eqnarray}
n_j\geq 1\;,~~n_b\geq 1\;,~~p_T(j,b)>25~{\rm GeV}\;,~~|\eta(j,b)|<2.5\;.
\label{eqn-pre-select}
\end{eqnarray}
The jets are reconstructed using the anti-kT algorithm with $R=0.4$ in FastJet~\cite{Cacciari:2011ma}.
The $jjW^\pm$, $b\bar{b}W^\pm$ and $t\bar{t}$ backgrounds can be further reduced by vetoing the charged lepton with
\begin{eqnarray}
   n_\ell = 0\;,~~\ell=e,\mu\;.
\end{eqnarray}
After pre-selection and considering the K factors, the cross sections of different SM backgrounds at LHC with $\sqrt{s}=13$ TeV are $\sigma_{jjZ}=31.29$ pb, $\sigma_{bbZ}=7.39$ pb, $\sigma_{jjW}=100.77$ pb, $\sigma_{bbW}=2.48$ pb and $\sigma_{t\Bar{t}}=2.14$ pb, respectively.

The application of decision trees in multivariate analysis is a highly effective and increasingly popular method to distinguish the signal and background events. It is expected to
perform better than the traditional cut analysis in discrimination. We employ the Boosted Decision Tree (BDT) algorithm~\cite{Roe:2004na} implemented in XGBoost~\cite{Chen:2016btl} to analyze the events passing the above pre-selection cuts.
The kinematic observables used to train the BDT classifier are as follows
\begin{itemize}
\item transverse momentum, pseudo-rapidity and azimuthal angle of the final $b$ jet and light jet $j$: $p_T(b)$, $p_T(j)$, $\eta(b)$, $\eta(j)$, $\phi(b)$, $\phi(j)$ \;;
\item the difference of the pseudo-rapidity, azimuthal angle and the separation in angular space between the $b$ jet and light jet: $\Delta \eta(b,j)$, $\Delta \phi(b,j)$, $\Delta R(b,j)$\;;
\item the invariant mass of the $b$ jet and light jet $m_{bj}$\;;
\item the missing transverse energy $\cancel{E}_T$\;;
\item the sum of the transverse momenta of final visible objects $H_T=p_T(b)+p_T(j)$\;.
\end{itemize}
\begin{figure}[th!]
\begin{center}
      \minigraph{6cm}{-0.05in}{}{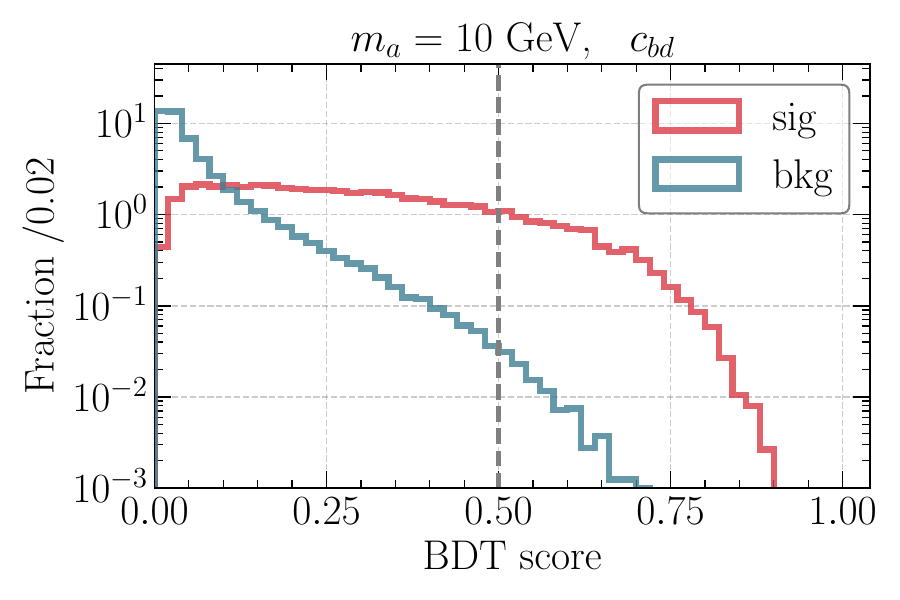}
      \minigraph{6cm}{-0.05in}{}{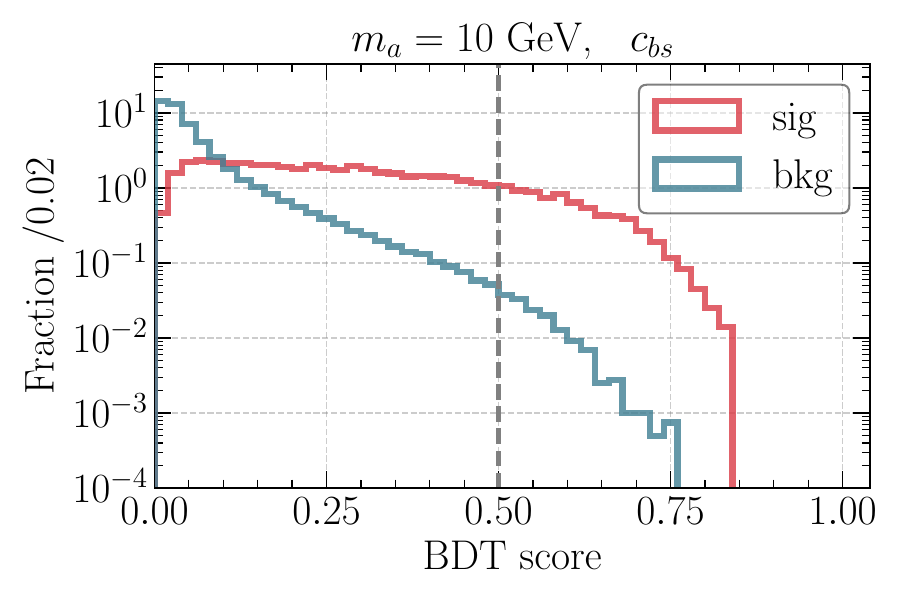}\\
      \minigraph{6cm}{-0.05in}{}{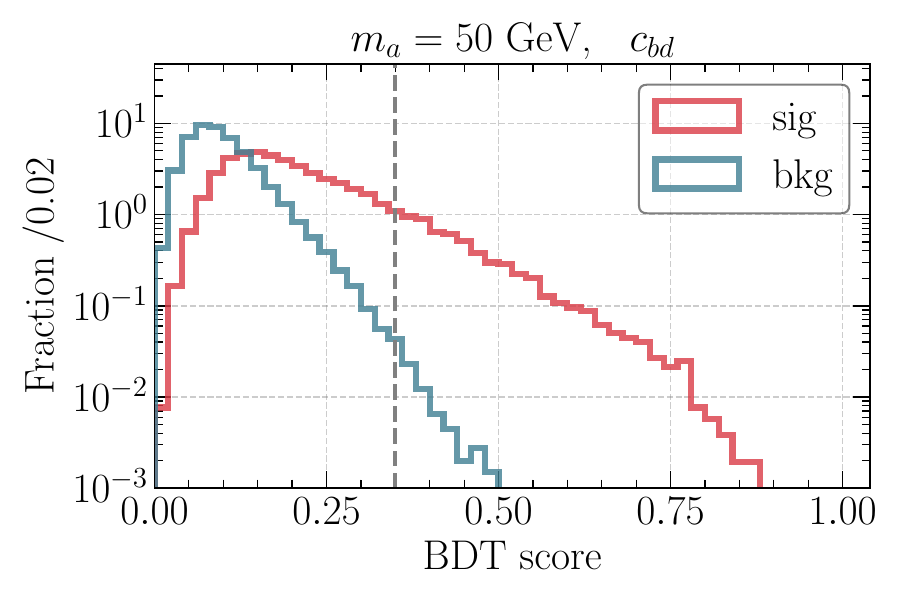}
      \minigraph{6cm}{-0.05in}{}{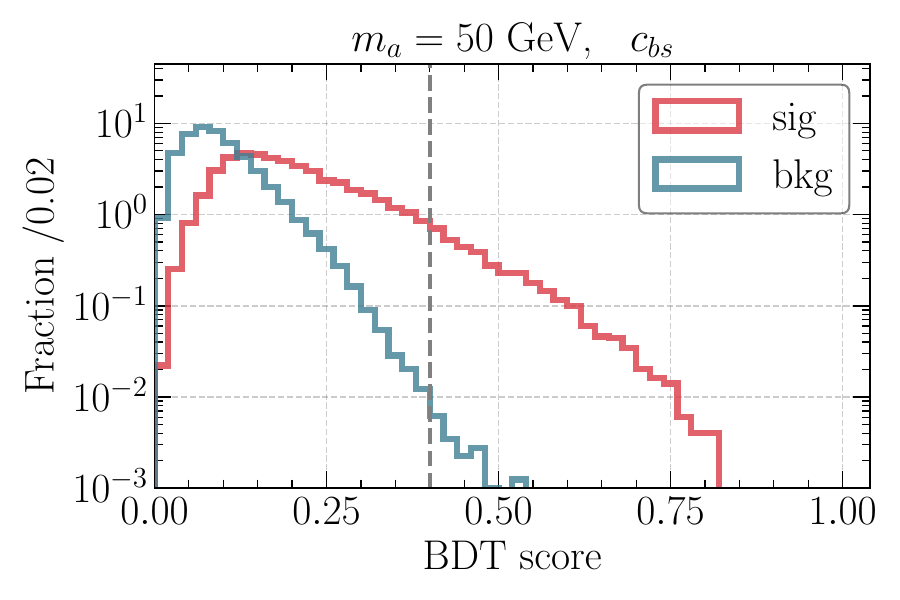}\\
      \minigraph{6cm}{-0.05in}{}{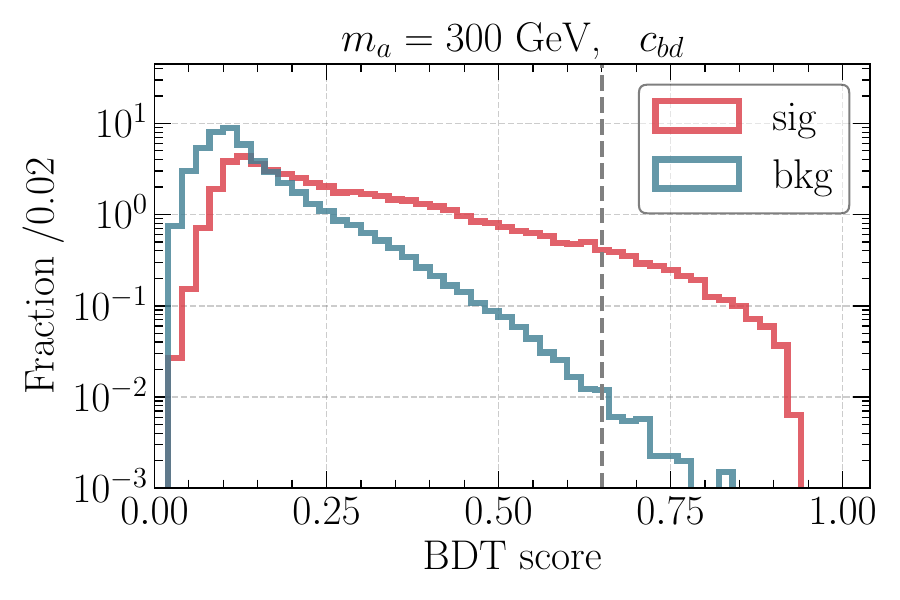}
      \minigraph{6cm}{-0.05in}{}{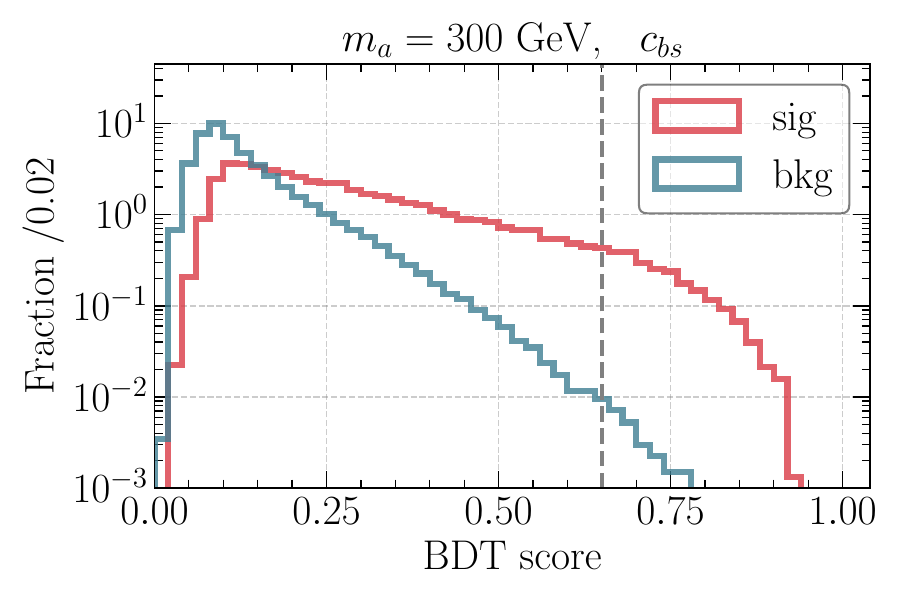}\\
      \minigraph{6cm}{-0.05in}{}{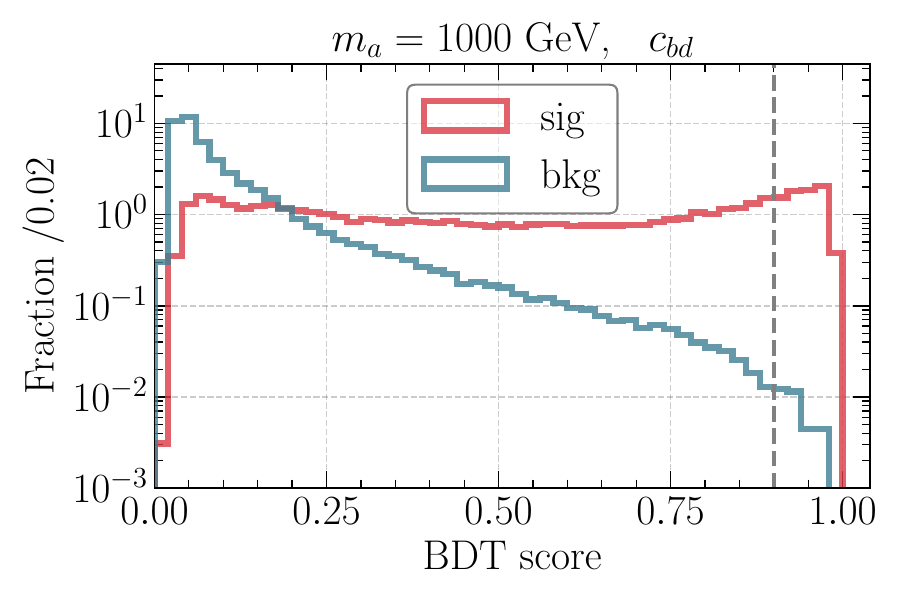}
      \minigraph{6cm}{-0.05in}{}{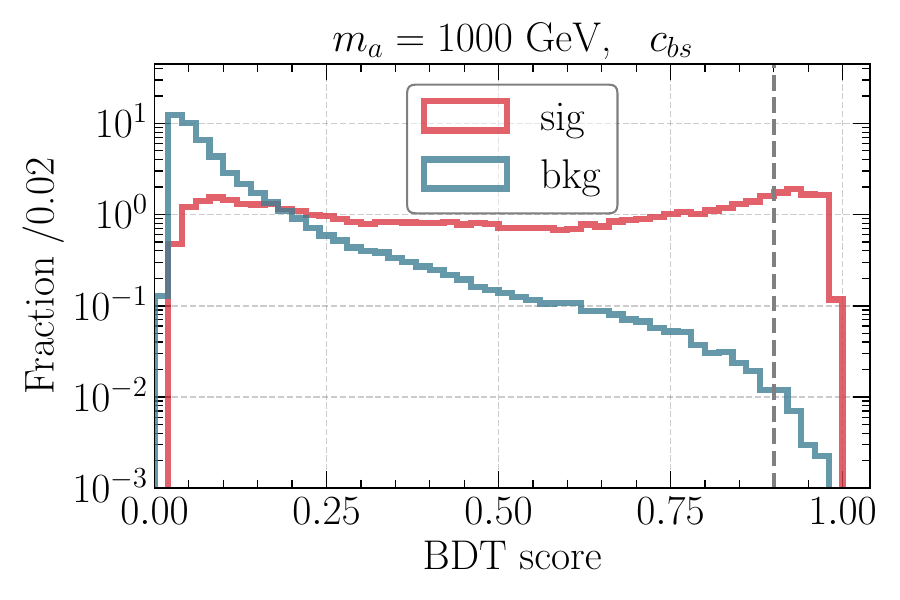}
\end{center}
\caption{ The BDT response score distribution of signal $bj+{\rm inv.}$ (red) and total SM background (blue) with $m_a=10,~50,~300$ and 1000 GeV (from top to bottom) at LHC with $\sqrt{s}=13$ TeV and $\mathcal{L}=300~{\rm fb}^{-1}$. The ``sig'' or ``bkg'' in the legend represents the signal or the sum of SM backgrounds. The grey dashed line indicates the BDT cut that maximizes the significance with fixed $|c_{bd}^{V(A)}|/f_a = 1~{\rm TeV}^{-1}$ (left four panels) or $|c_{bs}^{V(A)}|/f_a = 1~{\rm TeV}^{-1}$ (right four panels).}
\label{fig:invBDT-score}
\end{figure}
Then, we take the probability that the BDT algorithm classifies an event as signal as the BDT response score. The BDT response score distributions of the signal (red) and total SM background (blue) with c.m. energy $\sqrt{s}=13$ TeV and luminosity $\mathcal{L}=300~{\rm fb}^{-1}$ are shown for benchmark masses $m_a=10,~50,~300$ and $1000$ GeV, and $c_{bq}^{V(A)}/f_a=1~{\rm TeV}^{-1}$ in Fig.~\ref{fig:invBDT-score}. We obtain the BDT cut by maximizing the following significance~\cite{Zyla:2020zbs}
\begin{eqnarray}
\mathcal{S}=\sqrt{2\Big((s+b){\rm ln}\Big(1+{s\over b}\Big)-s\Big)}\;,
\label{eqn-significance}
\end{eqnarray}
where $s$ and $b$ are the signal and background event expectations, respectively. The obtained BDT cut, cut efficiencies of signal and backgrounds and the significance are collected in Table~\ref{tab:invBDT} for the above benchmarks. We put the results for HL-LHC and FCC-hh in Appendix~\ref{App:inv}. The BDT cut becomes more severe as the ALP mass increases except for low masses $m_a\simeq 10$ GeV.

\begin{table}[tb!]
\centering
\resizebox{\textwidth}{!}{
\begin{tabular}{c|c|c|c|c|c|c|c|c}
    \hline
     $m_a$ & BDT cut & $\epsilon_{\rm sig.}$ & $\epsilon_{jjZ}$ & $\epsilon_{bbZ}$ & $\epsilon_{jjW}$ & $\epsilon_{bbW}$ & $\epsilon_{t\bar{t}}$ &  $\mathcal{S}_{\rm max}$  \\
    \hline
    10   GeV& 0.50 & 1.57$\times10^{-1}$ & 2.86$\times10^{-3}$ & 2.94$\times10^{-3}$ & 1.95$\times10^{-3}$ & 2.29$\times10^{-3}$ & 8.12$\times10^{-5}$ &  1.64$\times10^{1}$\\
    50   GeV& 0.35 & 1.30$\times10^{-1}$ & 5.04$\times10^{-3}$& 1.17$\times10^{-3}$ & 4.43$\times10^{-3}$ & 1.11$\times10^{-3}$ & 1.22$\times10^{-4}$ & 2.06$\times10^{0}$  \\
    300  GeV& 0.65 & 6.49$\times10^{-2}$ & 3.15$\times10^{-3}$ & 6.05$\times10^{-4}$ & 7.09$\times10^{-4}$ & 1.81$\times10^{-4}$ & 4.47$\times10^{-4}$ &  5.50$\times10^{-2}$ \\
    1000 GeV& 0.90 & 1.57$\times10^{-1}$& 2.00$\times10^{-3}$  & 4.50$\times10^{-4}$ &  3.55$\times10^{-4}$ & 1.13$\times10^{-4}$ &  6.50$\times10^{-4}$ & 2.92$\times10^{-3}$ \\
    \hline
    \hline
    10   GeV& 0.50 & 1.80$\times10^{-1}$ & 3.78$\times10^{-3}$ & 3.24$\times10^{-3}$ &  1.24$\times10^{-3}$  & 2.67$\times10^{-3}$ & 4.06$\times10^{-5}$ &  8.19$\times10^{0}$\\
    50   GeV& 0.40 & 7.65$\times10^{-2}$ & 8.02$\times10^{-4}$ & 3.03$\times10^{-4}$ & 7.09$\times10^{-4}$ & 2.04$\times10^{-4}$& $-$  & 1.04$\times10^{0}$  \\
    300  GeV& 0.65 & 4.50$\times10^{-2}$ & 2.23$\times10^{-3}$ & 3.76$\times10^{-4}$ & 1.77$\times10^{-4}$ & 9.05$\times10^{-5}$ & 1.22$\times10^{-4}$ & 1.54$\times10^{-2}$ \\
    1000 GeV& 0.90 & 1.44$\times10^{-1}$& 1.94$\times10^{-3}$  & 4.04$\times10^{-4}$& 5.32$\times10^{-4}$ & 9.05$\times10^{-5}$ & 5.28$\times10^{-4}$ & 4.46$\times10^{-4}$ \\
    \hline
\end{tabular}}
\caption{The BDT cut, cut efficiencies and achieved maximal significance in Eq.~(\ref{eqn-significance}) for the signal $bj+{\rm invisible}$ and SM backgrounds at LHC with $\sqrt{s}=13$ TeV and $\mathcal{L}=300~{\rm fb}^{-1}$. The benchmark masses are $m_a=10,~50,~300$ and 1000 GeV and the parameter is fixed as $|c_{bd}^{V(A)}|/f_a = 1~{\rm TeV}^{-1}$ (above the double line) or $|c_{bs}^{V(A)}|/f_a = 1~{\rm TeV}^{-1}$ (below the double line). The label ``$-$'' denotes the background at negligible level.}
\label{tab:invBDT}
\end{table}

\begin{figure}[htb!]
\begin{center}
\minigraph{7.5cm}{-0.05in}{}{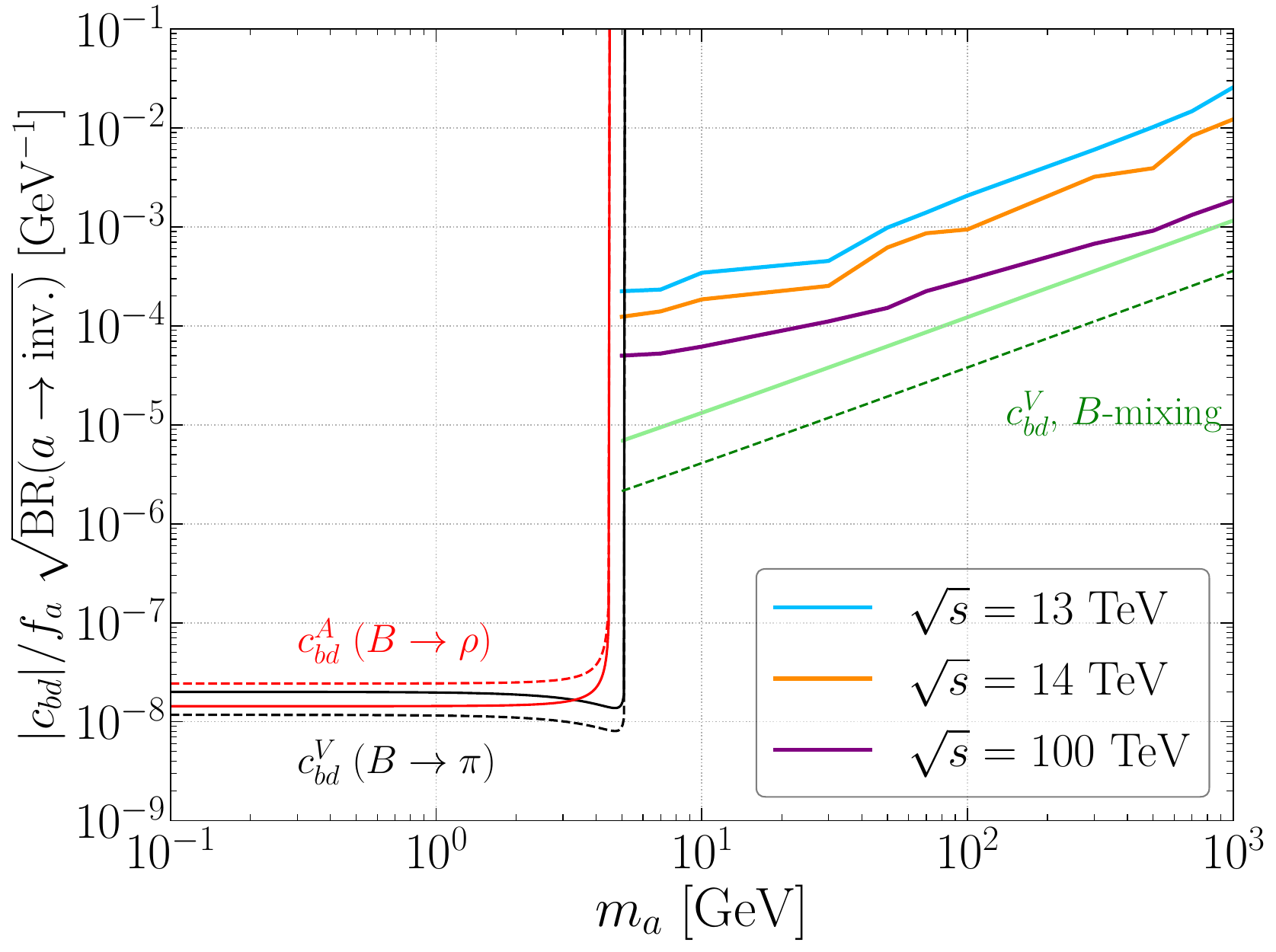}
\minigraph{7.5cm}{-0.05in}{}{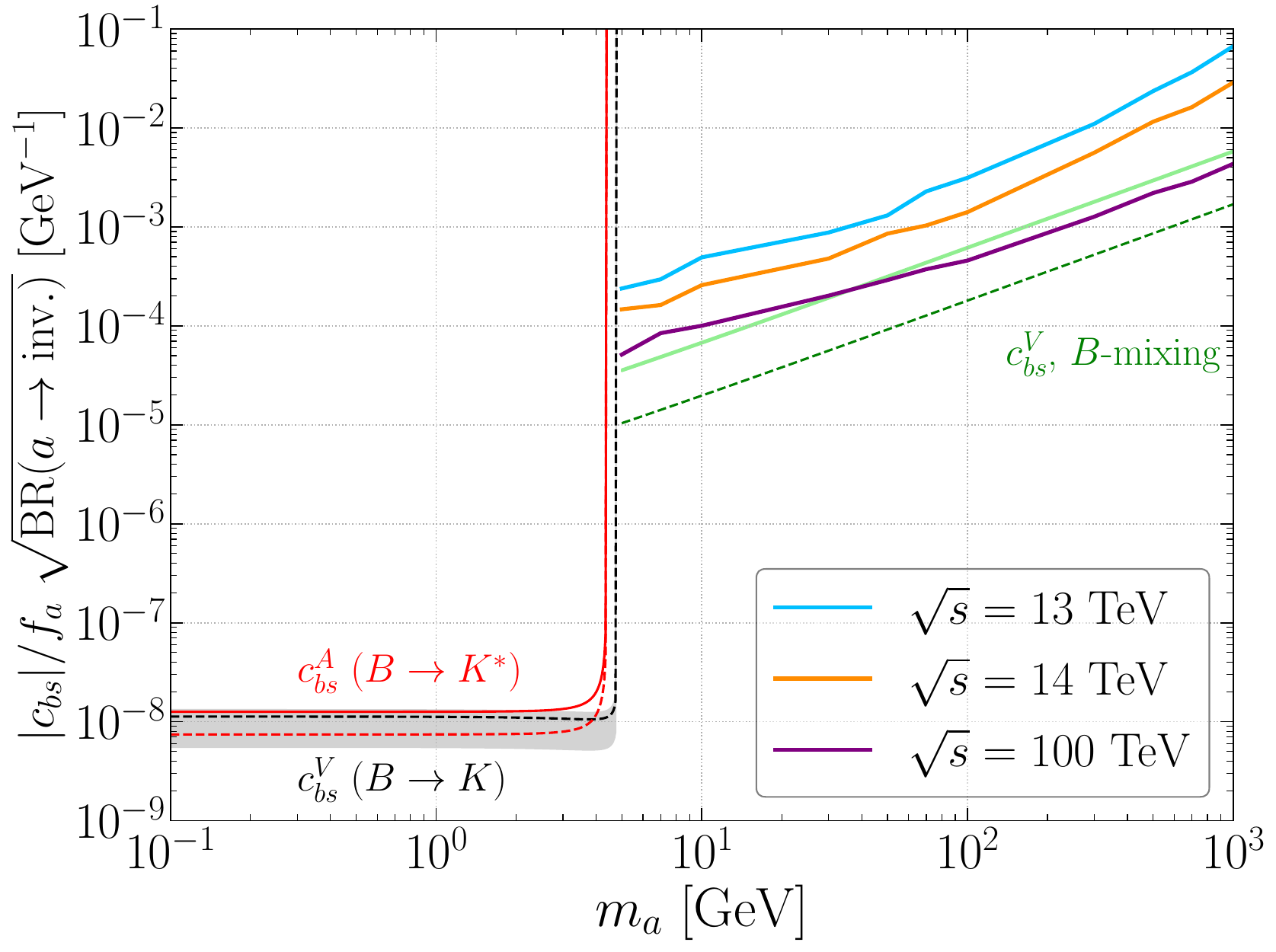}
\end{center}
\caption{The constraints from $B$ meson decays for $m_a\lesssim m_B$ and the 2$\sigma$ exclusion limits at hadron colliders for $m_a\gtrsim m_B$. The 2$\sigma$ exclusion limits for $|c_{bd}^{V(A)}|/f_a \sqrt{{\rm BR}(a\to {\rm inv.})}$ (left) and $|c_{bs}^{V(A)}|/f_a \sqrt{{\rm BR}(a\to {\rm inv.})}$ (right) at LHC with $\sqrt{s}=13$ TeV and $\mathcal{L}=300~{\rm fb}^{-1}$ (blue lines), HL-LHC with $\sqrt{s}=14$ TeV and $\mathcal{L}=3~{\rm ab}^{-1}$ (orange lines) or FCC-hh with $\sqrt{s}=100$ TeV and $\mathcal{L}=30~{\rm ab}^{-1}$ (purple lines). The $B$ meson decays include $B^+\to \pi^+ +{\rm inv.}$ (black solid curve), $B^+\to K^{\ast +}+{\rm inv.}$ or $B^+\to \rho^+ +{\rm inv.}$ (red solid curve), $B^0\to K^0+{\rm inv.}$ or $B^0\to \pi^0+{\rm inv.}$ (black dashed curve), $B^0\to K^{\ast 0}+{\rm inv.}$ or $B^0\to \rho^0 +{\rm inv.}$ (red dashed curve). The 2$\sigma$ region preferred by $B^+\to K^+ +{\rm inv.}$ is given by the gray band. The constraints of $B$ meson oscillations for the benchmark choices of $c_{bd}^V=e^{i\pi/8}$ and $c_{bs}^V=1$ (green band and dashed line as the upper boundary of preferred region) also maximize the interference with the SM.
}
\label{fig:invcva-limit}
\end{figure}

In Fig.~\ref{fig:invcva-limit}, for $m_a\gtrsim 5$ GeV, we show the 2$\sigma$ exclusion limits on $|c_{bq}^{V(A)}|/f_a \sqrt{{\rm BR}(a\to {\rm inv.})}$ with $q=d$ (left) and $q=s$ (right) at LHC (blue lines), HL-LHC (orange lines) and FCC-hh (purple lines). Here we assume either $|c_{bq}^{V}|$ or $|c_{bq}^{A}|$ is present in order to compare with the low-energy $B$ decay constraints for $m_a\lesssim 5$ GeV. The limits from LHC and HL-LHC are at the level of $10^{-4}~{\rm GeV}^{-1}$ and $10^{-3}~{\rm GeV}^{-1}$ for $m_a=10$ GeV and $m_a=100$ GeV, respectively. The FCC-hh can push the limits lower by 3 - 6 times. According to Sec.~\ref{sec:BKinv}, when $m_a\lesssim 5$ GeV, the preferred region or upper limits are also shown for $c^{V(A)}_{bs}/f_a \sqrt{{\rm BR}(a\to {\rm inv.})}$ by $B\to K/K^\ast+{\rm inv.}$ (left) and $c^{V(A)}_{bd}/f_a \sqrt{{\rm BR}(a\to {\rm inv.})}$ by $B\to \pi/\rho+{\rm inv.}$ (right). We assume that the mass of the invisible particle is smaller than $m_a/2$. The upper limit from $B^+\to \pi^+ +{\rm inv.}$ is shown as black solid curve, $B^+\to K^{\ast +}+{\rm inv.}$ or $B^+\to \rho^+ +{\rm inv.}$ as red solid curve, $B^0\to K^0+{\rm inv.}$ or $B^0\to \pi^0+{\rm inv.}$ as black dashed curve, $B^0\to K^{\ast 0}+{\rm inv.}$ or $B^0\to \rho^0 +{\rm inv.}$ as red dashed curve. The 2$\sigma$ region preferred by $B^+\to K^+ +{\rm inv.}$ is given by the gray band.
These low-energy constraints are at least four orders of magnitude more stringent than the collider bounds. Moreover, we also compare with the $B-\bar{B}$ mixing constraints on $c_{bq}/f_a$ for heavy ALP (green band and dashed line as the boundary of preferred region). Here, we only show the favored regions for the benchmark choices of $c_{bd}^V=e^{i\pi/8}$ and $c_{bs}^V=1$, which maximize the destructive interference with the SM. 
It turns out that the FCNC search for invisibly decaying ALP at hadron colliders is not able to reach the parameter space preferred by the $B$ meson oscillations for these benchmarks unless there is a cancellation between the vector and axial-vector ALP couplings to high precision.

\begin{figure}[htb!]
\begin{center}
      \minigraph{6.3cm}{-0.05in}{}{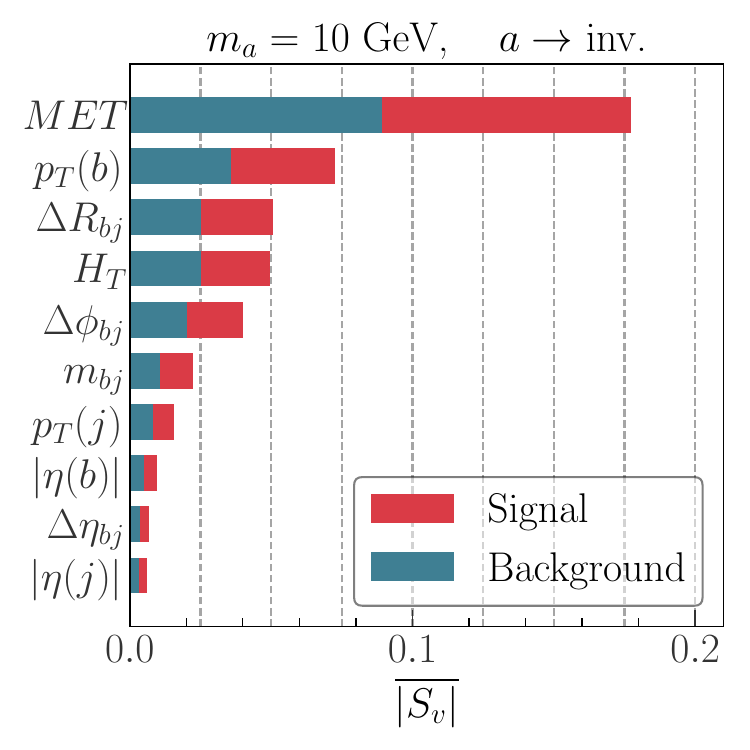}
      \minigraph{6.3cm}{-0.05in}{}{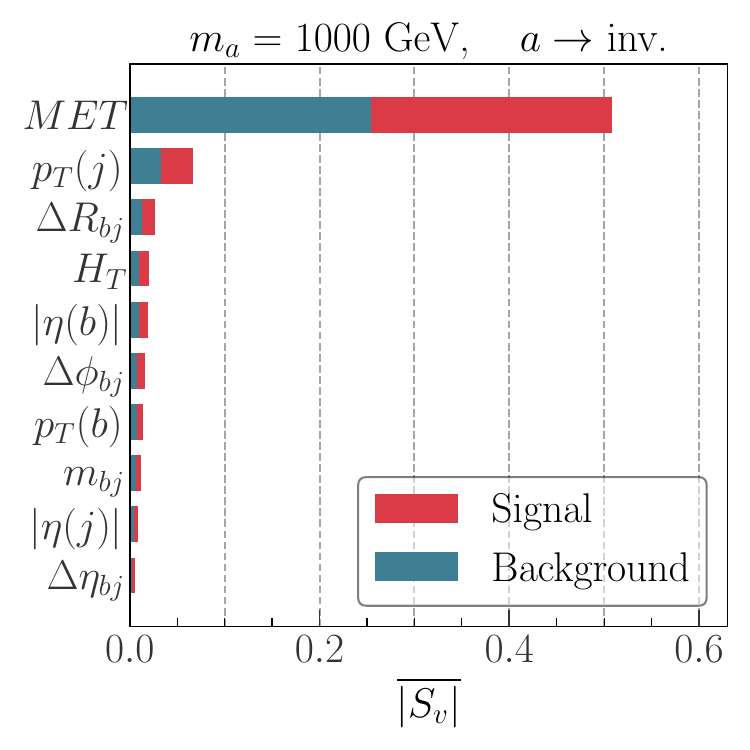}\\
      \minigraph{6.3cm}{-0.05in}{}{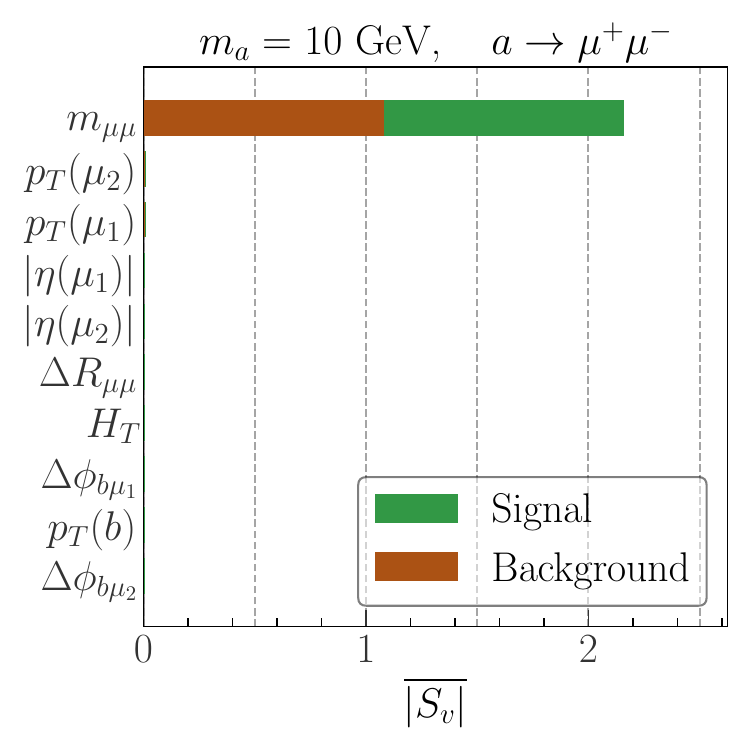}
      \minigraph{6.3cm}{-0.05in}{}{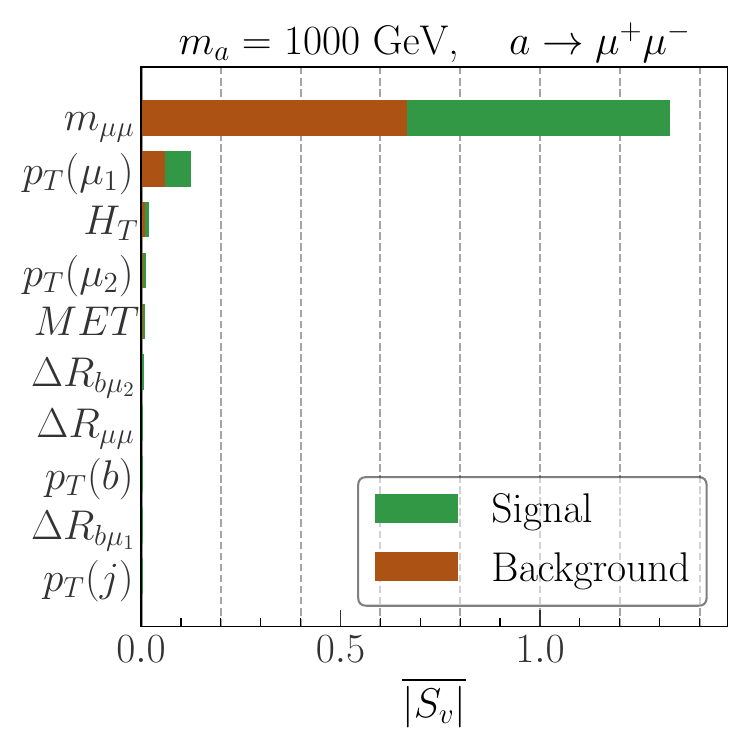}\\
      \minigraph{6.3cm}{-0.05in}{}{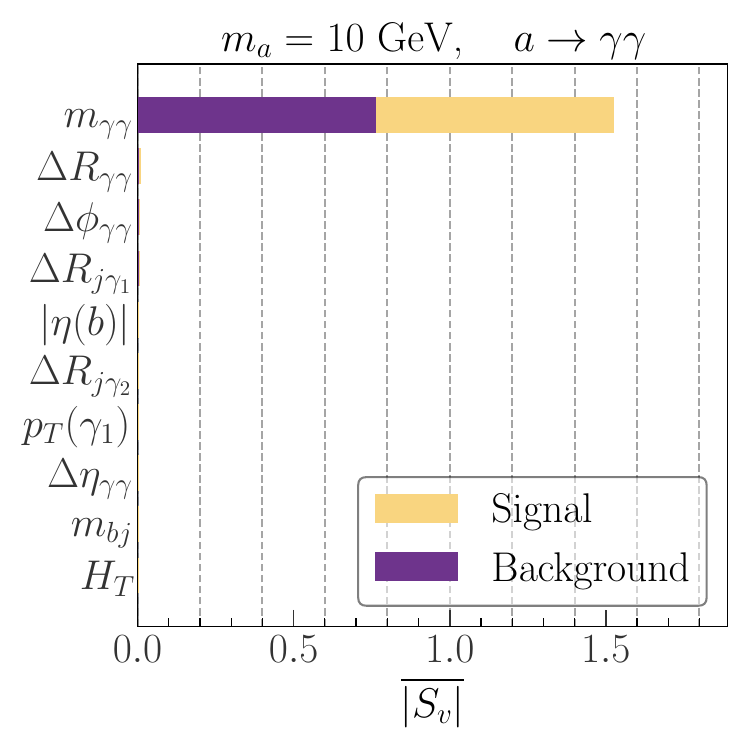}
      \minigraph{6.3cm}{-0.05in}{}{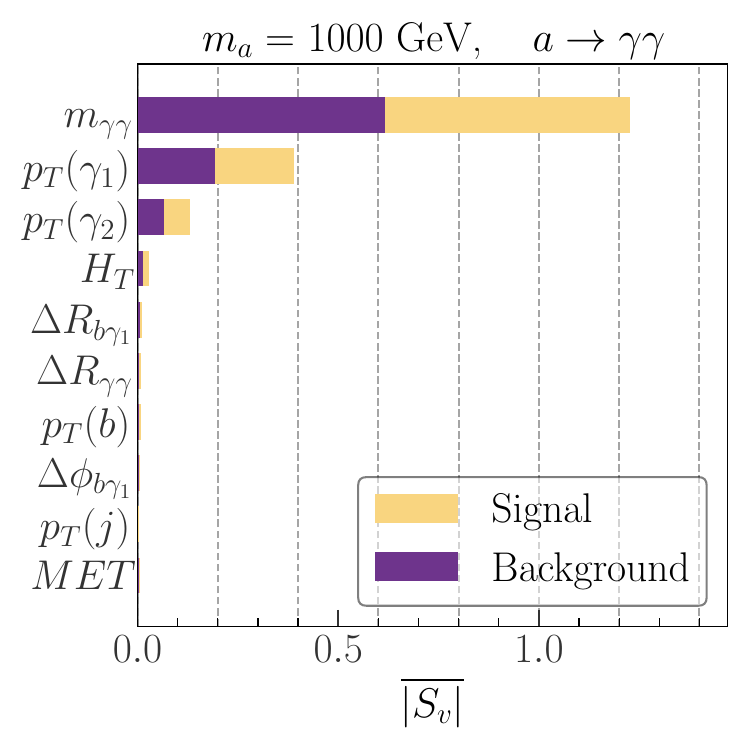}
\end{center}
\caption{The Shapley values of the kinematic observables which are considered to train the BDT classifier with the mass benchmarks $m_a=10$ GeV (left) and $1000$ GeV (right) for the decay channel $a\to$ inv. (top panels), $a \to \mu^+ \mu^-$ (middle panels) and $a\to \gamma\gamma$ (bottom panel) at LHC. Only ten observables with the highest Shapley values are shown. The ``${\rm MET}$'' represents the transverse missing energy.}
\label{fig:shap-value}
\end{figure}

When the machine learning is applied in the research of theoretical particle physics, it is often thought to be a black-box algorithm due to the lack of interpretability. However, the Shapley value can play as an effective quantity to better understand the application of machine learning in the phenomenological study of particle physics~\cite{Grojean:2020ech}.
As a well-known concept in cooperative game theory, the Shapley values~\cite{Shapley:1951} were introduced by Shapley in the 1950s to solve the problem of fairly allocating the payoffs to each player in a n-player cooperative game based on their respective contributions.
In a cooperative game characterized by $(v,N)$ with n players, $N=\{1,\cdots,n \}$ is a set of players in the game. $T$ as a subset of $N$ refers to a coalition and the largest coalition is $N$, i.e., $ T \subseteq N $.
$v$ is the characteristic function that maps each coalition to a real number. $v(T)$ refers to the payoff of this coalition and $v(T \cup \{ i \})-v(T)$ is the marginal contribution of the $i$-th player to the coalition $T$ not containing $i$. The Shapley value after considering all possible subsets $T$ not containing $i$, i.e. the payoff of the $i$-th player, is defined as~\cite{Christoph:2023,Alasfar:2022vqw}
\begin{align}
    S_v^i= \sum\limits_{T \subseteq N / \{ i\}} \frac{|T|!(n-|T|-1)!}{n!}(v (T \cup \{ i \})-v(T)), ~~i=1,\cdots,n\;,
\end{align}
where $|T|$ is the cardinality of coalition $T$.
For this work, the Shapley values can serve as a tool to easily understand which of the kinematic observables are more important for the separation of the signal from the SM background. For a single event with n kinematic variables, there are n associated Shapley values $S_v$. A positive $S_v$ indicates that the event is more likely to belong to a certain channel, while a negative value means that the event is less likely to belong to this channel. We compute the average of the absolute Shapley values $\overline{|S_v|}$ of all events for each kinematic observable and show ten of them with the highest Shapley values for benchmark masses $m_a=10$ (left panels) and $1000$ GeV (right panels) at LHC in Fig.~\ref{fig:shap-value}.
For the invisibly decaying ALPs (top panels), as expected, the missing transverse energy (MET) with the highest Shapley value is the most important observable to discriminate the signal and background events. The discrimination power of MET is dominant for large $m_a$ around and above 1000 GeV, for which the MET value is also large for the signal. For smaller $m_a$ around 10 GeV and relatively moderate MET value, additional observable such as $p_T(b)$ also helps to distinguish ALP from SM background such as $jj(bb)Z$.
For the decay channels $a \to \mu^+\mu^-$ (middle panels) and $a \to \gamma \gamma$ (bottom panels) to be analyzed in Sec.~\ref{sec:dilepton} and Sec.~\ref{sec:diphoton}, due to the narrow resonance of ALP,
it is evident that the invariant mass of the final muon $m_{\mu\mu}$ or photon pair $m_{\gamma\gamma}$ far surpasses the importance of other observable in discrimination power. With larger $m_a$ though, the nevertheless model-dependent total width of the ALP also grows, and secondary observable such as $p_T(\mu,\gamma)$ further helps to distinguish ALP from SM background.
The result of Shapley values is consistent with our expectations, indicating the importance of kinematic observables in each channel.

Finally, we comment on the validity of effective ALP theory. The approximation of EFT is valid for $f_a > m_a/4\pi$. The collider limits are 
compatible with EFT validity. The larger couplings $c_{bq}/f_a$ will violate the perturbative unitarity. Then, the EFT expansion breaks down and cannot represent a reliable description of an underlying theory. Moreover, the EFT validity at colliders further requires that the decay constant should be larger than the partonic c.m. energy of the subprocesses in the signal events, i.e., $f_a> \sqrt{\hat{s}}$. However, one is not able to directly measure $\sqrt{\hat{s}}$ at hadron colliders. We can approximately take the invariant mass of final visible states as $\sqrt{\hat{s}}$, for instance $M_{\rm inv.}=M_{bj\mu\mu}$ or $M_{bj\gamma\gamma}$ for the two jets and the visible decay products of ALP. Then, the c.m. energy scale highly depends on the ALP mass and increases as the mass gets larger. Although it is not practical to impose the cut $f_a> M_{\rm inv.}$ for the signal events, we should note that this cut effect would become weaker for larger $f_a$ and the collider limits would be relaxed for very heavy ALP mass. For a roughly quantitative illustration, as shown in the collider exclusion figures, we consider a horizontal line at $10^{-3}~{\rm GeV}^{-1}$ representing the typical reaction scale for LHC and HL-LHC, and $10^{-4}~{\rm GeV}^{-1}$ for FCC-hh. One can see that a majority of the collider limits in both dimuon and diphoton channels fulfill the EFT validity.

\clearpage
\section{FCNC search for ALP decaying to dimuon at colliders}
\label{sec:dilepton}

We suppose that the ALP promptly decays into a muon-antimuon pair, i.e., $a\to \mu^+\mu^-$.
We pre-select the events containing at least two muons and two jets with one of them tagged as $b$ jet satisfying
\begin{eqnarray}
n_\mu\geq 2\;,~n_j\geq 1\;,~n_b\geq 1\;,~p_T(j,b)>25~{\rm GeV}\;,~p_T(\mu)>10~{\rm GeV}\;,~|\eta(\mu,j,b)|<2.5\;.
\end{eqnarray}
We also set the maximal missing transverse energy as
\begin{eqnarray}
\cancel{E}_T<30~{\rm GeV}\;.
\end{eqnarray}
The major SM backgrounds include
\begin{eqnarray}
p~p\to jjZ/\gamma^\ast (\to \mu^+\mu^-)\;,~~bbZ/\gamma^\ast (\to \mu^+\mu^-)\;,~~jjWW\;,~~bbWW\;,~~t\bar{t}\;,
\end{eqnarray}
with $W$ boson or top quark's leptonic decay. The K factors for $jjWW$ and $bbWW$ are 1.19~\cite{Djamaa:2020afx} and 1.12~\cite{Denner:2012yc}, respectively.
After pre-selection and considering the K factors, the cross sections of different SM backgrounds at LHC with $\sqrt{s}=13$ TeV are $\sigma_{jjZ/\gamma}=1.47$ pb, $\sigma_{bbZ/\gamma}=0.45$ pb, $\sigma_{jjWW}=5.02\times10^{-4}$ pb, $\sigma_{bbW}=0.16$ pb and $\sigma_{t\Bar{t}}=0.24$ pb, respectively.
After including the final muons from axion decay, the kinematic observables considered to train the BDT classifier are as follows
\begin{itemize}
\item transverse momentum, pseudo-rapidity and azimuthal angle of the final $b$ jet, light jet $j$ and muons $\mu$ : $p_T(b)$, $p_T(j)$, $p_T(\mu)$, $\eta(b)$, $\eta(j)$, $\eta(\mu)$, $\phi(b)$, $\phi(j)$, $\phi(\mu)$\;;
\item the difference of the pseudo-rapidity, azimuthal angle and the separation in angular space between each two of the four final states $\Delta \eta_{ij}$, $\Delta \phi_{ij}$, $\Delta R_{ij}$, where $i,j= b,~j,~\mu_1,~\mu_2~~(i\neq j)$\;;
\item the invariant mass of the $b$ jet and light jet $m_{bj}$, and the two final muons $m_{\mu\mu}$ \;;
\item the missing transverse energy $\cancel{E}_T$\;;
\item the sum of the transverse momenta of final visible objects $H_T=p_T(b)+p_T(j)+p_T(\mu_1)+p_T(\mu_2)$\;.
\end{itemize}

\begin{figure}[t!]
\begin{center}
      \minigraph{6cm}{-0.05in}{}{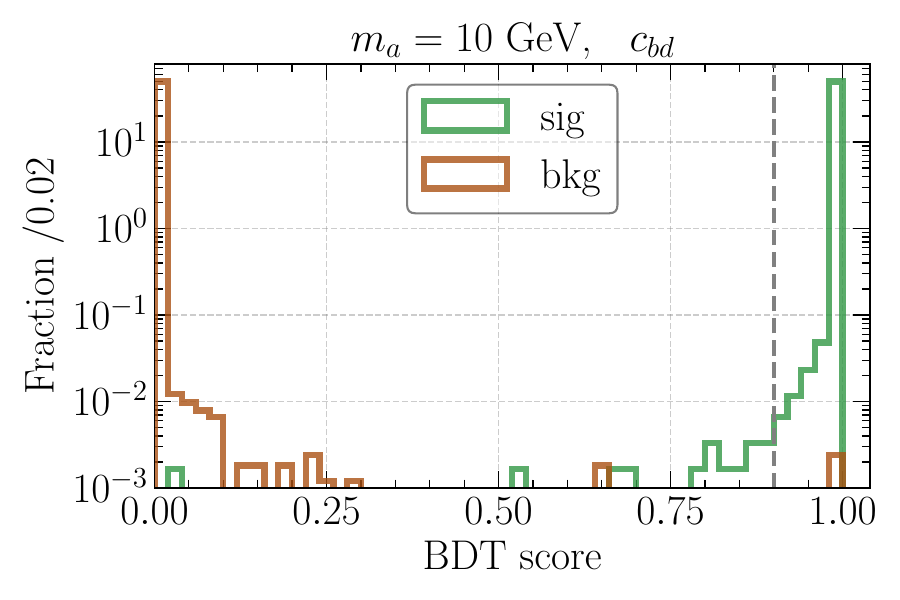}
      \minigraph{6cm}{-0.05in}{}{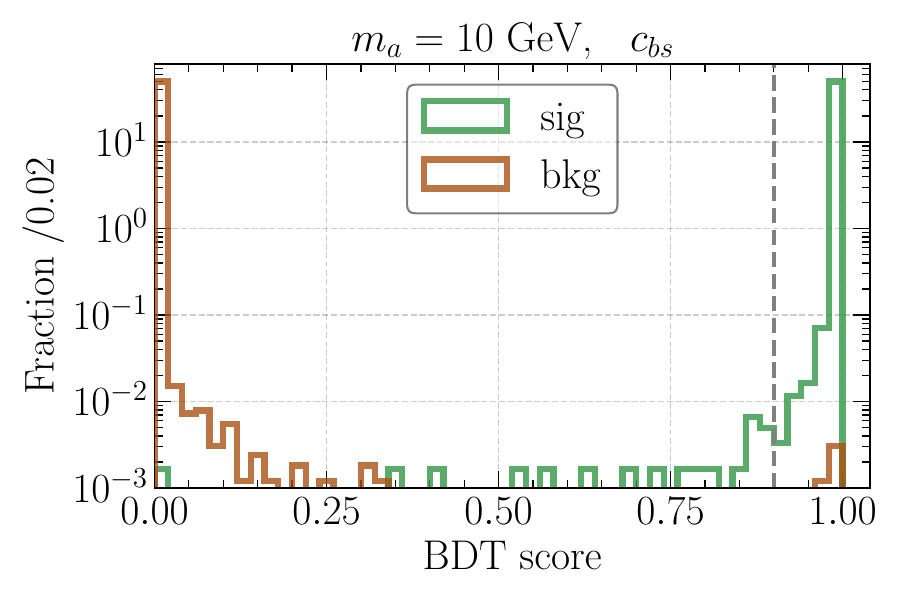}\\
      \minigraph{6cm}{-0.05in}{}{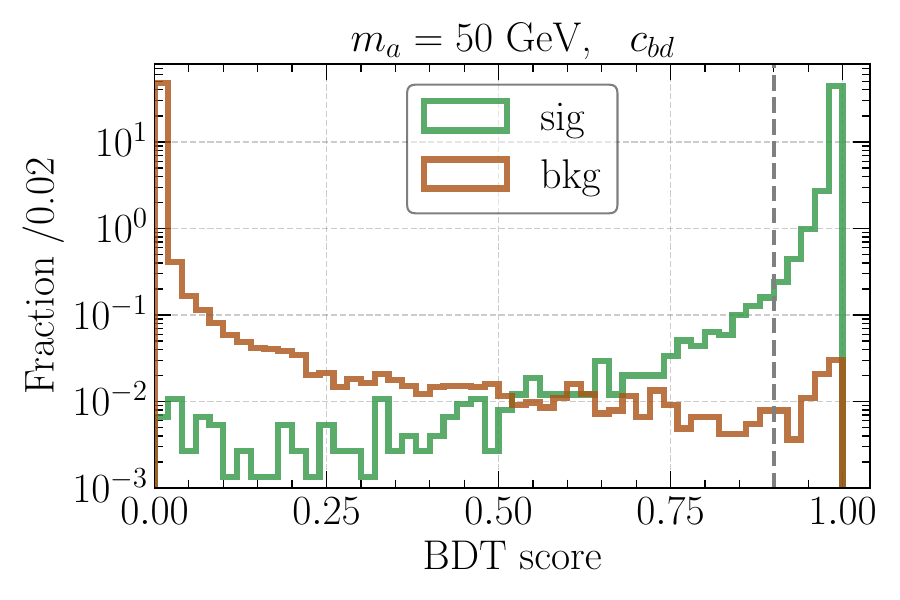}
      \minigraph{6cm}{-0.05in}{}{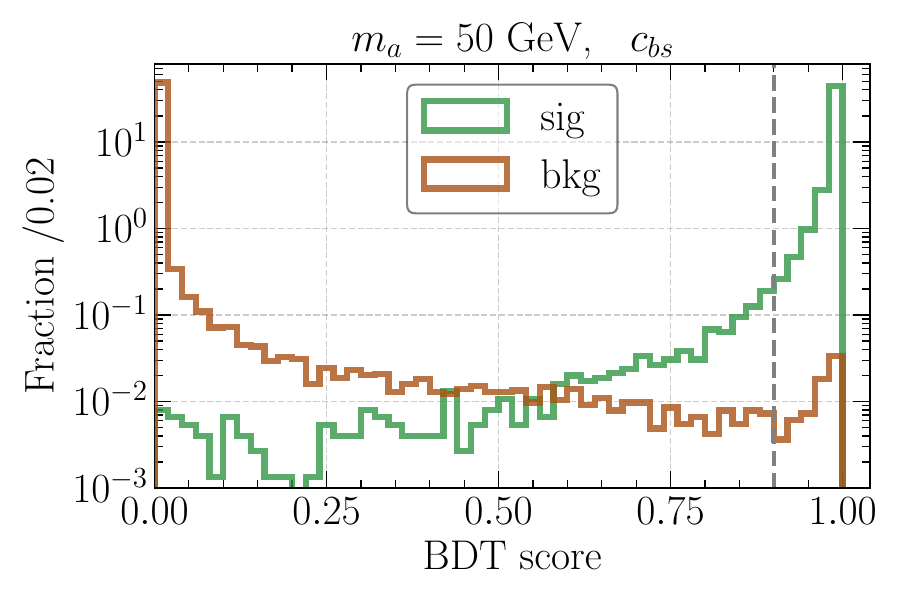}\\
      \minigraph{6cm}{-0.05in}{}{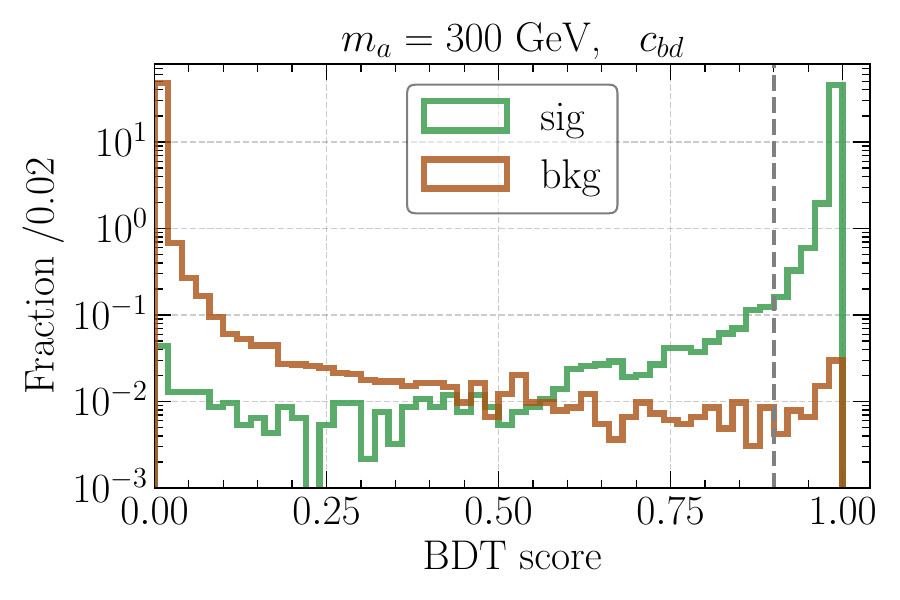}
      \minigraph{6cm}{-0.05in}{}{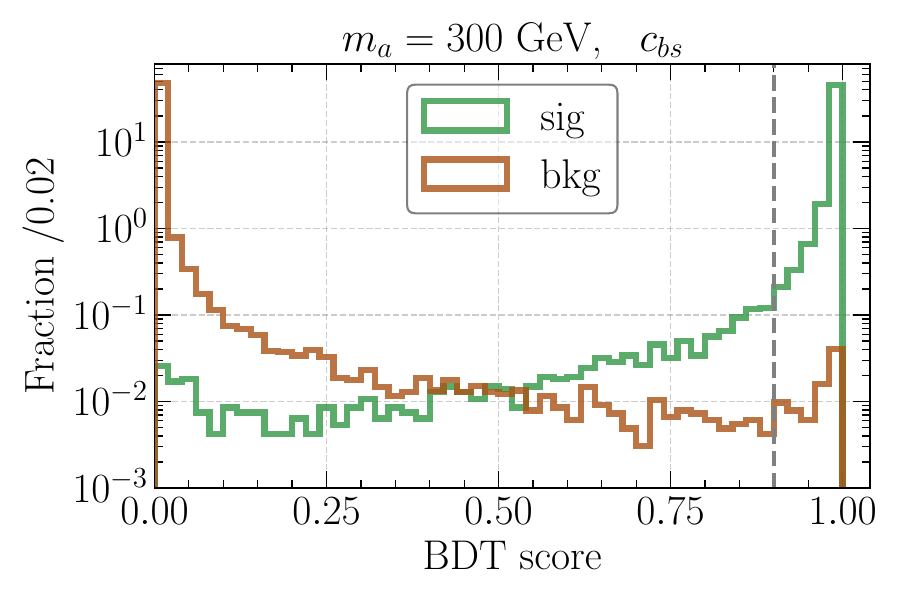}\\
      \minigraph{6cm}{-0.05in}{}{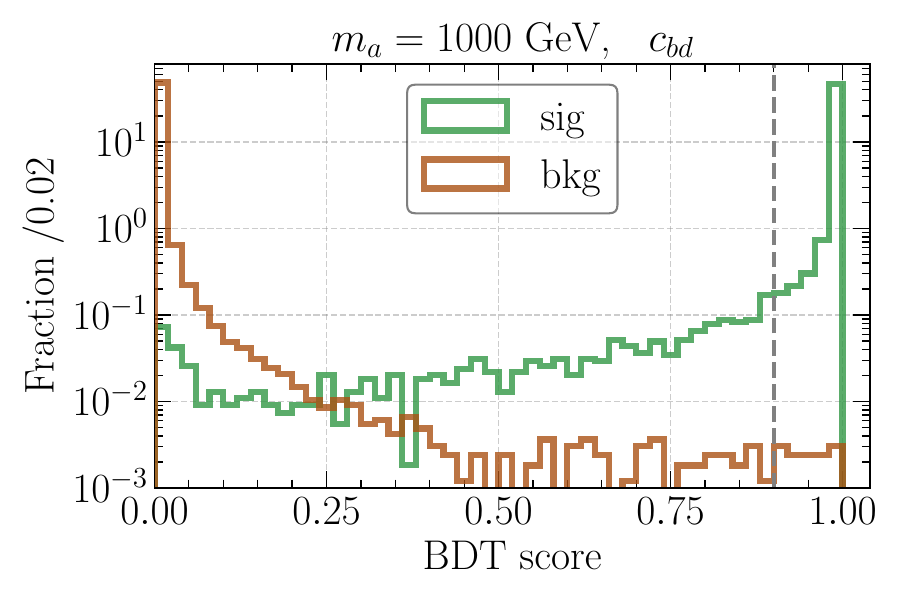}
      \minigraph{6cm}{-0.05in}{}{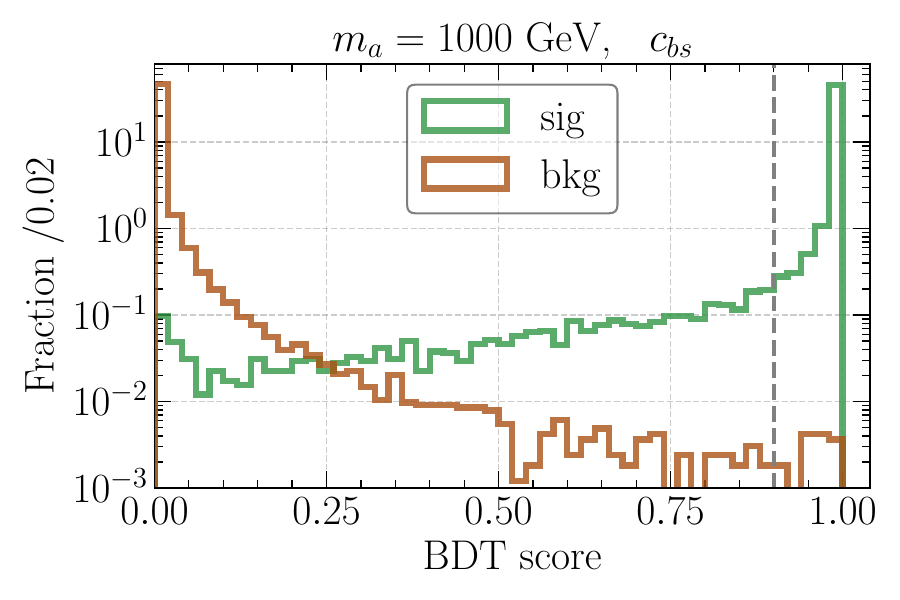}
\end{center}
\caption{ The BDT response score distribution of signal $bj\mu\mu$ (green) and total SM background (brown) with $m_a=10,~50,~300$ and 1000 GeV (from top to bottom) at LHC with $\sqrt{s}=13$ TeV and $\mathcal{L}=300~{\rm fb}^{-1}$. The grey dashed line indicates the BDT cut that maximizes the significance with fixed $|c_{bd}^{V(A)}|/f_a = 1~{\rm TeV}^{-1}$ (left four panels) or $|c_{bs}^{V(A)}|/f_a = 1~{\rm TeV}^{-1}$ (right four panels).
}
\label{fig:mumuBDT-score}
\end{figure}

Since the final muons are from different decay processes in the signal and SM backgrounds, the kinematic observables associated with muons allow for a significant distinction between them.
We also show the distributions of BDT response score of the signal (green) and total SM background (brown) with c.m. energy $\sqrt{s}=13$ TeV and luminosity $\mathcal{L}=300~{\rm fb}^{-1}$ for benchmark masses $m_a=10,~50,~300$ and $1000$ GeV, and $c_{bq}^{V(A)}/f_a=1~{\rm TeV}^{-1}$ in Fig.~\ref{fig:mumuBDT-score}. We can see that the signal and backgrounds are better separated than the invisible channel, due to the reconstructable ALP resonance. Through maximizing the significance in Eq.~(\ref{eqn-significance}), we can obtain the BDT cut. The information related to the signal for the above benchmarks and backgrounds are provided in Table~\ref{tab:mumuBDT}. We put the results for HL-LHC and FCC-hh in Appendix~\ref{App:mumu}. It turns out that in this case one can achieve a good distinction between the signal and the backgrounds. The remaining fraction of the signal events is more than 95\%. The BDT cut efficiency of backgrounds is at the level of $10^{-4}\sim 10^{-3}$. The BDT cut is universal for all benchmark masses.

\begin{figure}[tbh!]
\begin{center}
      \minigraph{7.5cm}{-0.05in}{}{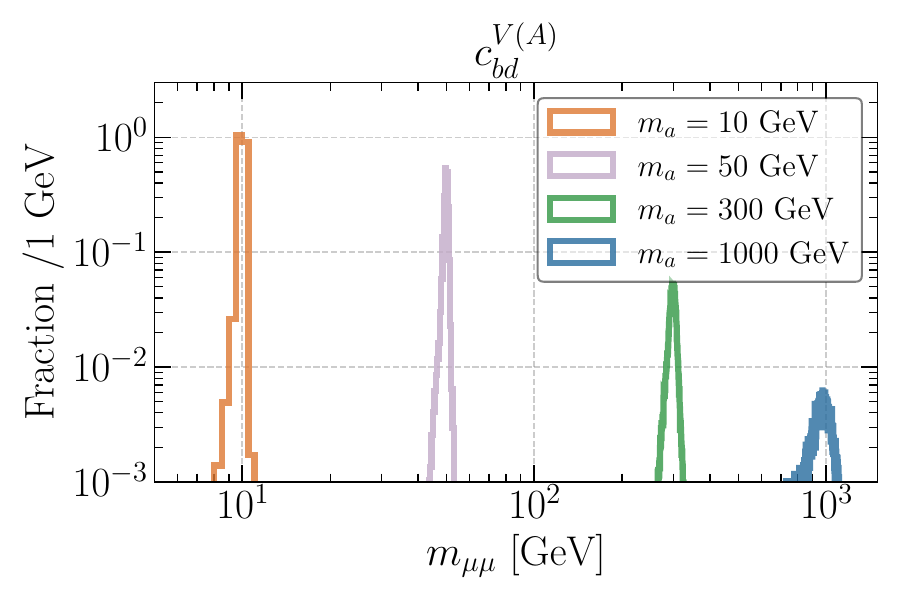}
      \minigraph{7.5cm}{-0.05in}{}{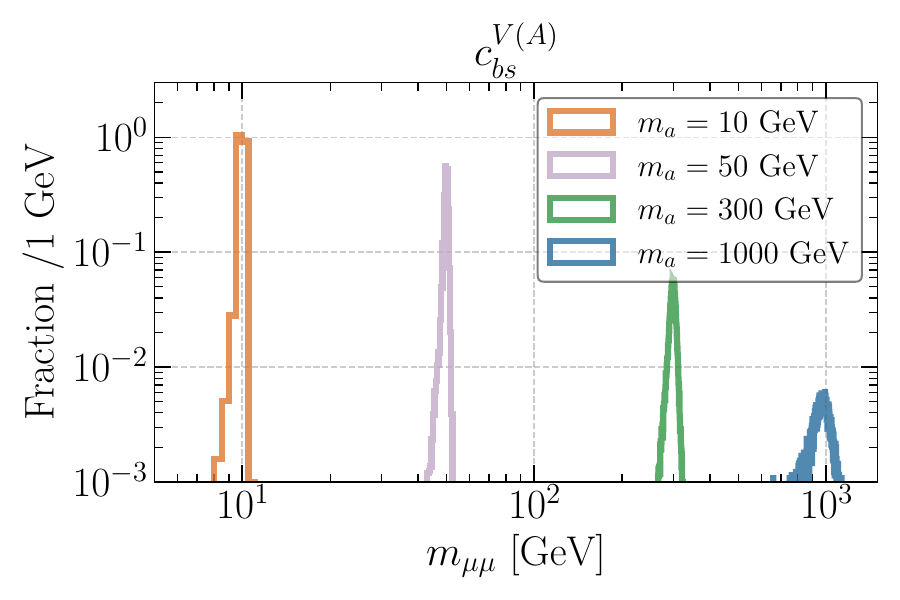}\\
\end{center}
\caption{The distribution of the invariant mass of the muon pair in $pp\to bj \mu^+\mu^-$ after applying the BDT cut for the benchmarks $m_a=10,~50,~300$ and 1000 GeV at LHC with $\sqrt{s}=13$ TeV and $\mathcal{L}=300~{\rm fb}^{-1}$. The parameter is fixed as $|c_{bd}^{V(A)}|/f_a = 1~{\rm TeV}^{-1}$ (left) or $|c_{bs}^{V(A)}|/f_a = 1~{\rm TeV}^{-1}$ (right).
}
\label{fig:mumu-invmass}
\end{figure}

\begin{figure}[t!]
\begin{center}
\minigraph{7.5cm}{-0.05in}{}{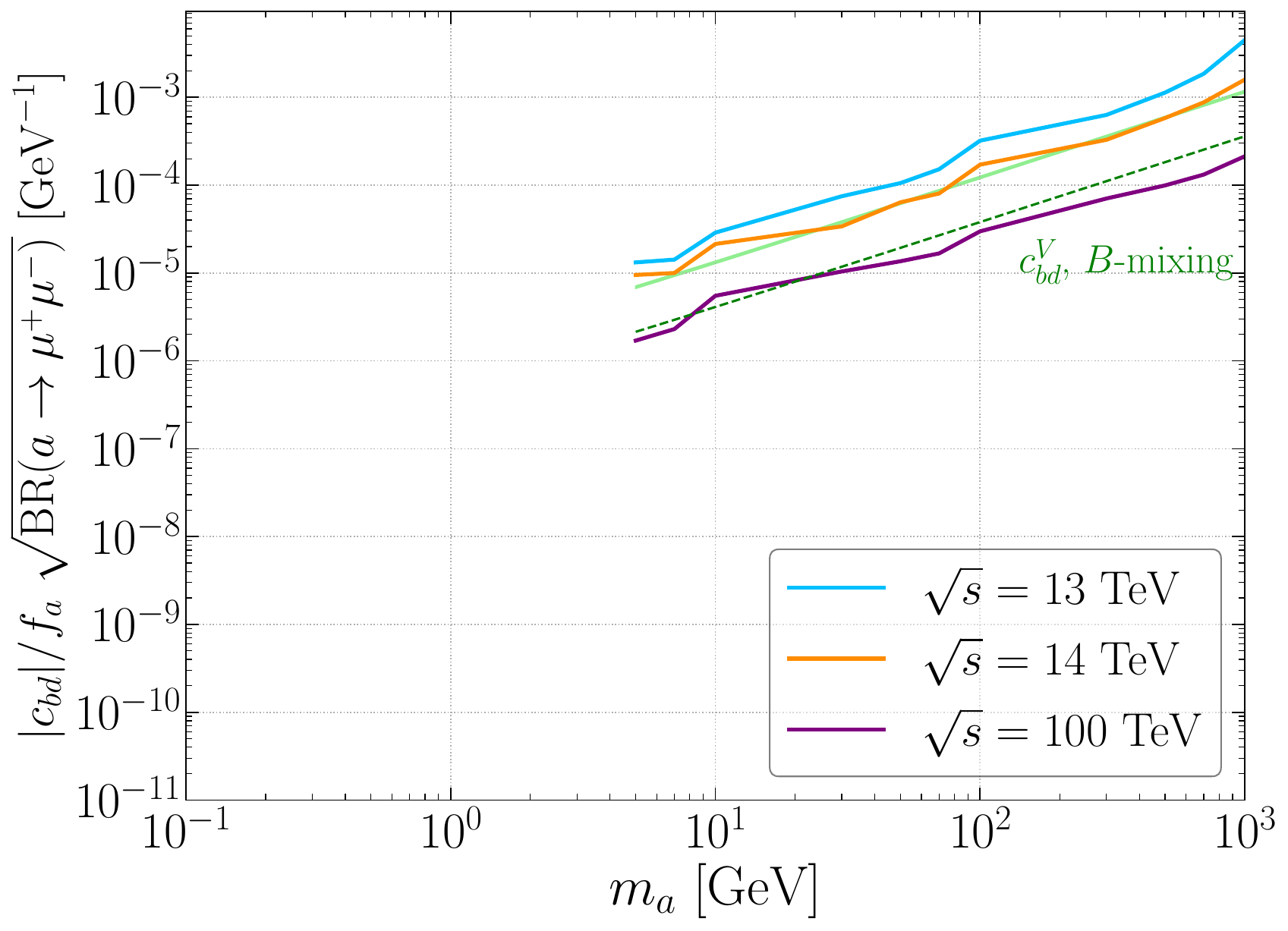}
\minigraph{7.5cm}{-0.05in}{}{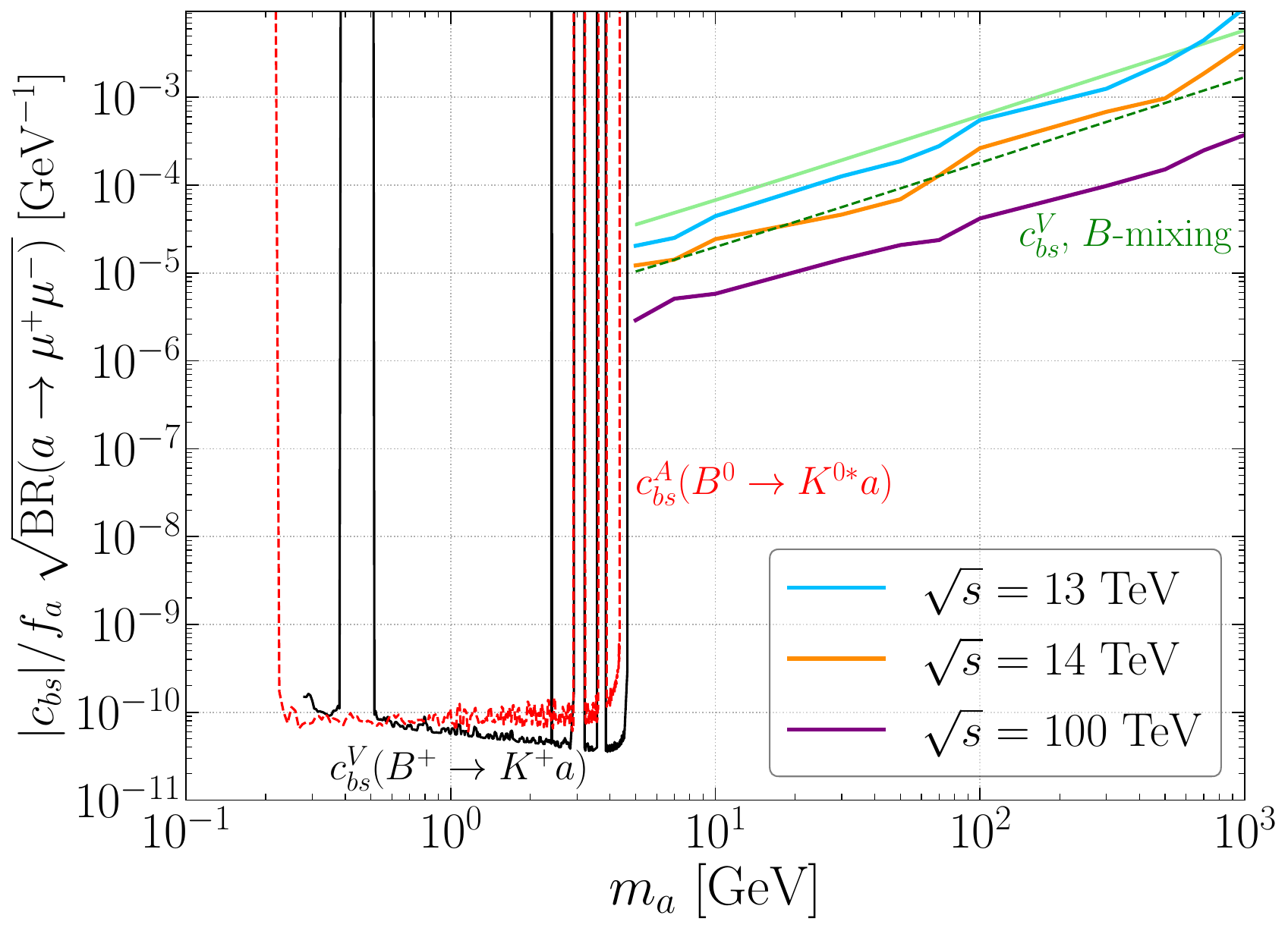}
\end{center}
\caption{The 2$\sigma$ exclusion limits for $|c_{bd}^{V(A)}|/f_a \sqrt{{\rm BR}(a\to \mu \mu)}$ (left) and $|c_{bs}^{V(A)}|/f_a \sqrt{{\rm BR}(a\to \mu \mu)}$ (right) at LHC with $\sqrt{s}=13$ TeV and $\mathcal{L}=300~{\rm fb}^{-1}$ (blue lines), HL-LHC with $\sqrt{s}=14$ TeV and $\mathcal{L}=3~{\rm ab}^{-1}$ (orange lines) or FCC-hh with $\sqrt{s}=100$ TeV and $\mathcal{L}=30~{\rm ab}^{-1}$ (purple lines).
The low-mass constraints from $B$ meson decay include $B^+\to K^+ a (\to \mu^+\mu^-)$ for $|c_{bs}^{V}|/f_a \sqrt{{\rm BR}(a\to \mu \mu)}$ (black solid curve) and $B^0\to K^{\ast 0} a (\to \mu^+\mu^-)$ for $|c_{bs}^{A}|/f_a \sqrt{{\rm BR}(a\to \mu \mu)}$ (red dashed curve). These limits correspond to 1 mm decay length of ALP. The $B$ mixing constraints are labeled as in Fig.~\ref{fig:invcva-limit}.
}
\label{fig:mumucva-limit}
\end{figure}

To better demonstrate the distinction between the signal and background, we show the invariant mass distribution of final muon pair in the signal after applying BDT cut in Fig.~\ref{fig:mumu-invmass}. The 2$\sigma$ exclusion limits on $|c_{bq}^{V(A)}|/f_a \sqrt{{\rm BR}(a\to \mu^+\mu^-)}$ are displayed for $m_a\gtrsim 5$ GeV in Fig.~\ref{fig:mumucva-limit}, with $q=d$ (left) and $q=s$ (right) at LHC, HL-LHC and FCC-hh. The bounds are more stringent than those from $a\to {\rm inv.}$ channel by one order of magnitude. As mentioned in Sec.~\ref{sec:BKmumu}, we also show the low-energy constraints on $|c_{bs}^{V(A)}|/f_a \sqrt{{\rm BR}(a\to \mu^+\mu^-)}$ from $B\to K^{(*)}a(\to\mu^+\mu^-)$ at LHCb. We take the limits by assuming short lifetime of ALP ($c\tau_a=1$ mm). They are more severe than the collider bounds by at least five orders of magnitude. Under the assumption of ${\rm BR}(a\to \mu^+\mu^-)=1$, the FCC-hh can probe the parameter space of both $c_{bd}^V/f_a$ and $c_{bs}^V/f_a$ favored by the $B$ meson oscillation for the two benchmarks. The LHC and HL-LHC are able to reach the favored parameter space of $c_{bs}^V/f_a$ for $m_a\lesssim 1$ TeV.

\begin{table}[htb!]
\centering
\resizebox{\textwidth}{!}{
\begin{tabular}{c|c|c|c|c|c|c|c|c}
\hline
     $m_a$ & BDT cut & $\epsilon_{\rm sig.}$ & $\epsilon_{jjZ/\gamma}$ & $\epsilon_{bbZ/\gamma}$ & $\epsilon_{jjWW}$ & $\epsilon_{bbWW}$ & $\epsilon_{t\bar{t}}$ &  $\mathcal{S}_{\rm max}$  \\
    \hline
    10   GeV& 0.90 & 9.99$\times10^{-1}$ & 2.13$\times10^{-4}$ & 5.95$\times10^{-5}$ & $-$ & 7.71$\times10^{-5}$ & 7.67$\times10^{-5}$ &  4.82$\times10^{2}$\\
    50   GeV& 0.90 & 9.81$\times10^{-1}$ & 6.38$\times10^{-4}$& 5.75$\times10^{-4}$ & 1.80$\times10^{-3}$ & 3.08$\times10^{-3}$ & 3.45$\times10^{-3}$ & 1.12$\times10^{2}$  \\
    300  GeV& 0.90 & 9.79$\times10^{-1}$ & 2.13$\times10^{-4}$ & $-$  & 1.35$\times10^{-2}$ & 3.55$\times10^{-3}$ & 2.99$\times10^{-3}$ &  4.92$\times10^{0}$ \\
    1000 GeV& 0.90 & 9.69$\times10^{-1}$& $-$  & 1.98$\times10^{-5}$ &  4.50$\times10^{-3}$ & 1.08$\times10^{-3}$ &  3.07$\times10^{-4}$ & 1.02$\times10^{-1}$ \\
    \hline
    \hline
    10   GeV& 0.90 & 9.99$\times10^{-1}$ & 2.13$\times10^{-4}$ & 7.94$\times10^{-5}$ &  $-$  & 7.71$\times10^{-5}$ & 7.67$\times10^{-5}$ &  2.83$\times10^{2}$\\
    50   GeV& 0.90 & 9.82$\times10^{-1}$ & 1.28$\times10^{-3}$ & 4.37$\times10^{-4}$ & 2.70$\times10^{-3}$ & 2.70$\times10^{-3}$& 2.92$\times10^{-3}$  & 4.58$\times10^{1}$  \\
    300  GeV& 0.90 & 9.79$\times10^{-1}$ & 4.25$\times10^{-4}$ & 1.98$\times10^{-5}$ & 1.80$\times10^{-2}$ & 4.32$\times10^{-3}$ & 4.37$\times10^{-3}$ & 1.15$\times10^{0}$ \\
    1000 GeV& 0.90 & 9.47$\times10^{-1}$ & $-$ & 3.97$\times10^{-5}$& 4.49$\times10^{-3}$ & 7.71$\times10^{-4}$ & 5.37$\times10^{-4}$ & 1.85$\times10^{-2}$ \\
    \hline
\end{tabular}}
\caption{The BDT cut, cut efficiencies and achieved maximal significance in Eq.~(\ref{eqn-significance}) for the signal $bj\mu\mu$ and SM backgrounds at LHC with $\sqrt{s}=13$ TeV and $\mathcal{L}=300~{\rm fb}^{-1}$. The benchmark masses are $m_a=10,~50,~300$ and 1000 GeV and the parameter is fixed as $|c_{bd}^{V(A)}|/f_a = 1~{\rm TeV}^{-1}$ (above the double line) or $|c_{bs}^{V(A)}|/f_a = 1~{\rm TeV}^{-1}$ (below the double line). The label ``$-$'' denotes the background at negligible level.}
\label{tab:mumuBDT}
\end{table}

\clearpage
\section{FCNC search for ALP decaying to diphoton at colliders}
\label{sec:diphoton}

We assume that the ALP promptly decays into diphoton.
We pre-select the events containing at least two photons and two jets with one of them tagged as $b$ jet satisfying
\begin{eqnarray}
n_\gamma\geq 2\;,~n_j\geq 1\;,~n_b\geq 1\;,~p_T(j,b)>25~{\rm GeV}\;,~p_T(\gamma)>10~{\rm GeV}\;,~|\eta(\gamma,j,b)|<2.5\;.
\label{eqn-pre-select}
\end{eqnarray}
The major SM backgrounds include
\begin{eqnarray}
p~p\to jj\gamma\gamma\;,~~bb\gamma\gamma\;.
\end{eqnarray}
The K factors for $jj\gamma\gamma$ and $bb\gamma\gamma$ are 1.3~\cite{Gehrmann:2013bga,Badger:2013ava} and 1.36~\cite{Fah:2017wlf}, respectively.
After pre-selection and considering the K factors, the cross sections of different SM backgrounds at LHC with $\sqrt{s}=13$ TeV are $\sigma_{jj\gamma\gamma}=1.26$ pb and $\sigma_{bb\gamma\gamma}=0.06$ pb.

\begin{figure}[th!]
\begin{center}
      \minigraph{6cm}{-0.05in}{}{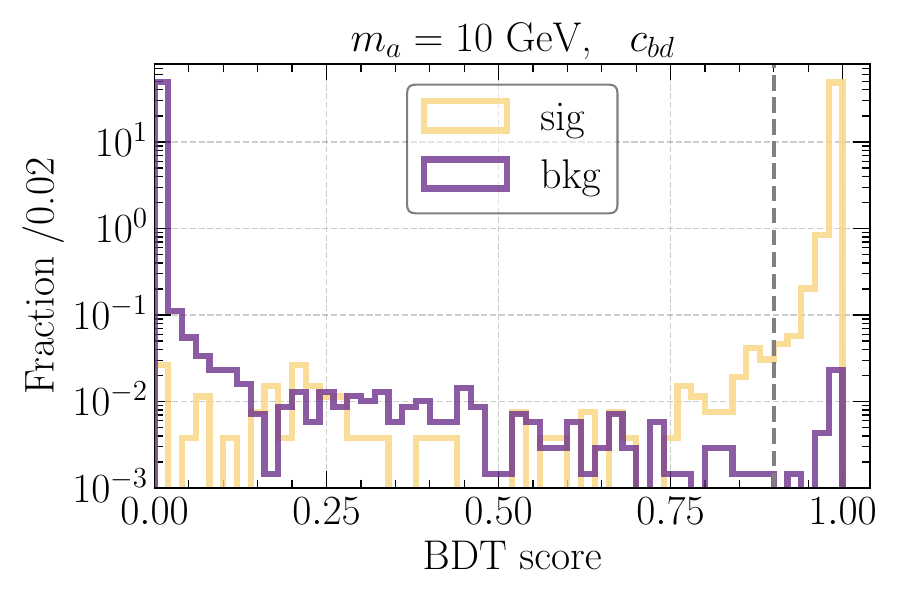}
      \minigraph{6cm}{-0.05in}{}{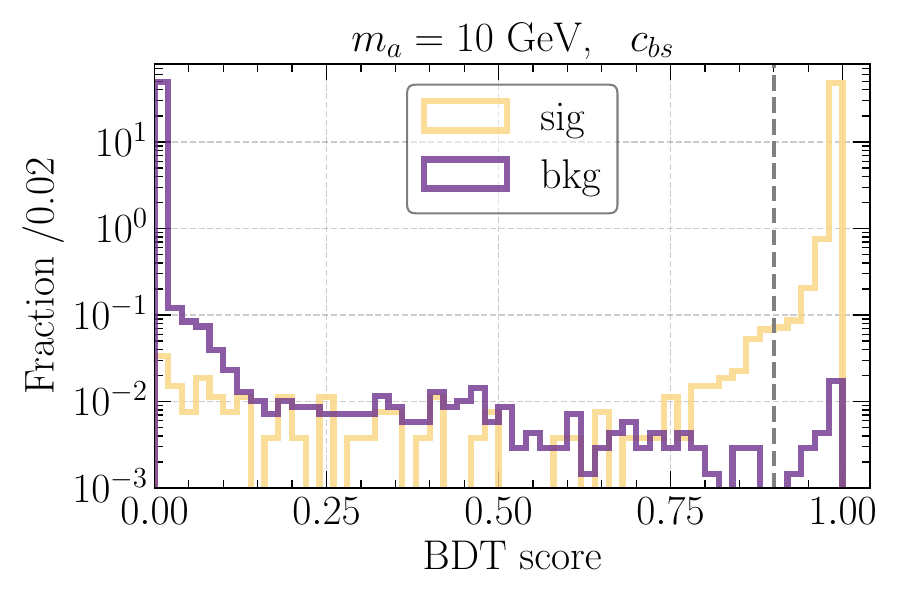}\\
      \minigraph{6cm}{-0.05in}{}{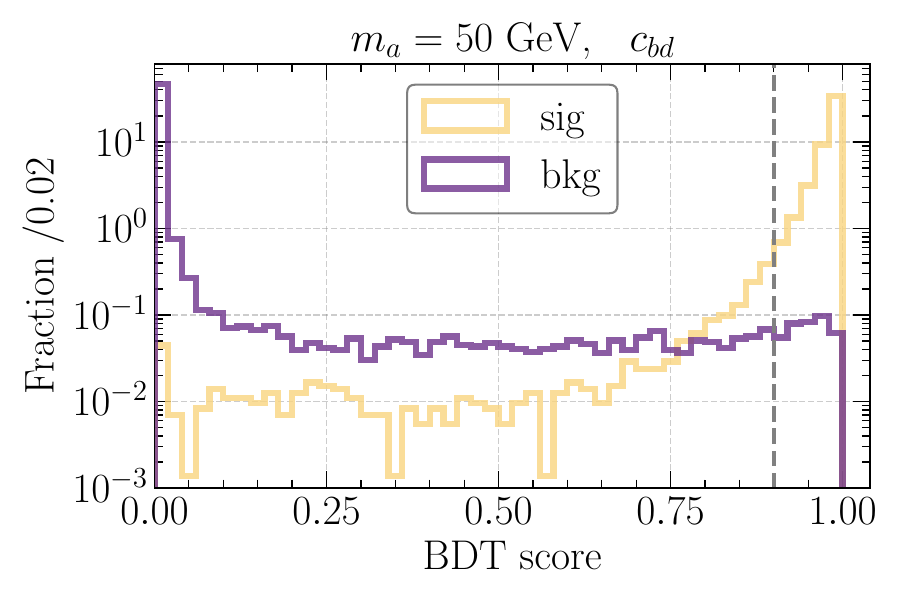}
      \minigraph{6cm}{-0.05in}{}{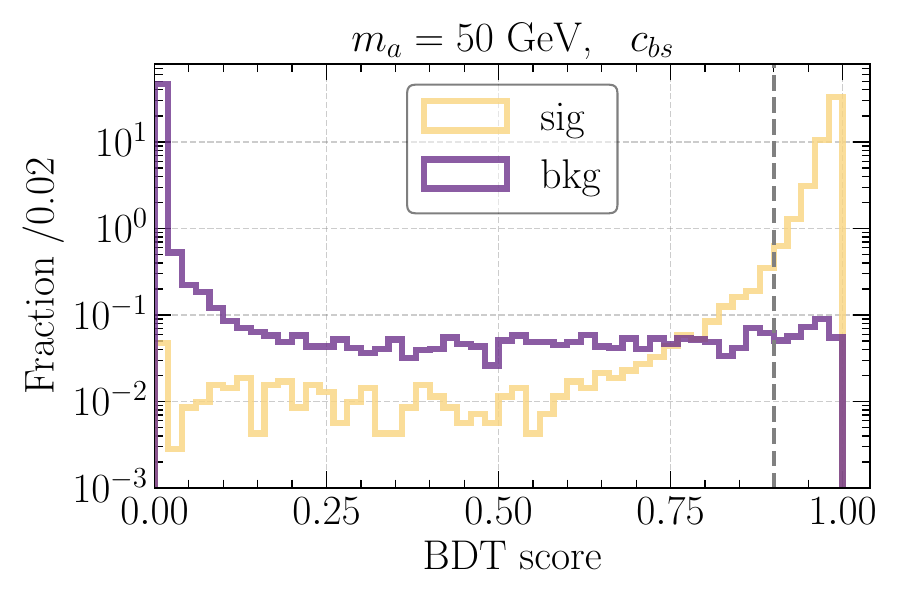}\\
      \minigraph{6cm}{-0.05in}{}{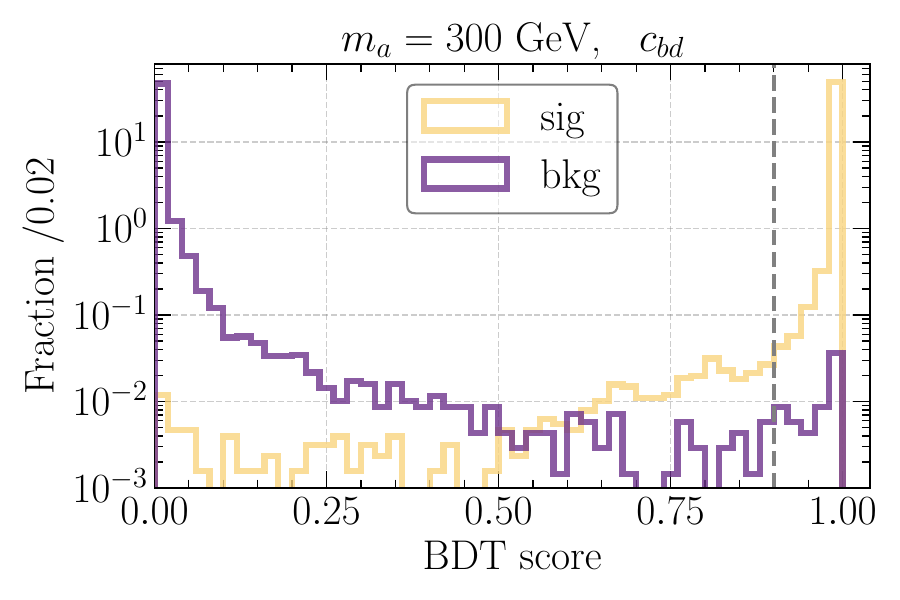}
      \minigraph{6cm}{-0.05in}{}{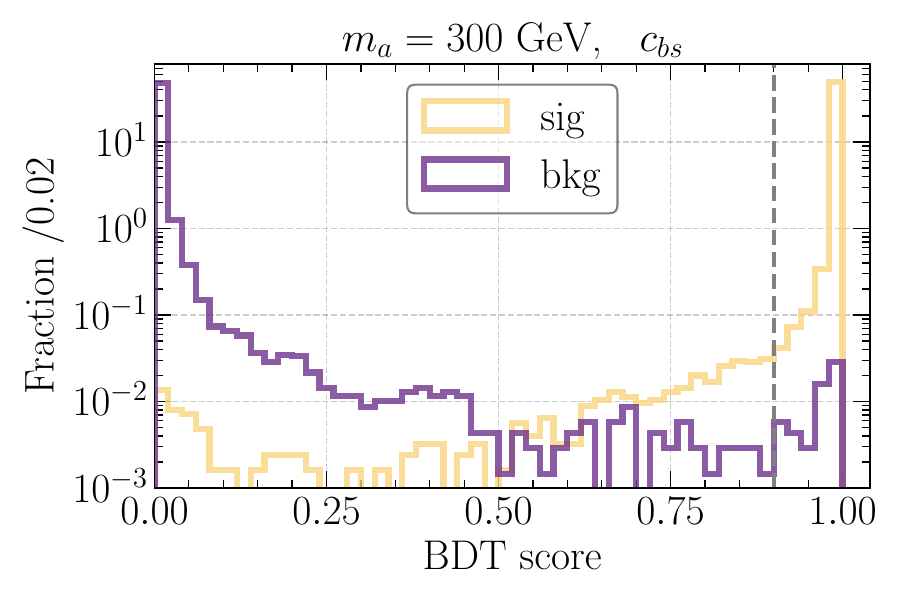}\\
      \minigraph{6cm}{-0.05in}{}{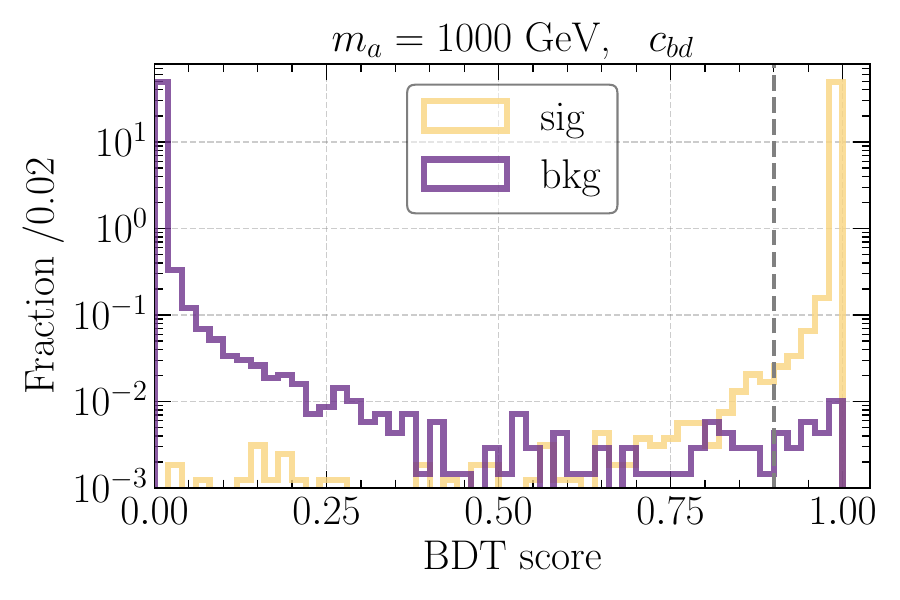}
      \minigraph{6cm}{-0.05in}{}{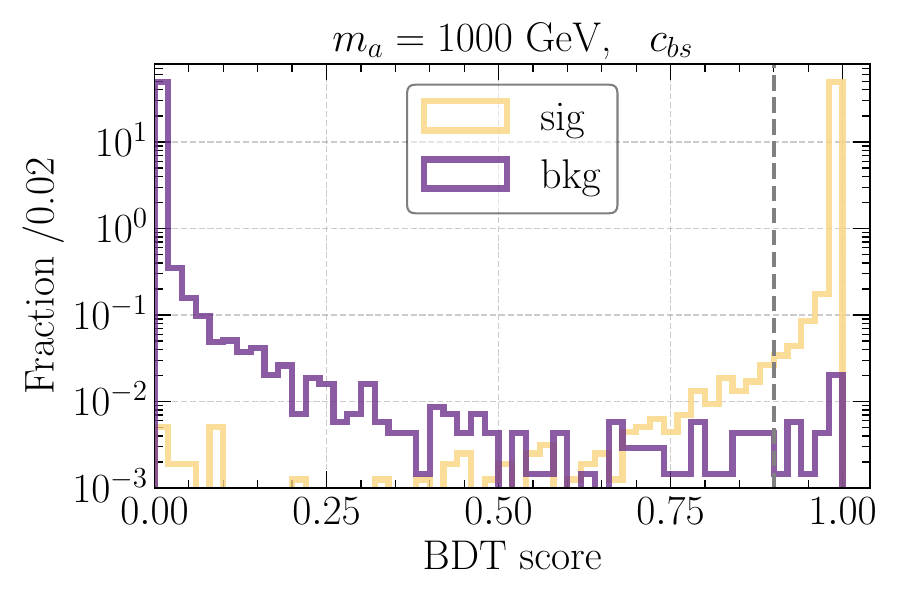}
\end{center}
\caption{ The BDT response score distribution of signal $bj\gamma\gamma$ (yellow) and total SM background (purple) with $m_a=10,~50,~300$ and 1000 GeV (from top to bottom) at LHC with $\sqrt{s}=13$ TeV and $\mathcal{L}=300~{\rm fb}^{-1}$. The grey dashed line indicates the BDT cut that maximizes the significance with fixed $|c_{bd}^{V(A)}|/f_a = 1~{\rm TeV}^{-1}$ (left four panels) or $|c_{bs}^{V(A)}|/f_a = 1~{\rm TeV}^{-1}$ (right four panels).}
\label{fig:aaBDT-score}
\end{figure}

\begin{figure}[th!]
\begin{center}
      \minigraph{7.5cm}{-0.05in}{}{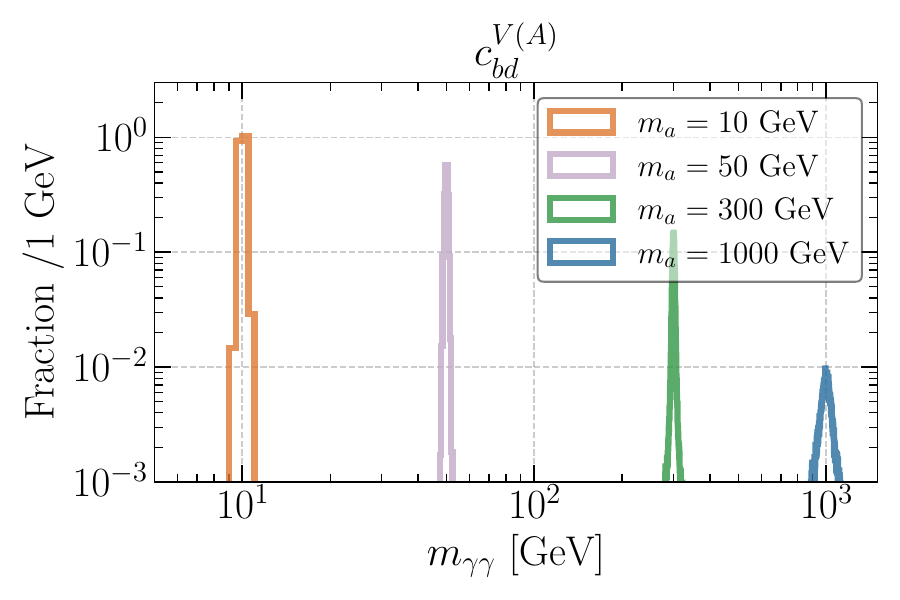}
      \minigraph{7.5cm}{-0.05in}{}{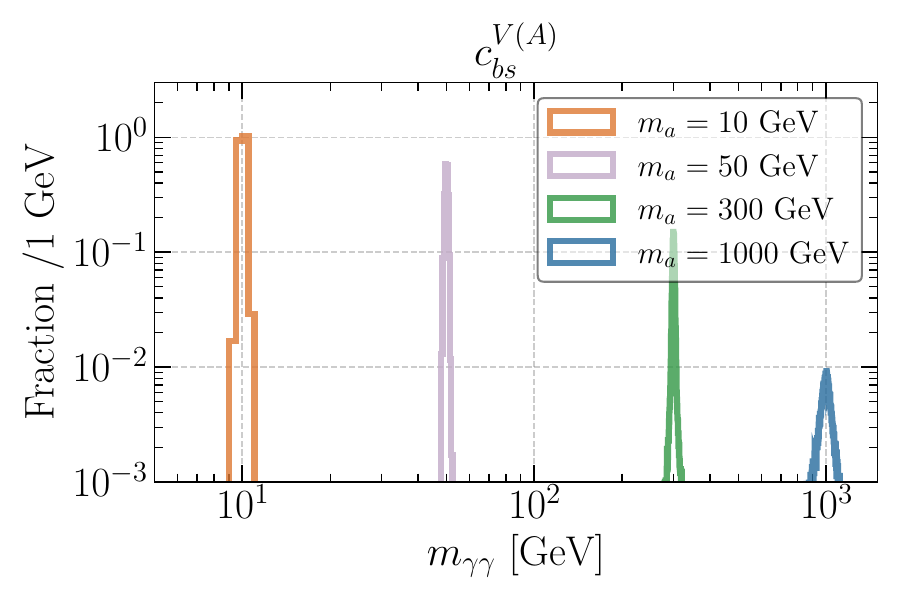}\\
\end{center}
\caption{The distribution of the invariant mass of the muon pair in $pp\to bj \gamma\gamma$ after applying the BDT cut for the benchmarks $m_a=10,~50,~300$ and 1000 GeV at LHC with $\sqrt{s}=13$ TeV and $\mathcal{L}=300~{\rm fb}^{-1}$. The parameter is fixed as $|c_{bd}^{V(A)}|/f_a = 1~{\rm TeV}^{-1}$ (left) or $|c_{bs}^{V(A)}|/f_a = 1~{\rm TeV}^{-1}$ (right).
}
\label{fig:aa-invmass}
\end{figure}

For the kinematic observables, which are considered to train the BDT classifier, are as follows
\begin{itemize}
\item transverse momentum, pseudo-rapidity and azimuthal angle of the final $b$ jet, light jet $j$ and photons $\gamma$ : $p_T(b)$, $p_T(j)$, $p_T(\gamma)$, $\eta(b)$, $\eta(j)$, $\eta(\gamma)$, $\phi(b)$, $\phi(j)$, $\phi(\gamma)$\;;
\item the difference of the pseudo-rapidity, azimuthal angle and the separation in angular space between each two of the four final states $\Delta \eta_{ij}$, $\Delta \phi_{ij}$, $\Delta R_{ij}$, where $i,j= b,~j,~\gamma_1,~\gamma_2~~(i\neq j)$\;;
\item the invariant mass of the $b$ jet and light jet $m_{bj}$, and the two final photons $m_{\gamma\gamma}$ \;;
\item the missing transverse energy $\cancel{E}_T$\;;
\item the sum of the transverse momenta of final visible objects $H_T=p_T(b)+p_T(j)+p_T(\gamma_1)+p_T(\gamma_2)$\;.
\end{itemize}
In Fig.~\ref{fig:aaBDT-score}, we show the distributions of BDT response score of the signal (yellow) and total SM background (purple) for LHC with $\sqrt{s}=13$ TeV and luminosity $\mathcal{L}=300~{\rm fb}^{-1}$. The signal and backgrounds are also well separated, similar to the dimuon channel. The BDT cut and cut efficiency for LHC with $\sqrt{s}=13$ TeV are collected in Table~\ref{tab:aaBDT}. The results for HL-LHC and FCC-hh are put in Appendix~\ref{App:gaga}. The BDT cut $0.9$ is also universal for all ALP mass benchmarks and colliders. Similar to the $a\to \mu^+\mu^-$ channel, the invariant mass of diphoton in the signal can be well reconstructed after applying the BDT cut as shown in Fig.~\ref{fig:aa-invmass}.

\begin{figure}[h!]
\begin{center}
\minigraph{7.5cm}{-0.05in}{}{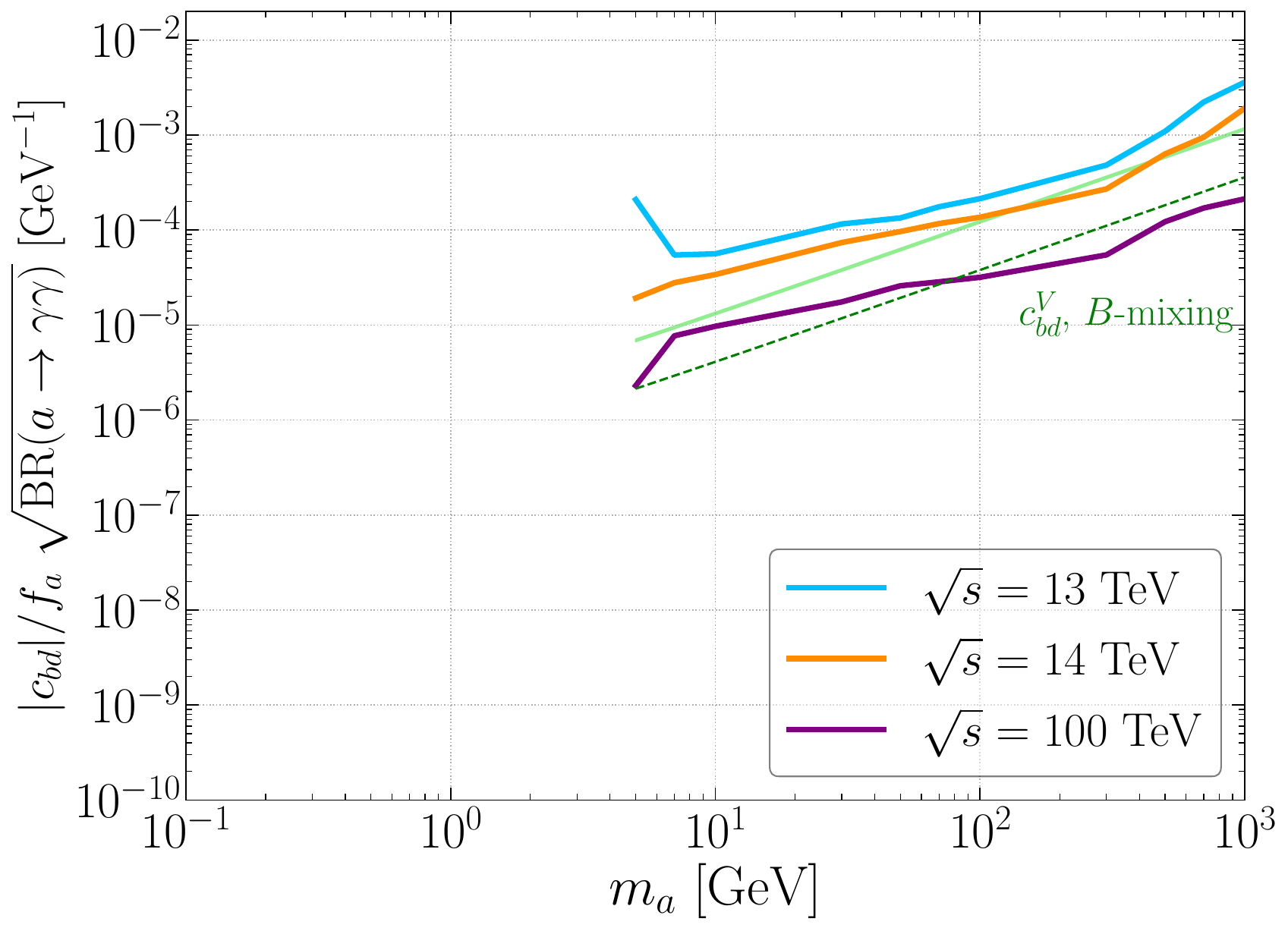}
\minigraph{7.5cm}{-0.05in}{}{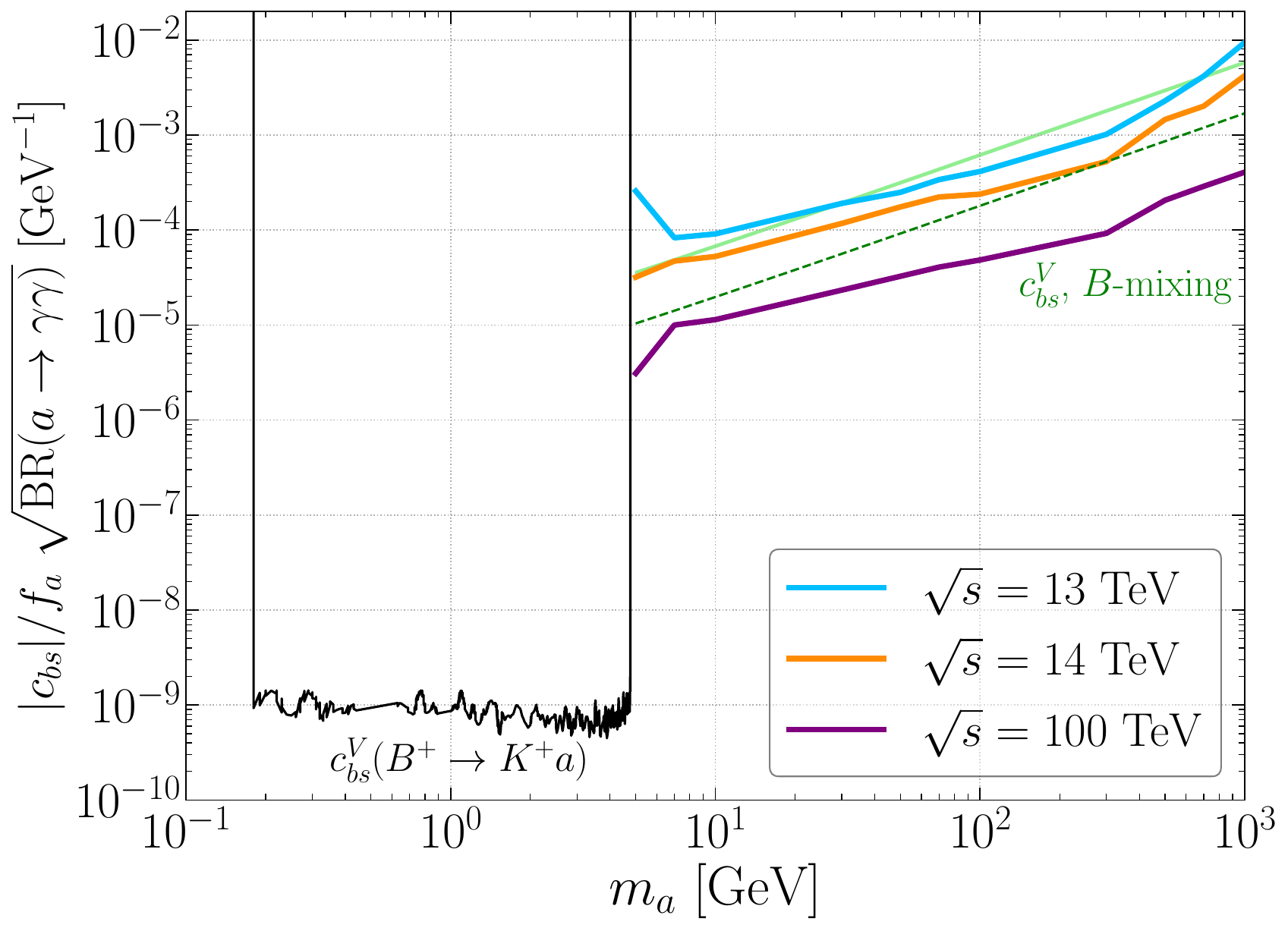}
\end{center}
\caption{The 2$\sigma$ exclusion limits for $|c_{bd}^{V(A)}|/f_a \sqrt{{\rm BR}(a\to \gamma \gamma)}$ (left) and $|c_{bs}^{V(A)}|/f_a \sqrt{{\rm BR}(a\to \gamma \gamma )}$ (right) at LHC with $\sqrt{s}=13$ TeV and $\mathcal{L}=300~{\rm fb}^{-1}$ (blue lines), HL-LHC with $\sqrt{s}=14$ TeV and $\mathcal{L}=3~{\rm ab}^{-1}$ (orange lines) or FCC-hh with $\sqrt{s}=100$ TeV and $\mathcal{L}=30~{\rm ab}^{-1}$ (purple lines).
The low-mass constraint is from $B^+\to K^+ a (\to \gamma\gamma)$ for $|c_{bs}^{V}|/f_a \sqrt{{\rm BR}(a\to \gamma \gamma)}$ (black solid curve). This limit corresponds to the decay length of ALP $c\tau_a\leq 1$ mm. The $B$ mixing constraints are labeled as in Fig.~\ref{fig:invcva-limit}.
}
\label{fig:gagacva-limit}
\end{figure}

The 2$\sigma$ exclusion limits on $|c_{bq}^{V(A)}|/f_a \sqrt{{\rm BR}(a\to \gamma\gamma)}$ are displayed for $m_a\gtrsim 5$ GeV in Fig.~\ref{fig:gagacva-limit}, with $q=d$ (left) and $q=s$ (right) at LHC, HL-LHC and FCC-hh.
These limits are close to those of $a\to \mu^+\mu^-$ channel. The low-energy constraints are also shown for comparison.
The low-energy bound from $B^+\to K^+ a(\to \gamma\gamma)$ is more severe than the collider bounds by at least four orders of magnitude. Under the assumption of ${\rm BR}(a\to \gamma\gamma)=1$, only the FCC-hh (all LHC, HL-LHC and FCC-hh) can reach the parameter space of both $c_{bd}^V/f_a$ ($c_{bs}^V/f_a$) favored by the $B$ meson oscillations.

\begin{table}[htb!]
\centering

\begin{tabular}{c|c|c|c|c|c}
\hline
     $m_a$ & BDT cut & $\epsilon_{\rm sig.}$ & $\epsilon_{jj\gamma\gamma}$ & $\epsilon_{bb\gamma\gamma}$ &  $\mathcal{S}_{\rm max}$  \\
    \hline
    10   GeV& 0.90 & 9.94$\times10^{-1}$ & 7.84$\times10^{-4}$ & 6.45$\times10^{-4}$ & 2.47$\times10^{2}$\\
    50   GeV& 0.90 & 9.69$\times10^{-1}$ & 5.29$\times10^{-3}$ & 5.77$\times10^{-3}$ & 7.94$\times10^{1}$  \\
    300  GeV& 0.90 & 9.93$\times10^{-1}$ & 7.84$\times10^{-4}$ & 1.32$\times10^{-3}$ & 8.11$\times10^{0}$ \\
    1000 GeV& 0.90 & 9.97$\times10^{-1}$ & 7.84$\times10^{-4}$ & 7.13$\times10^{-4}$ & 1.55$\times10^{-1}$ \\
    \hline
    \hline
    10   GeV& 0.90 & 9.94$\times10^{-1}$ & 9.80$\times10^{-4}$ & 9.16$\times10^{-4}$ & 1.32$\times10^{2}$\\
    50   GeV& 0.90 & 9.68$\times10^{-1}$ & 6.66$\times10^{-3}$ & 8.04$\times10^{-3}$ & 2.86$\times10^{1}$  \\
    300  GeV& 0.90 & 9.93$\times10^{-1}$ & 1.18$\times10^{-3}$ & 1.46$\times10^{-3}$ & 1.91$\times10^{0}$ \\
    1000 GeV& 0.90 & 9.97$\times10^{-1}$ & 1.18$\times10^{-3}$ & 6.45$\times10^{-4}$ & 2.27$\times10^{-2}$ \\
    \hline
\end{tabular}
\caption{The BDT cut, cut efficiencies and achieved maximal significance in Eq.~(\ref{eqn-significance}) for the signal $bj\gamma\gamma$ and SM backgrounds at LHC with $\sqrt{s}=13$ TeV and $\mathcal{L}=300~{\rm fb}^{-1}$. The benchmark masses are $m_a=10,~50,~300$ and 1000 GeV and the parameter is fixed as $|c_{bd}^{V(A)}|/f_a = 1~{\rm TeV}^{-1}$ (above the double line) or $|c_{bs}^{V(A)}|/f_a = 1~{\rm TeV}^{-1}$ (below the double line). 
}
\label{tab:aaBDT}
\end{table}

\clearpage
\section{Conclusions}
\label{sec:Sum}

In this paper, we studied the bottom quark flavor-violating interactions of the ALP through the low-energy FCNC processes of $B$ mesons and at high-energy hadron colliders. We investigate the low-energy constraints from recent $B$ meson decay measurements and the mixing of neutral $B$ mesons. We also explore the search potential of the flavor-violating couplings of heavy ALP to bottom quark at LHC and its upgrades. Our main results are summarized as follows.
\begin{itemize}
\item For light ALP with $m_a\lesssim m_B$, assuming promptly decaying ALP, the $B$ meson FCNC decays constrain the product $c_{bq}^{V,A}/f_a \sqrt{{\rm BR}(a\to X)}$ as low as the level of $10^{-8}$ ($10^{-10}$) [$10^{-9}$] ${\rm GeV}^{-1}$ for $X={\rm invisible}$ ($\mu^+\mu^-$) [$\gamma\gamma$].
\item The $B$ meson oscillations constrains the heavy ALP regime. The bounds highly depend on the assumption of the $c_{bq}^{V,A}$ parameters. The $B_d-\bar{B}_d$ mixing places more stringent constraints on $c_{bd}^{V,A}$ than the $B_s-\bar{B}_s$ mixing for $c_{bs}^{V,A}$.
\item The FCNC search for invisibly decaying ALP at hadron colliders cannot reach the parameter space of $c_{bq}^{V,A}/f_a$ and $m_a$ preferred by the $B$ meson oscillations.
\item The exclusion limits of the FCNC search for the ALP decaying to dimuon at hadron colliders are more stringent than those from $a\to {\rm inv.}$ channel by one order of magnitude. The HL-LHC and FCC-hh are able to probe the parameter space preferred by the $B$ meson oscillations. Especially, for $m_B\lesssim m_a\lesssim 1$ TeV region, the preferred FCNC $c_{bs}^V$ coupling should already be probed or excluded by the search of $bj+$ dimuon at current 13 TeV LHC.
\item The exclusion limits of the FCNC search for the ALP decaying to diphoton at hadron colliders are close to those from the dimuon channel.
\end{itemize}

\section*{ACKNOWLEDGMENTS}
T.~L. is supported by the National Natural Science Foundation of China (Grant No. 12375096, 12035008, 11975129) and ``the Fundamental Research Funds for the Central Universities'', Nankai University (Grant No. 63196013).
M.~S. acknowledges support by the Australian Research Council Discovery Project DP200101470.

\appendix
\clearpage

\section{The BDT score distribution and cut efficiency at HL-LHC and FCC-hh}

\subsection{$a\to {\rm inv.}$}
\label{App:inv}

\begin{figure}[th!]
\begin{center}
      \minigraph{6.5cm}{-0.05in}{}{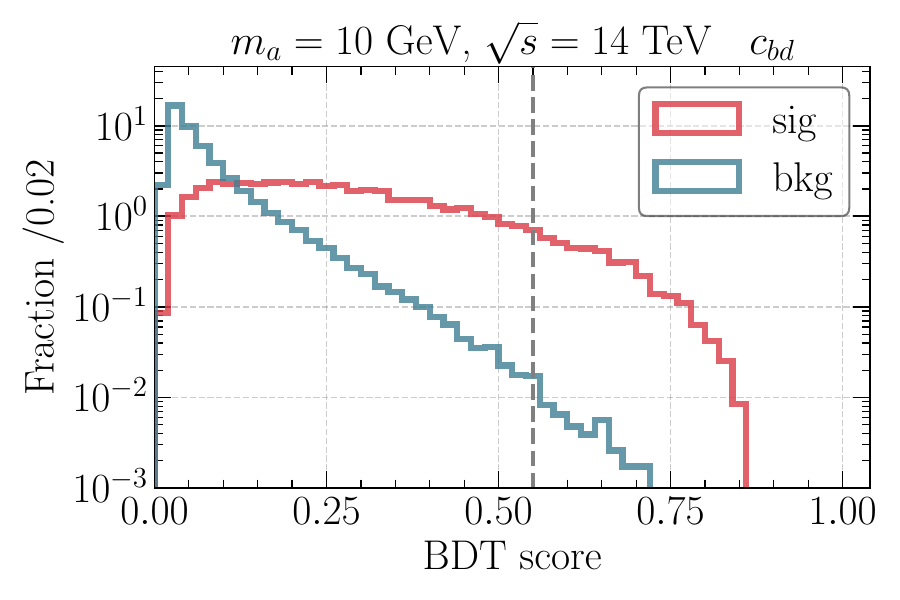}
      \minigraph{6.5cm}{-0.05in}{}{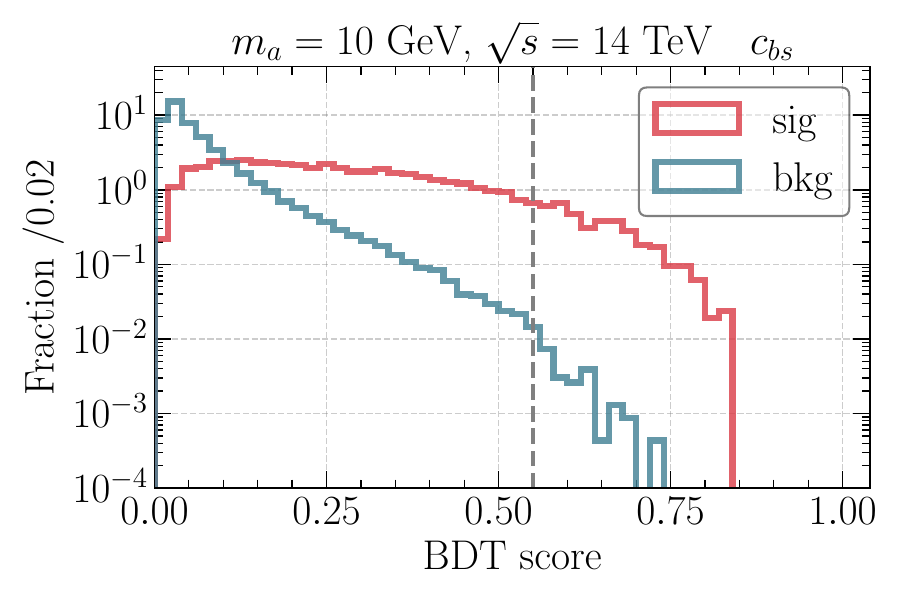}\\
      \minigraph{6.5cm}{-0.05in}{}{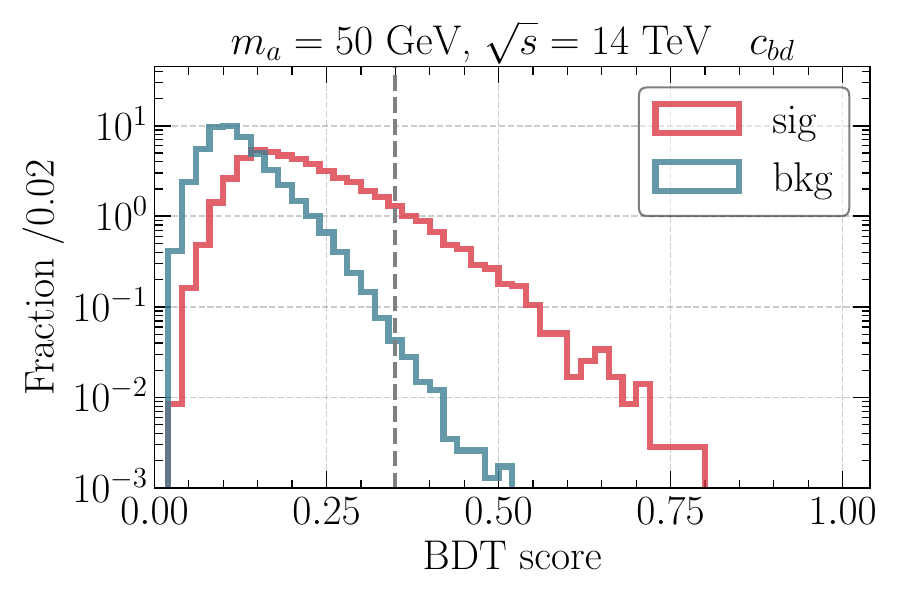}
      \minigraph{6.5cm}{-0.05in}{}{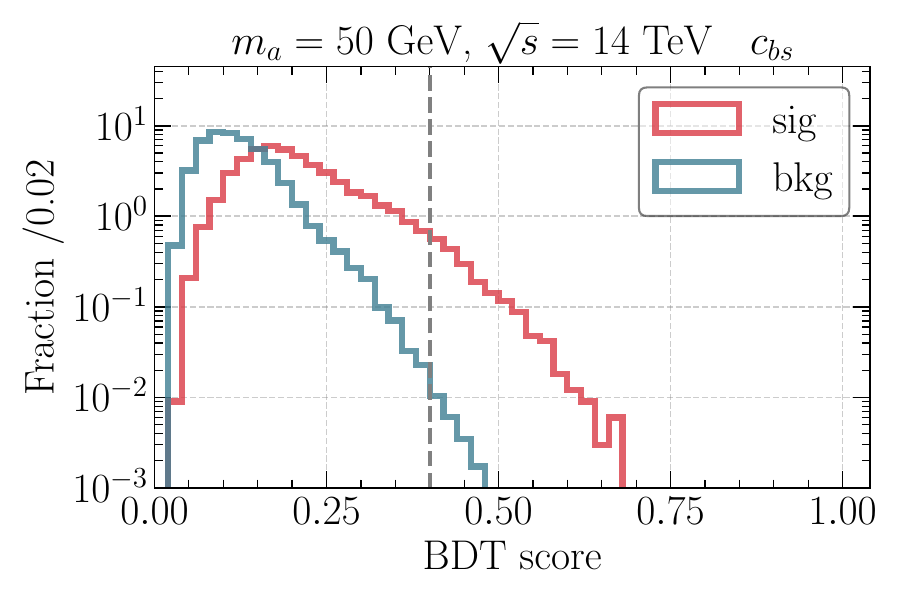}\\
      \minigraph{6.5cm}{-0.05in}{}{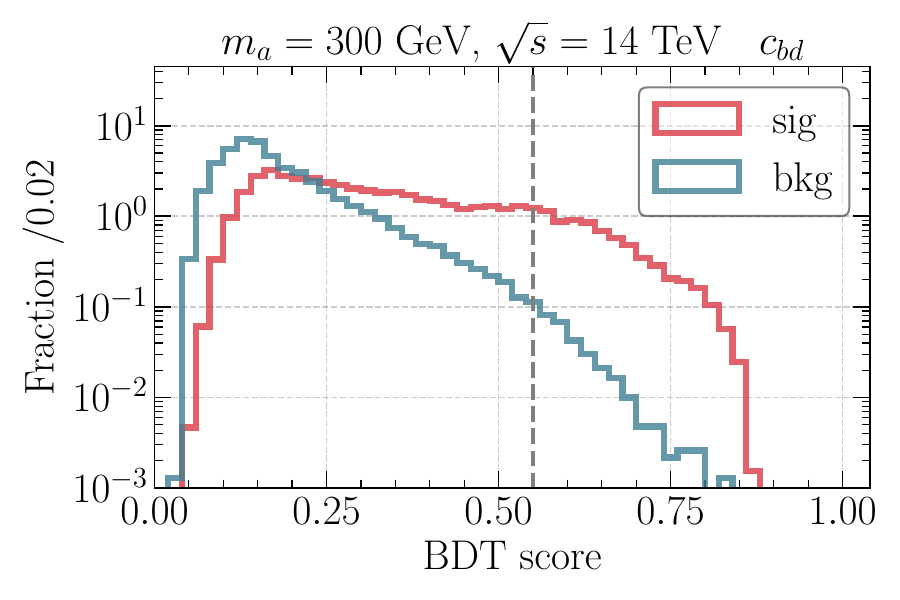}
      \minigraph{6.5cm}{-0.05in}{}{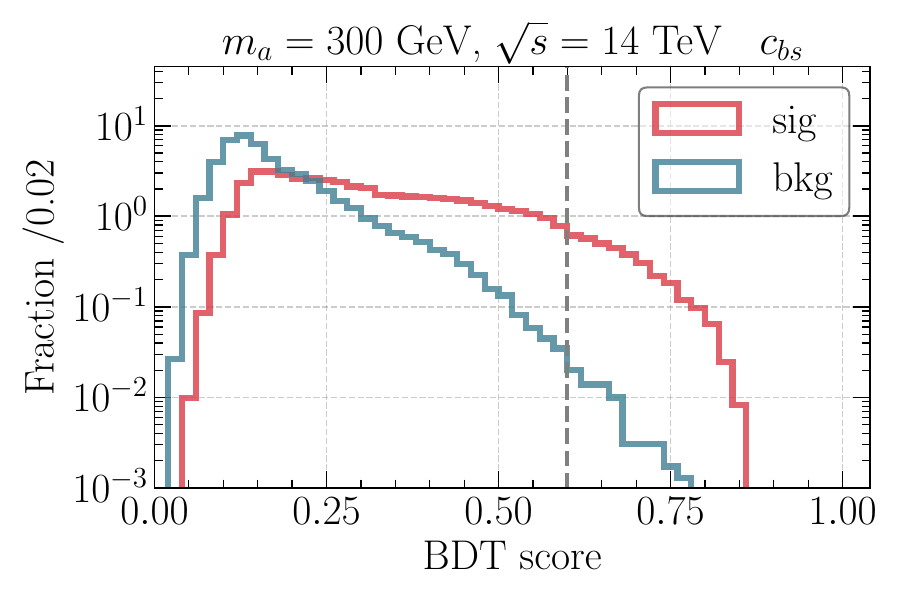}\\
      \minigraph{6.5cm}{-0.05in}{}{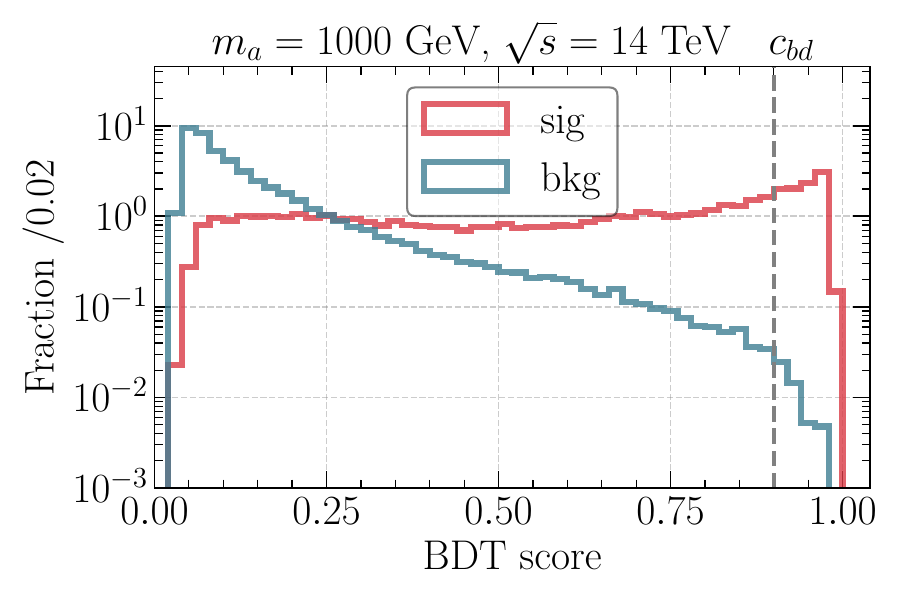}
      \minigraph{6.5cm}{-0.05in}{}{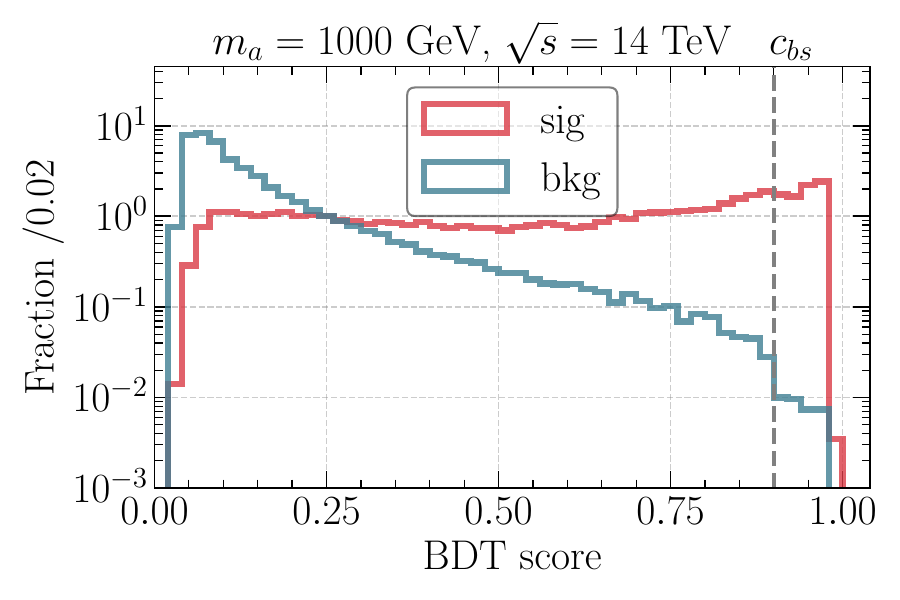}
\end{center}
\caption{ The BDT response score distribution of signal $bj+{\rm inv.}$ (red) and total SM background (blue) with $m_a=10,~50,~300$ and 1000 GeV at HL-LHC with $\sqrt{s}=14$ TeV and $\mathcal{L}=3~{\rm ab}^{-1}$. The grey dash line is the BDT cut that maximizes the significance with fixed $|c_{bd}^{V(A)}|/f_a = 1~{\rm TeV}^{-1}$ (left four panels) or $|c_{bs}^{V(A)}|/f_a = 1~{\rm TeV}^{-1}$ (right four panels).}
\label{fig:invBDT-score-HLLHC}
\end{figure}

\begin{figure}[th!]
\begin{center}
      \minigraph{6.5cm}{-0.05in}{}{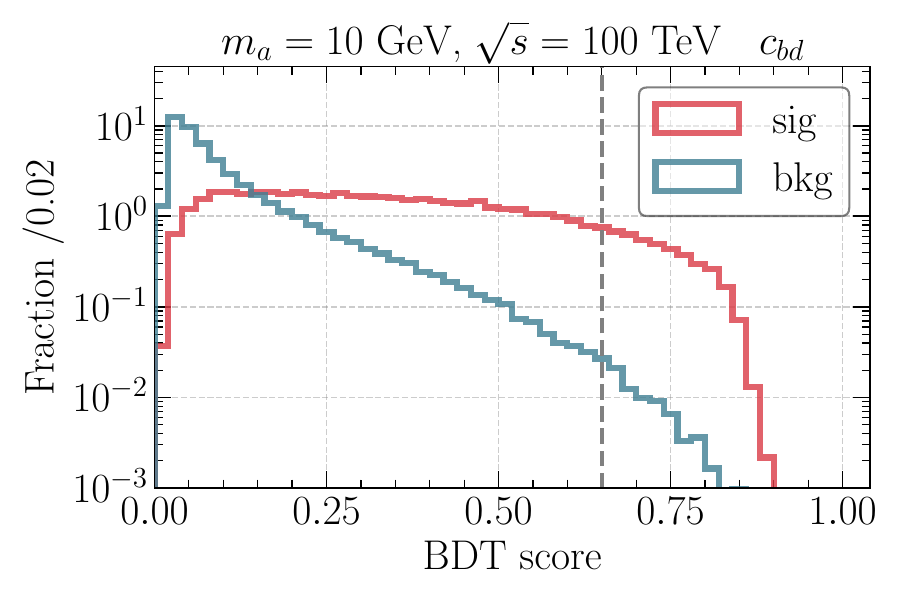}
      \minigraph{6.5cm}{-0.05in}{}{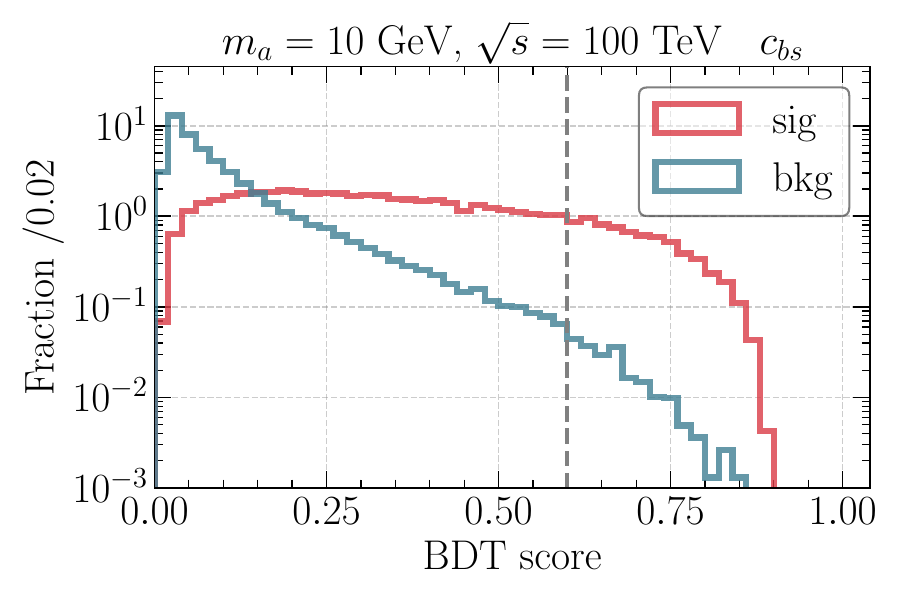}\\
      \minigraph{6.5cm}{-0.05in}{}{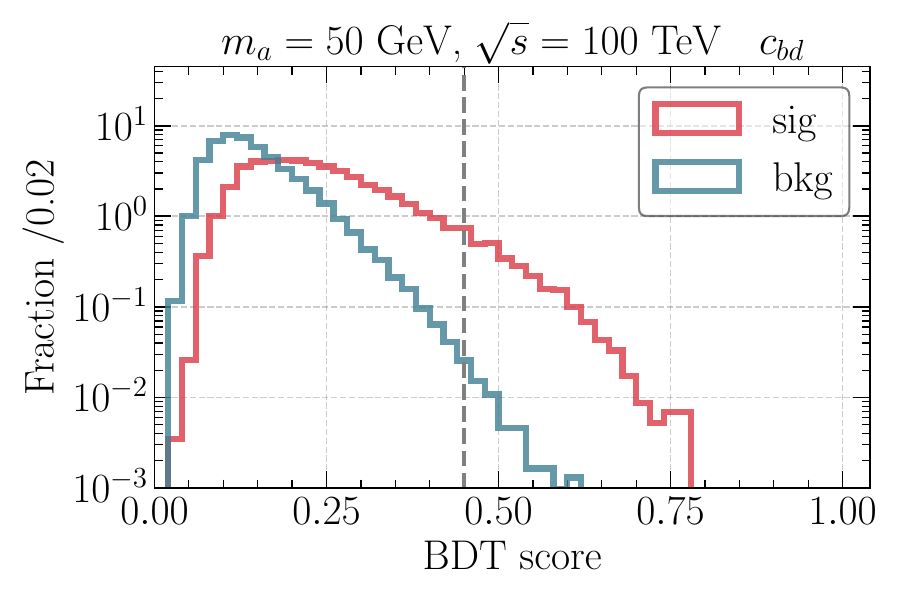}
      \minigraph{6.5cm}{-0.05in}{}{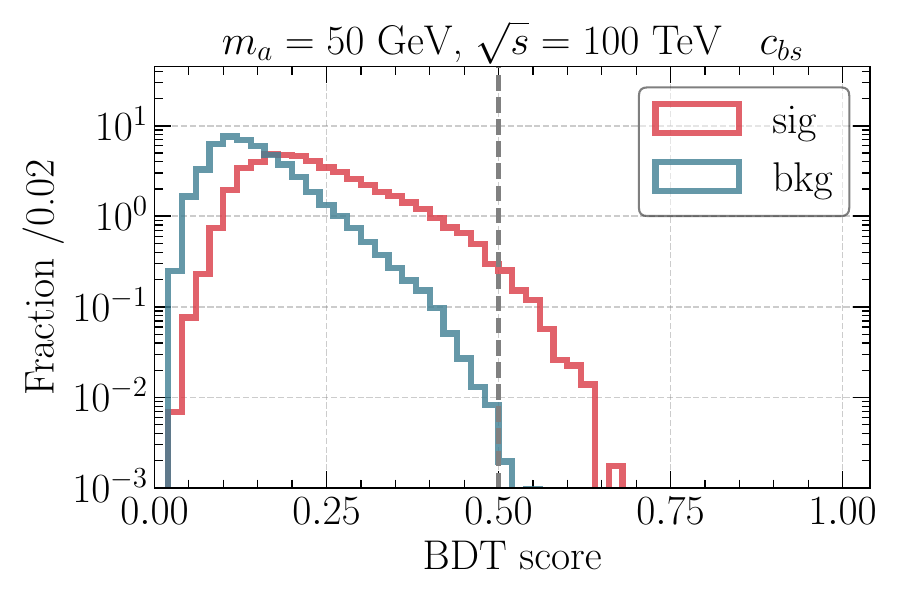}\\
      \minigraph{6.5cm}{-0.05in}{}{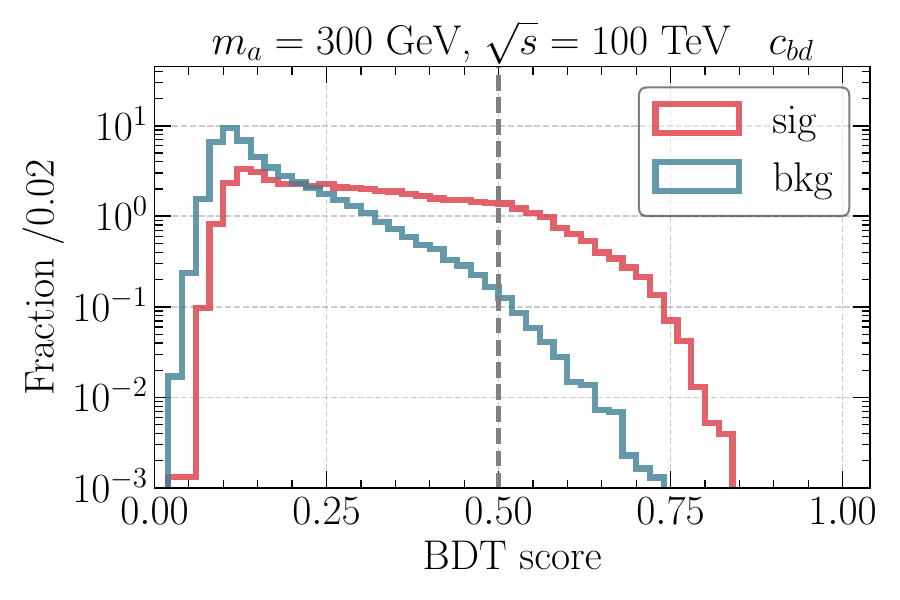}
      \minigraph{6.5cm}{-0.05in}{}{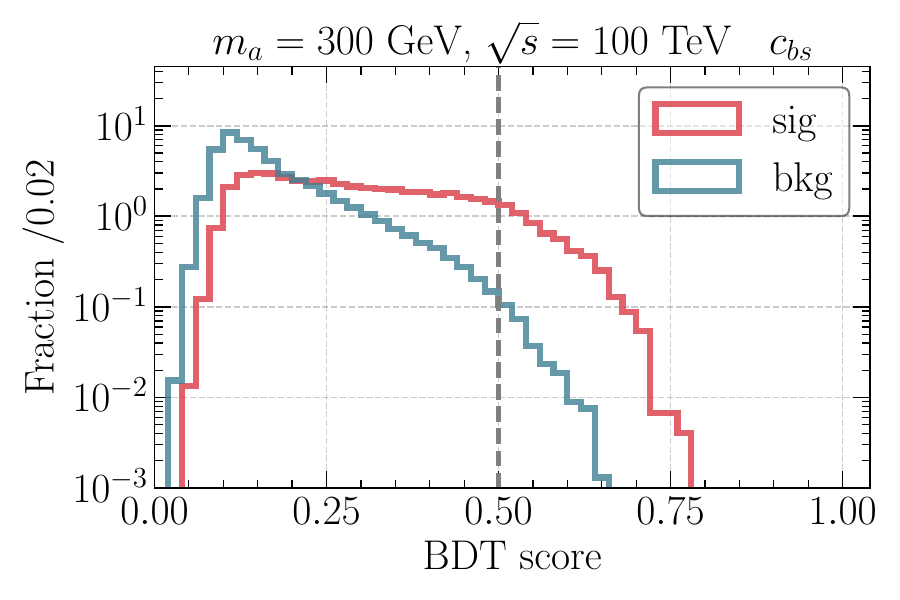}\\
      \minigraph{6.5cm}{-0.05in}{}{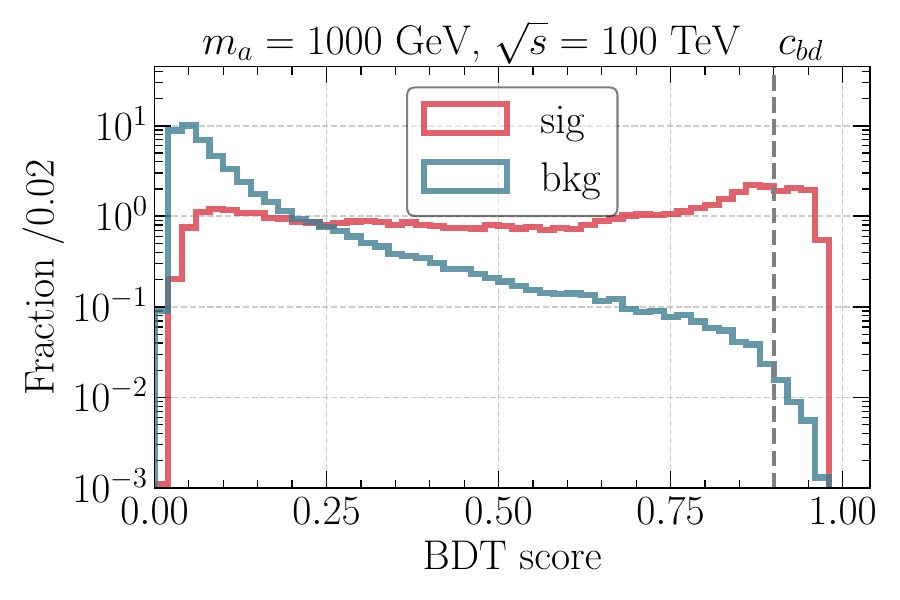}
      \minigraph{6.5cm}{-0.05in}{}{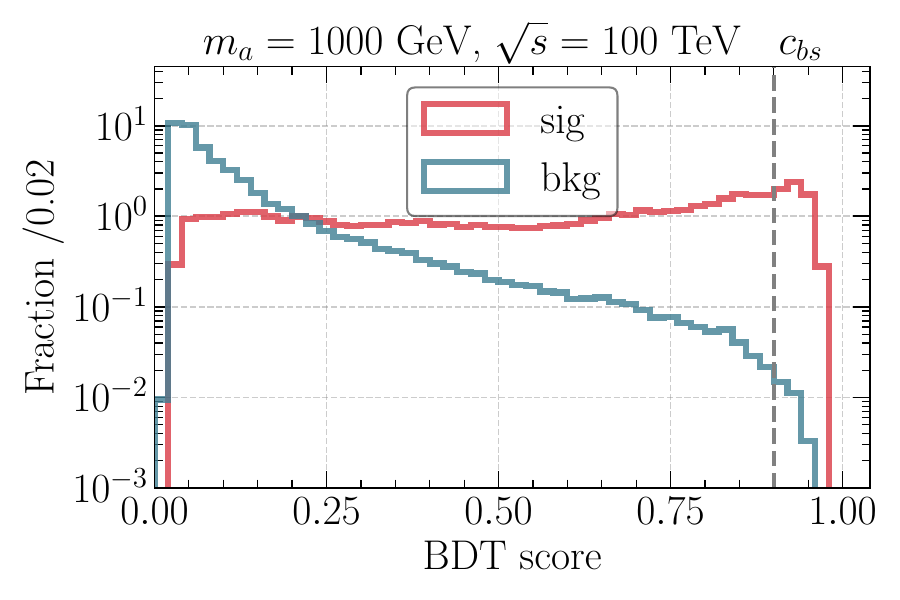}
\end{center}
\caption{ The BDT response score distribution of signal $bj+{\rm inv.}$ (red) and total SM background (blue) with $m_a=10,~50,~300$ and 1000 GeV at HL-LHC with $\sqrt{s}=100$ TeV and $\mathcal{L}=30~{\rm ab}^{-1}$. The grey dash line is the BDT cut that maximizes the significance with fixed $|c_{bd}^{V(A)}|/f_a = 1~{\rm TeV}^{-1}$ (left four panels) or $|c_{bs}^{V(A)}|/f_a = 1~{\rm TeV}^{-1}$ (right four panels).}
\label{fig:invBDT-score-FCC}
\end{figure}

For $pp\to jba$ with $a\to {\rm invisible}$, as the illustrations of HL-LHC ($\sqrt{s}=14$ TeV, $\mathcal{L}=3$ ab$^{-1}$) and FCC-hh ($\sqrt{s}=100$ TeV, $\mathcal{L}=30$ ab$^{-1}$), we also select four benchmarks ($m_a=10,~ 50, ~300$ and $1000$ GeV) to show the distribution of BDT-score in Fig.~\ref{fig:invBDT-score-HLLHC} (HL-LHC) and Fig.~\ref{fig:invBDT-score-FCC} (FCC-hh). The efficiencies of signal and backgrounds after applying the BDT-cut are shown in Table~\ref{tab:invBDT-HLLHC} (HL-LHC) and Table~\ref{tab:invBDT-FCC} (FCC-hh).

\begin{table}[htb!]
\centering
\resizebox{\textwidth}{!}{
\begin{tabular}{c|c|c|c|c|c|c|c|c}
\hline
     $m_a$ & BDT cut & $\epsilon_{\rm sig.}$ & $\epsilon_{jjZ}$ & $\epsilon_{bbZ}$ & $\epsilon_{jjW}$ & $\epsilon_{bbW}$ & $\epsilon_{t\bar{t}}$ &  $\mathcal{S}_{\rm max}$  \\
    \hline
    10   GeV& 0.55 & 9.11$\times10^{-2}$ & 1.24$\times10^{-3}$ & 8.39$\times10^{-4}$ & $-$ & 1.36$\times10^{-3}$ & 7.27$\times10^{-5}$ &  5.49$\times10^{1}$\\
    50   GeV& 0.35 & 9.68$\times10^{-2}$ & 3.27$\times10^{-3}$& 1.72$\times10^{-3}$ & 2.97$\times10^{-3}$  & 1.87$\times10^{-3}$ & 2.91$\times10^{-4}$ & 5.18$\times10^{0}$  \\
    300  GeV& 0.55 & 1.42$\times10^{-1}$ & 1.30$\times10^{-2}$ & 7.60$\times10^{-3}$ & 2.68$\times10^{-3}$ & 2.48$\times10^{-3}$ & 5.67$\times10^{-3}$ &  1.91$\times10^{-1}$ \\
    1000 GeV& 0.90 & 1.87$\times10^{-1}$& 2.82$\times10^{-3}$  & 8.24$\times10^{-4}$ & 2.97$\times10^{-4}$ & 1.87$\times10^{-4}$ &  7.99$\times10^{-4}$ & 1.20$\times10^{-2}$ \\
    \hline
    \hline
    10   GeV& 0.55 & 8.42$\times10^{-2}$ & 4.52$\times10^{-4}$ & 1.13$\times10^{-3}$ &  $-$  & 8.43$\times10^{-4}$ & $-$ &  2.85$\times10^{1}$\\
    50   GeV& 0.40 & 3.91$\times10^{-2}$ &  $-$ & 3.09$\times10^{-4}$ & 2.97$\times10^{-4}$ & 1.41$\times10^{-4}$ & $-$  & 2.72$\times10^{0}$  \\
    300  GeV& 0.60 & 7.03$\times10^{-1}$ & 2.37$\times10^{-3}$ & 1.28$\times10^{-3}$ & 2.97$\times10^{-4}$ & 3.75$\times10^{-4}$ & 7.27$\times10^{-4}$ & 6.29$\times10^{-2}$ \\
    1000 GeV& 0.90 & 1.64$\times10^{-1}$& 2.14$\times10^{-3}$  & 9.86$\times10^{-4}$& $-$ & 3.74$\times10^{-4}$ & 9.45$\times10^{-4}$ & 2.38$\times10^{-3}$ \\
    \hline
\end{tabular}}
\caption{The BDT cut, cut efficiencies and achieved maximal significance in Eq.~(\ref{eqn-significance}) for the signal $bj+{\rm invisible}$ and SM backgrounds at HL-LHC with $\sqrt{s}=14$ TeV and $\mathcal{L}=3~{\rm ab}^{-1}$. The benchmark masses are $m_a=10,~50,~300$ and 1000 GeV and the parameter is fixed as $|c_{bd}^{V(A)}|/f_a = 1~{\rm TeV}^{-1}$ (above the double line) or $|c_{bs}^{V(A)}|/f_a = 1~{\rm TeV}^{-1}$ (below the double line). The label ``$-$'' denotes the background at negligible level.}
\label{tab:invBDT-HLLHC}
\end{table}

\begin{table}[htb!]
\centering
\resizebox{\textwidth}{!}{
\begin{tabular}{c|c|c|c|c|c|c|c|c}
\hline
     $m_a$ & BDT cut & $\epsilon_{\rm sig.}$ & $\epsilon_{jjZ}$ & $\epsilon_{bbZ}$ & $\epsilon_{jjW}$ & $\epsilon_{bbW}$ & $\epsilon_{t\bar{t}}$ &  $\mathcal{S}_{\rm max}$  \\
    \hline
    10   GeV& 0.65 & 8.74$\times10^{-2}$ & 1.76$\times10^{-3}$ & 1.56$\times10^{-3}$ & 5.13$\times10^{-4}$ & 1.78$\times10^{-3}$& $-$ &  5.16$\times10^{2}$\\
    50   GeV& 0.45 & 5.54$\times10^{-2}$ & 1.29$\times10^{-3}$& 7.75$\times10^{-4}$ & 1.54$\times10^{-3}$  & 6.11$\times10^{-4}$  & 2.46$\times10^{-4}$ & 7.81$\times10^{1}$  \\
    300  GeV& 0.50 & 1.62$\times10^{-1}$ & 1.37$\times10^{-2}$ & 6.91$\times10^{-3}$ &  4.88$\times10^{-3}$  & 3.21$\times10^{-3}$   & 1.17$\times10^{-2}$  &  4.36$\times10^{0}$ \\
    1000 GeV& 0.90 & 1.28$\times10^{-1}$& 1.54$\times10^{-3}$  & 4.95$\times10^{-4}$ &  $-$ & 3.56$\times10^{-4}$ &  7.37$\times10^{-4}$ & 5.63$\times10^{-1}$ \\
    \hline
    \hline
    10   GeV& 0.60 & 1.49$\times10^{-1}$ & 5.02$\times10^{-3}$ & 4.22$\times10^{-3}$ &  2.05$\times10^{-3}$  & 4.68$\times10^{-3}$ & 1.23$\times10^{-4}$ &  1.97$\times10^{2}$\\
    50   GeV& 0.50 & 1.58$\times10^{-2}$ & 2.51$\times10^{-4}$ & 1.40$\times10^{-4}$ & $-$ & 5.09$\times10^{-5}$& $-$  & 2.34$\times10^{1}$  \\
    300  GeV& 0.50 & 1.16$\times10^{-1}$ & 9.50$\times10^{-3}$ & 4.95$\times10^{-3}$ & 7.70$\times10^{-4}$ & 1.98$\times10^{-3}$ & 4.18$\times10^{-3}$ & 1.11$\times10^{0}$ \\
    1000 GeV& 0.90 & 1.27$\times10^{-1}$& 1.51$\times10^{-3}$  & 5.17$\times10^{-4}$& 5.13$\times10^{-4}$ & 3.05$\times10^{-4}$ & 1.23$\times10^{-4}$ & 1.05$\times10^{-1}$ \\
    \hline
\end{tabular}}
\caption{The BDT cut, cut efficiencies and achieved maximal significance in Eq.~(\ref{eqn-significance}) for the signal $bj+{\rm invisible}$ and SM backgrounds at FCC-hh with $\sqrt{s}=100$ TeV and $\mathcal{L}=30~{\rm ab}^{-1}$. The benchmark masses are $m_a=10,~50,~300$ and 1000 GeV and the parameter is fixed as $|c_{bd}^{V(A)}|/f_a = 1~{\rm TeV}^{-1}$ (above the double line) or $|c_{bs}^{V(A)}|/f_a = 1~{\rm TeV}^{-1}$ (below the double line). The label ``$-$'' denotes the background at negligible level.}
\label{tab:invBDT-FCC}
\end{table}

\clearpage
\subsection{$a\to \mu^+ \mu^-$}
\label{App:mumu}

\begin{figure}[th!]
\begin{center}
      \minigraph{6.5cm}{-0.05in}{}{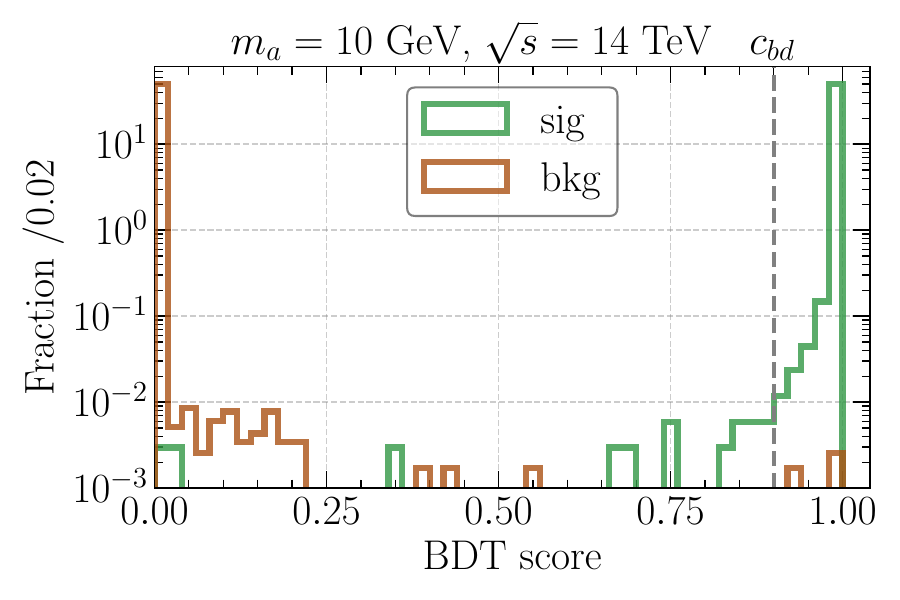}
      \minigraph{6.5cm}{-0.05in}{}{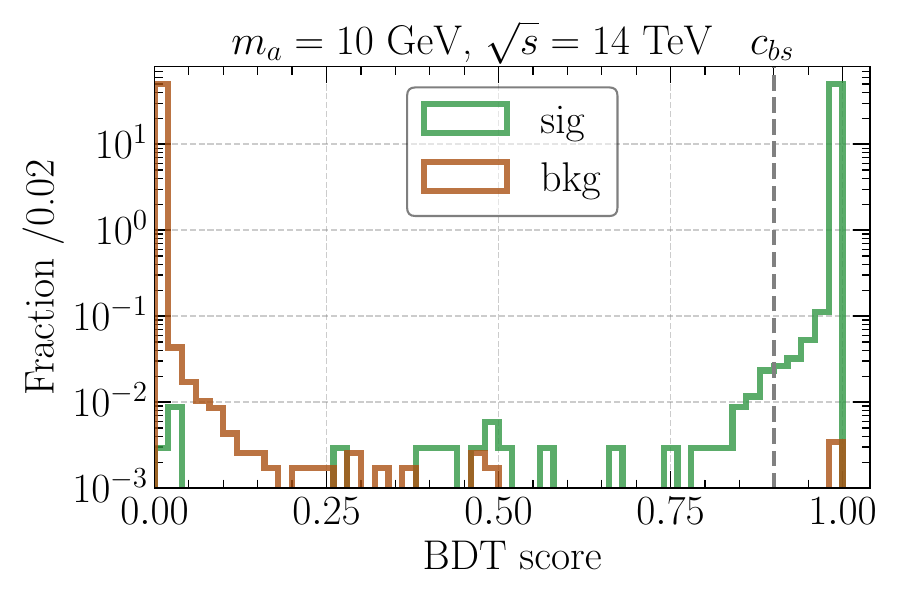}\\
      \minigraph{6.5cm}{-0.05in}{}{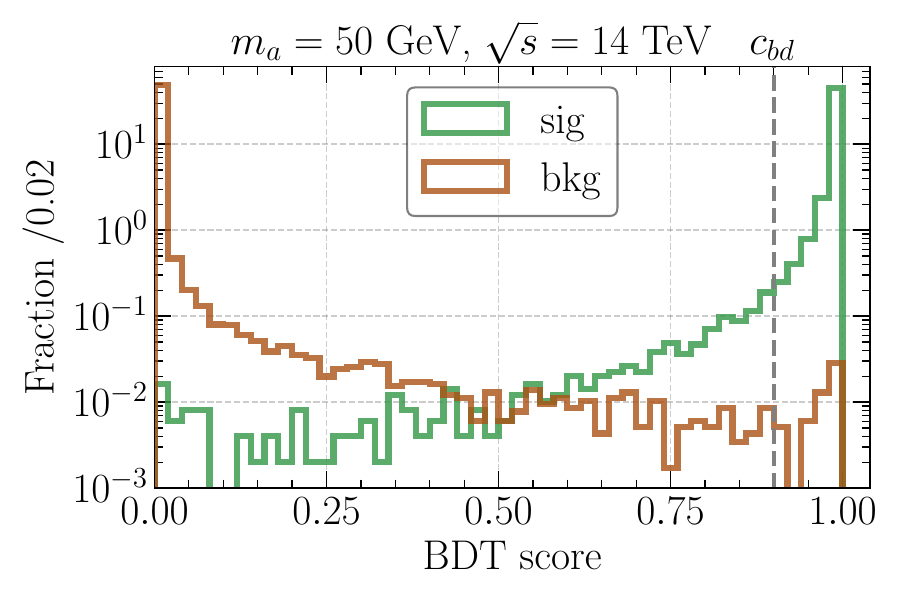}
      \minigraph{6.5cm}{-0.05in}{}{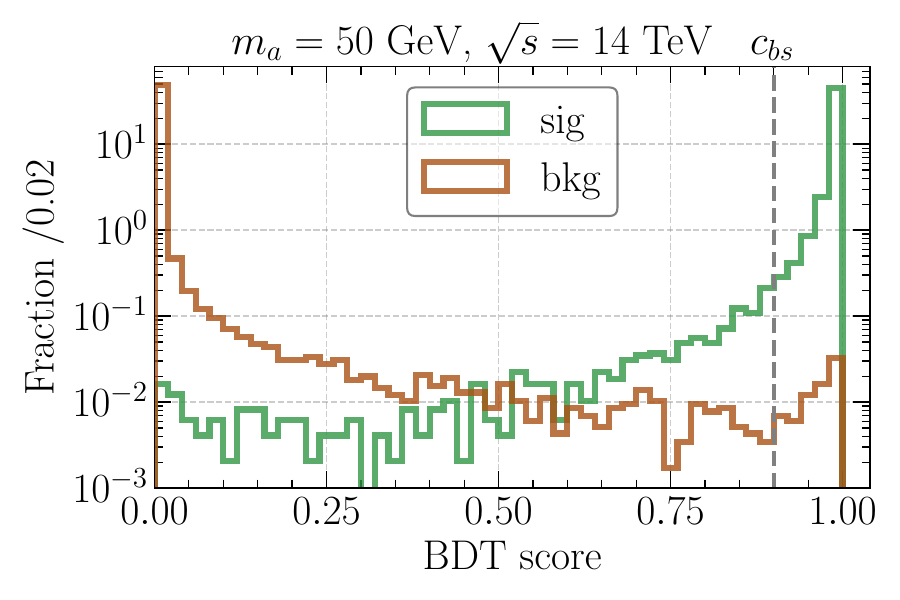}\\
      \minigraph{6.5cm}{-0.05in}{}{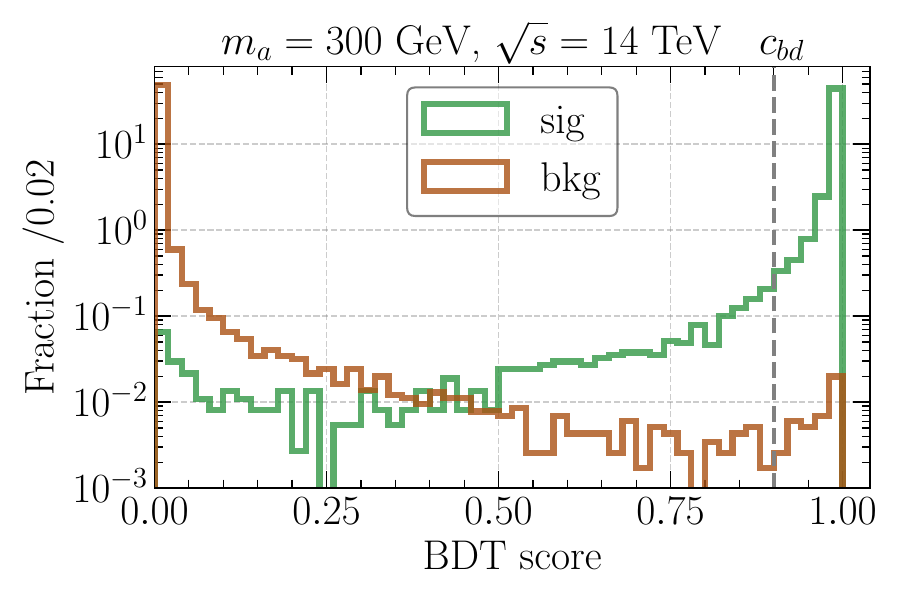}
      \minigraph{6.5cm}{-0.05in}{}{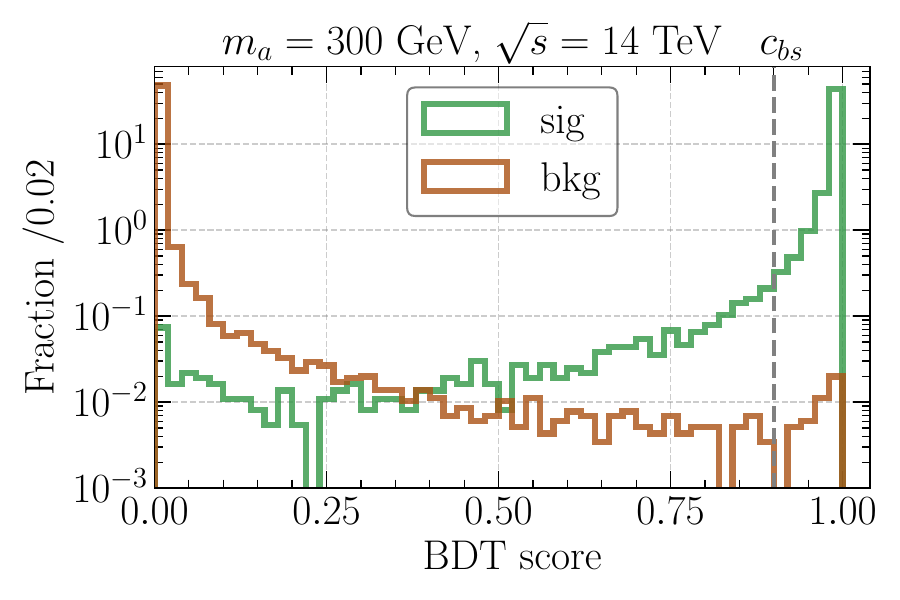}\\
      \minigraph{6.5cm}{-0.05in}{}{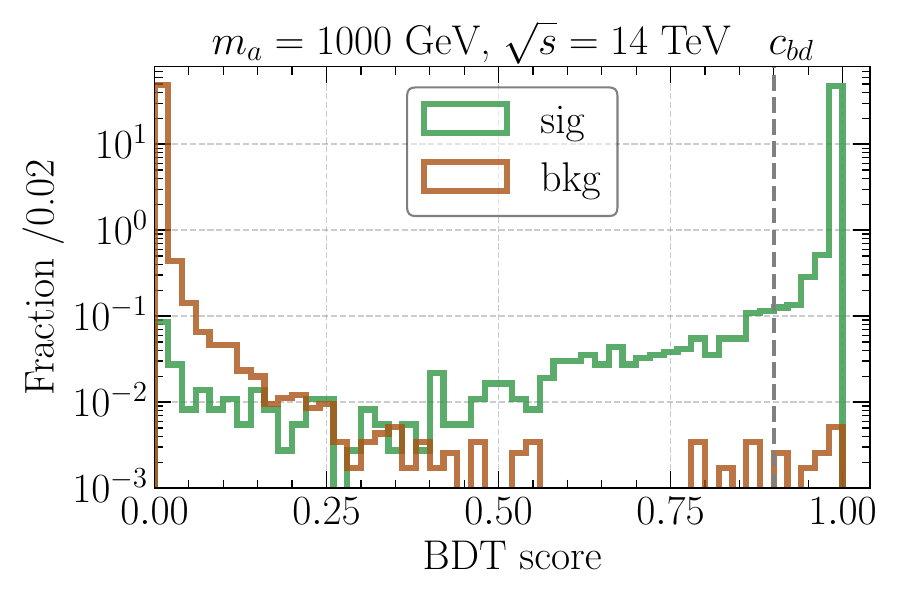}
      \minigraph{6.5cm}{-0.05in}{}{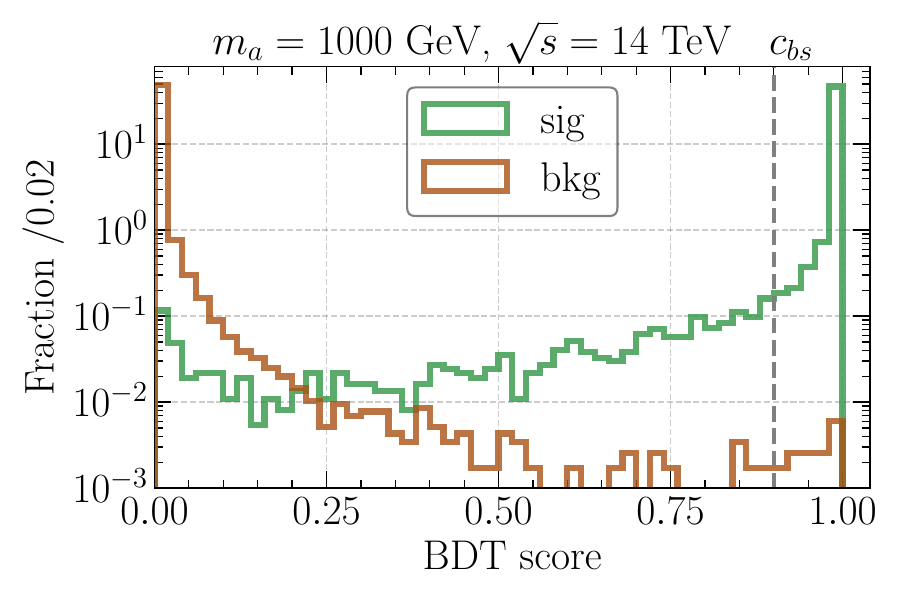}
\end{center}
\caption{ The BDT response score distribution of signal $bj\mu\mu$ (green) and total SM background (brown) with $m_a=10,~50,~300$ and 1000 GeV at HL-LHC with $\sqrt{s}=14$ TeV and $\mathcal{L}=3~{\rm ab}^{-1}$. The grey dash line is the BDT cut that maximizes the significance with fixed $|c_{bd}^{V(A)}|/f_a = 1~{\rm TeV}^{-1}$ (left four panels) or $|c_{bs}^{V(A)}|/f_a = 1~{\rm TeV}^{-1}$ (right four panels).}
\label{fig:mumuBDT-score-HLLHC}
\end{figure}

\begin{figure}[th!]
\begin{center}
      \minigraph{6.5cm}{-0.05in}{}{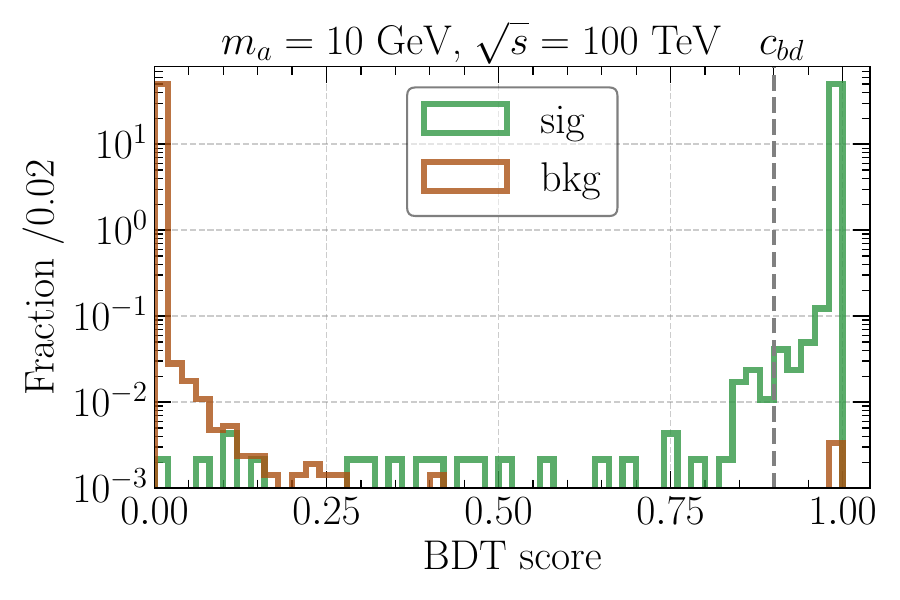}
      \minigraph{6.5cm}{-0.05in}{}{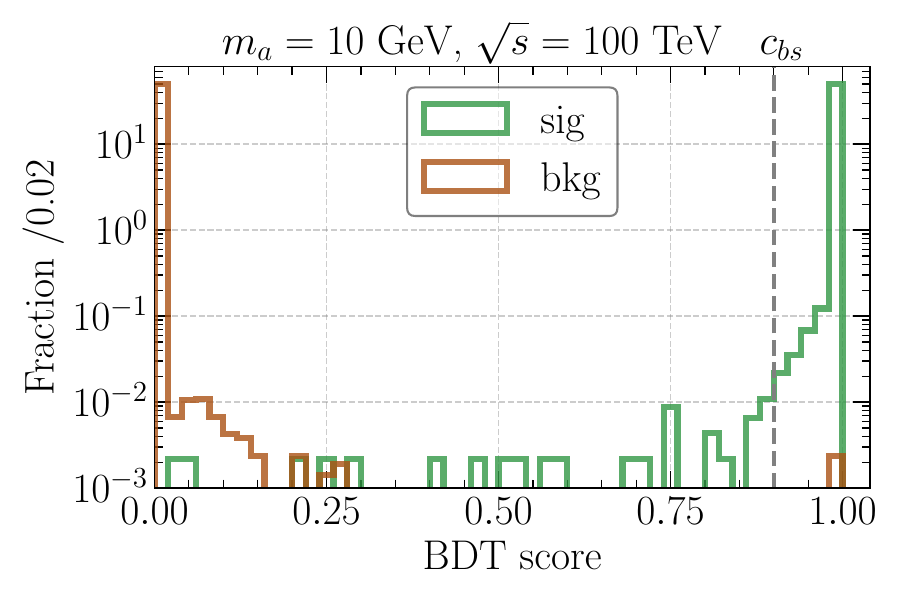}\\
      \minigraph{6.5cm}{-0.05in}{}{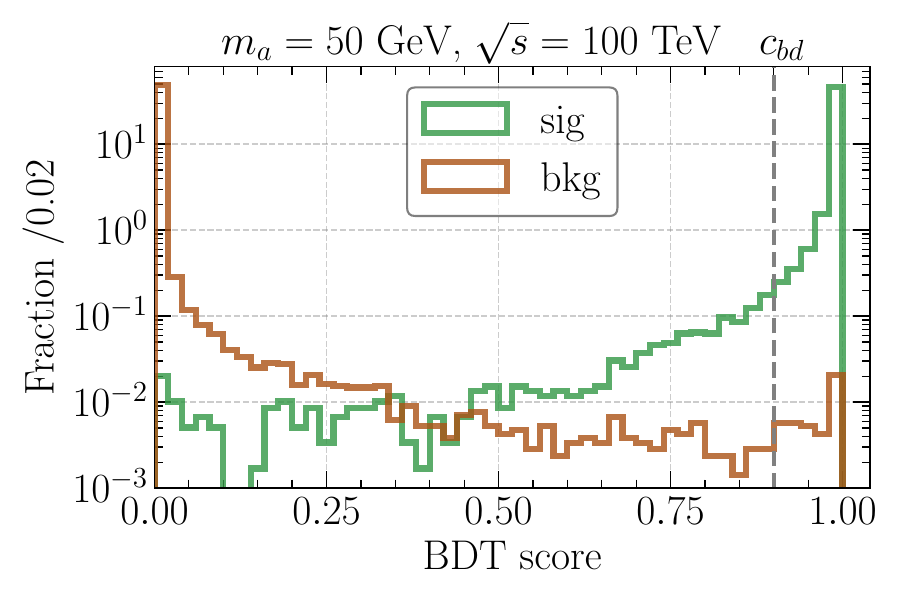}
      \minigraph{6.5cm}{-0.05in}{}{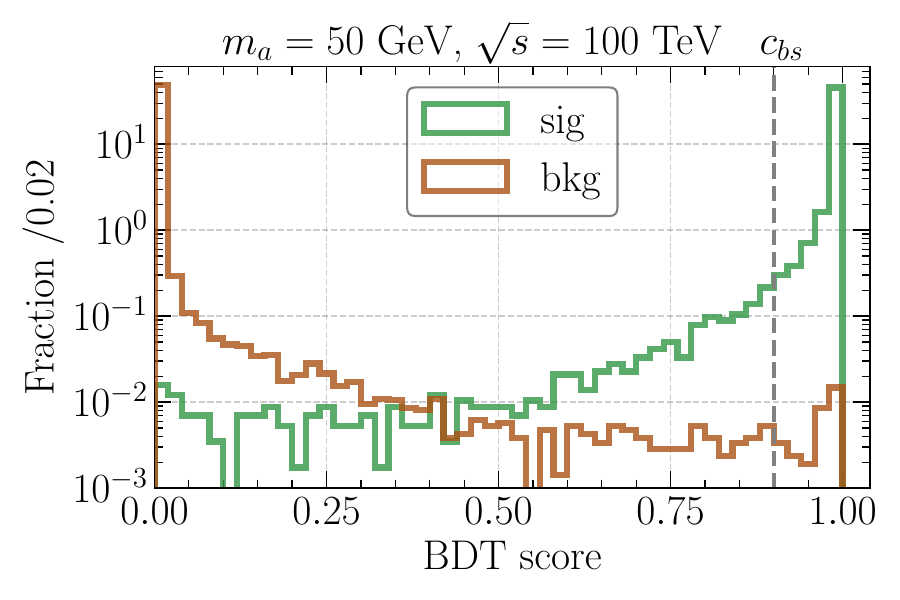}\\
      \minigraph{6.5cm}{-0.05in}{}{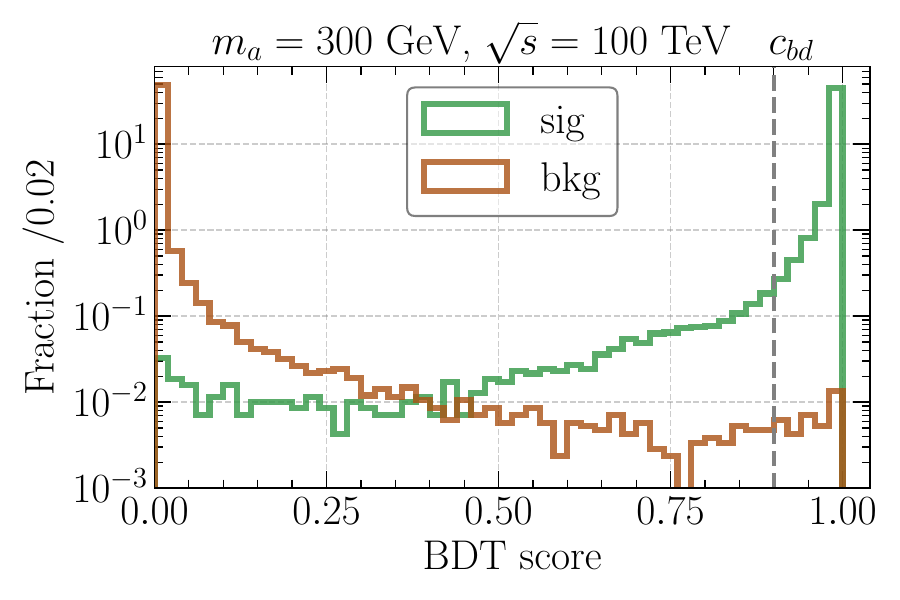}
      \minigraph{6.5cm}{-0.05in}{}{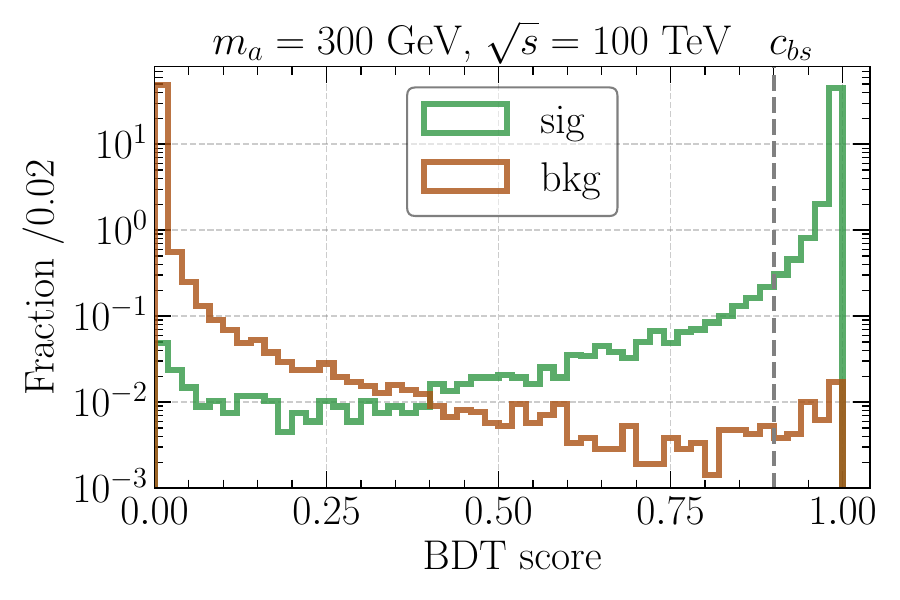}\\
      \minigraph{6.5cm}{-0.05in}{}{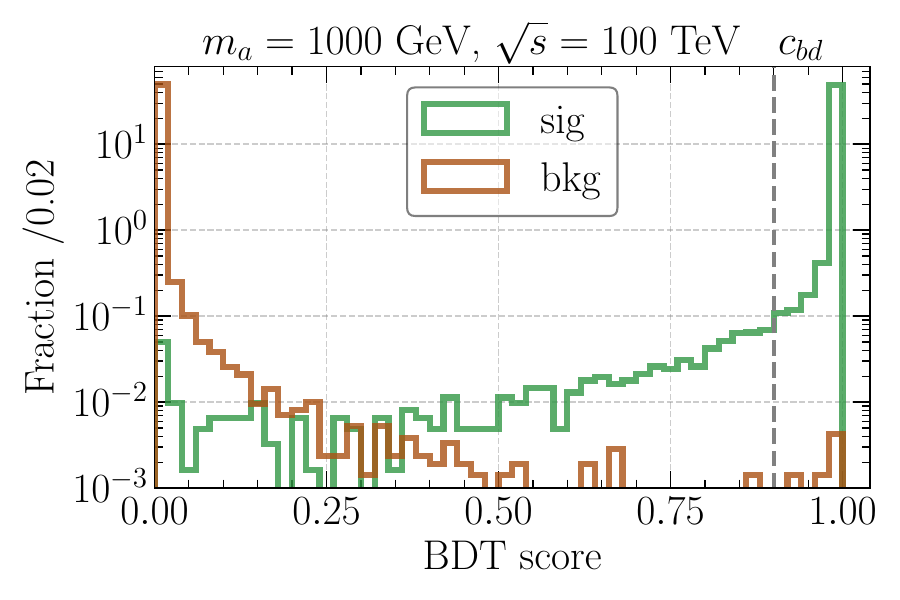}
      \minigraph{6.5cm}{-0.05in}{}{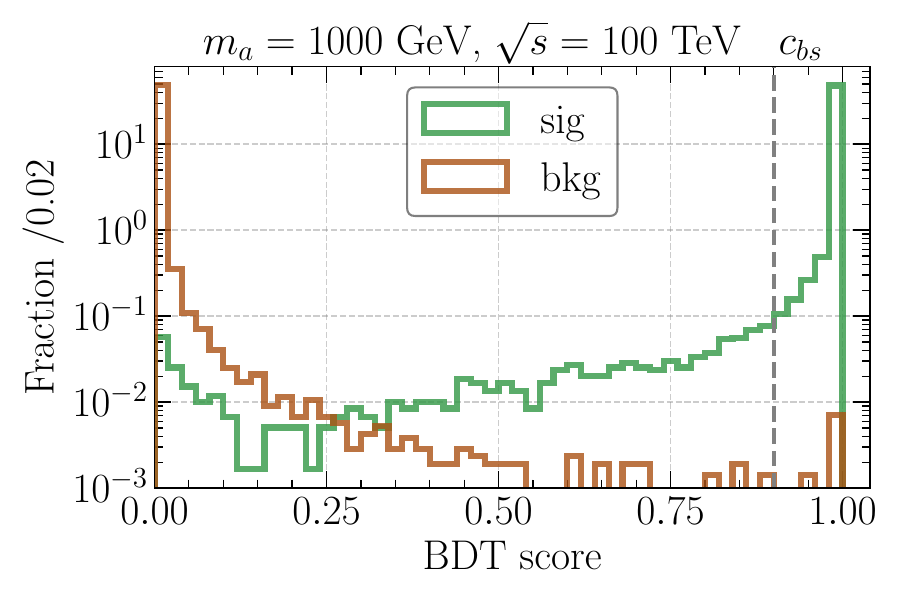}
\end{center}
\caption{ The BDT response score distribution of signal $bj\mu\mu$ (green) and total SM background (brown) with $m_a=10,~50,~300$ and 1000 GeV at HL-LHC with $\sqrt{s}=100$ TeV and $\mathcal{L}=30~{\rm ab}^{-1}$. The grey dash line is the BDT cut that maximizes the significance with fixed $|c_{bd}^{V(A)}|/f_a = 1~{\rm TeV}^{-1}$ (left four panels) or $|c_{bs}^{V(A)}|/f_a = 1~{\rm TeV}^{-1}$ (right four panels).}
\label{fig:mumuBDT-score-FCC}
\end{figure}

For $pp\to jba$ with $a\to \mu^+ \mu^-$, as the illustrations of HL-LHC ($\sqrt{s}=14$ TeV, $\mathcal{L}=3$ ab$^{-1}$) and FCC-hh ($\sqrt{s}=100$ TeV, $\mathcal{L}=30$ ab$^{-1}$), we also select four benchmarks ($m_a=10,~ 50, ~300$ and $1000$ GeV) to show the distribution of BDT-score in Fig.~\ref{fig:mumuBDT-score-HLLHC} (HL-LHC) and Fig.~\ref{fig:mumuBDT-score-FCC} (FCC-hh). The efficiencies of signal and backgrounds after applying the BDT-cut are shown in Table~\ref{tab:mumuBDT-HLLHC} (HL-LHC) and Table~\ref{tab:mumuBDT-FCC} (FCC-hh).

\begin{table}[htb!]
\centering
\resizebox{\textwidth}{!}{
\begin{tabular}{c|c|c|c|c|c|c|c|c}
\hline
     $m_a$ & BDT cut & $\epsilon_{\rm sig.}$ & $\epsilon_{jjZ}$ & $\epsilon_{bbZ}$ & $\epsilon_{jjW}$ & $\epsilon_{bbW}$ & $\epsilon_{t\bar{t}}$ &  $\mathcal{S}_{\rm max}$  \\
    \hline
    10   GeV& 0.90 & 9.98$\times10^{-1}$ & 2.57$\times10^{-4}$ & 6.26$\times10^{-5}$ & 1.04$\times10^{-3}$  & 4.73$\times10^{-4}$ & 9.36$\times10^{-5}$ & 1.10$\times10^{3}$\\
    50   GeV& 0.90 & 9.79$\times10^{-1}$ & 5.16$\times10^{-4}$ & 3.13$\times10^{-4}$ & 6.24$\times10^{-3}$  & 2.93$\times10^{-3}$ & 1.97$\times10^{-3}$ & 3.10$\times10^{2}$  \\
    300  GeV& 0.90 & 9.70$\times10^{-1}$ & 2.58$\times10^{-4}$ & 3.13$\times10^{-5}$  & 4.16$\times10^{-3}$ & 1.99$\times10^{-3}$ & 1.59$\times10^{-3}$ &  1.75$\times10^{1}$ \\
    1000 GeV& 0.90 & 9.79$\times10^{-1}$& $-$  & 3.13$\times10^{-5}$ &  5.20$\times10^{-3}$ & 9.46$\times10^{-5}$ &  4.68$\times10^{-4}$ & 7.34$\times10^{-1}$ \\
    \hline
    \hline
    10   GeV& 0.90 & 9.98$\times10^{-1}$ & 5.16$\times10^{-4}$ & 6.26$\times10^{-5}$ &  $-$  & 4.73$\times10^{-4}$ & 9.36$\times10^{-5}$ &  1.03$\times10^{3}$\\
    50   GeV& 0.90 & 9.77$\times10^{-1}$ & 7.74$\times10^{-4}$ & 3.44$\times10^{-4}$ & 3.12$\times10^{-3}$ & 2.46$\times10^{-3}$& 3.18$\times10^{-3}$  & 2.84$\times10^{2}$  \\
    300  GeV& 0.90 & 9.67$\times10^{-1}$ & 2.58$\times10^{-4}$ & 1.25$\times10^{-4}$ & 3.12$\times10^{-3}$ & 2.74$\times10^{-3}$ & 2.81$\times10^{-3}$ & 4.25$\times10^{0}$ \\
    1000 GeV& 0.90 & 9.66$\times10^{-1}$& $-$  & 3.13$\times10^{-5}$& 1.04$\times10^{-3}$ & 2.84$\times10^{-4}$ & 3.74$\times10^{-4}$ & 1.35$\times10^{-1}$ \\
    \hline
\end{tabular}}
\caption{The BDT cut, cut efficiencies and achieved maximal significance in Eq.~(\ref{eqn-significance}) for the signal $bj\mu\mu$ and SM backgrounds at HL-LHC with $\sqrt{s}=14$ TeV and $\mathcal{L}=3~{\rm ab}^{-1}$. The benchmark masses are $m_a=10,~50,~300$ and 1000 GeV and the parameter is fixed as $|c_{bd}^{V(A)}|/f_a = 1~{\rm TeV}^{-1}$ (above the double line) or $|c_{bs}^{V(A)}|/f_a = 1~{\rm TeV}^{-1}$ (below the double line). The label ``$-$'' denotes the background at negligible level.}
\label{tab:mumuBDT-HLLHC}
\end{table}

\begin{table}[htb!]
\centering
\resizebox{\textwidth}{!}{
\begin{tabular}{c|c|c|c|c|c|c|c|c}
\hline
     $m_a$ & BDT cut & $\epsilon_{\rm sig.}$ & $\epsilon_{jjZ}$ & $\epsilon_{bbZ}$ & $\epsilon_{jjW}$ & $\epsilon_{bbW}$ & $\epsilon_{t\bar{t}}$ &  $\mathcal{S}_{\rm max}$  \\
    \hline
    10   GeV& 0.90 & 9.98$\times10^{-1}$ & 2.64$\times10^{-4}$ & 6.04$\times10^{-5}$ & $-$ & 2.40$\times10^{-4}$ & $-$ &  1.69$\times10^{4}$\\
    50   GeV& 0.90 & 9.79$\times10^{-1}$ & 3.51$\times10^{-4}$ & 2.56$\times10^{-4}$ & 2.38$\times10^{-3}$ & 1.28$\times10^{-3}$ & 1.90$\times10^{-3}$ & 5.48$\times10^{3}$  \\
    300  GeV& 0.90 & 9.71$\times10^{-1}$ & 1.76$\times10^{-4}$ & 3.02$\times10^{-5}$  & 8.57$\times10^{-3}$ & 3.28$\times10^{-3}$ & 2.53$\times10^{-3}$ &  3.80$\times10^{2}$ \\
    1000 GeV& 0.90 & 9.87$\times10^{-1}$ & $-$  & 1.51$\times10^{-5}$ &  2.86$\times10^{-3}$ & 4.80$\times10^{-4}$ &  3.17$\times10^{-4}$ & 4.22$\times10^{1}$ \\
    \hline
    \hline
    10   GeV& 0.90 & 9.99$\times10^{-1}$ & 8.78$\times10^{-5}$ & 9.05$\times10^{-5}$ & 4.76$\times10^{-4}$ & 3.20$\times10^{-4}$ & 7.91$\times10^{-5}$ &  1.40$\times10^{4}$\\
    50   GeV& 0.90 & 9.77$\times10^{-1}$ & 8.77$\times10^{-4}$ & 2.41$\times10^{-4}$ & 2.86$\times10^{-3}$ & 8.79$\times10^{-4}$ & 1.66$\times10^{-3}$  & 3.24$\times10^{3}$  \\
    300  GeV& 0.90 & 9.66$\times10^{-1}$ & 1.76$\times10^{-4}$ & 1.06$\times10^{-4}$ & 3.33$\times10^{-3}$ & 2.24$\times10^{-3}$ & 2.06$\times10^{-3}$ & 2.01$\times10^{2}$ \\
    1000 GeV& 0.90 & 9.83$\times10^{-1}$ & $-$  & $-$& 4.29$\times10^{-3}$ & 4.00$\times10^{-4}$ & 5.54$\times10^{-4}$ & 1.41$\times10^{1}$ \\
    \hline
\end{tabular}}
\caption{The BDT cut, cut efficiencies and achieved maximal significance in Eq.~(\ref{eqn-significance}) for the signal $bj\mu\mu$ and SM backgrounds at FCC-hh with $\sqrt{s}=100$ TeV and $\mathcal{L}=30~{\rm ab}^{-1}$. The benchmark masses are $m_a=10,~50,~300$ and 1000 GeV and the parameter is fixed as $|c_{bd}^{V(A)}|/f_a = 1~{\rm TeV}^{-1}$ (above the double line) or $|c_{bs}^{V(A)}|/f_a = 1~{\rm TeV}^{-1}$ (below the double line). The label ``$-$'' denotes the background at negligible level.}
\label{tab:mumuBDT-FCC}
\end{table}

\clearpage


\subsection{$a\to \gamma \gamma$}
\label{App:gaga}

\begin{figure}[th!]
\begin{center}
      \minigraph{6.5cm}{-0.05in}{}{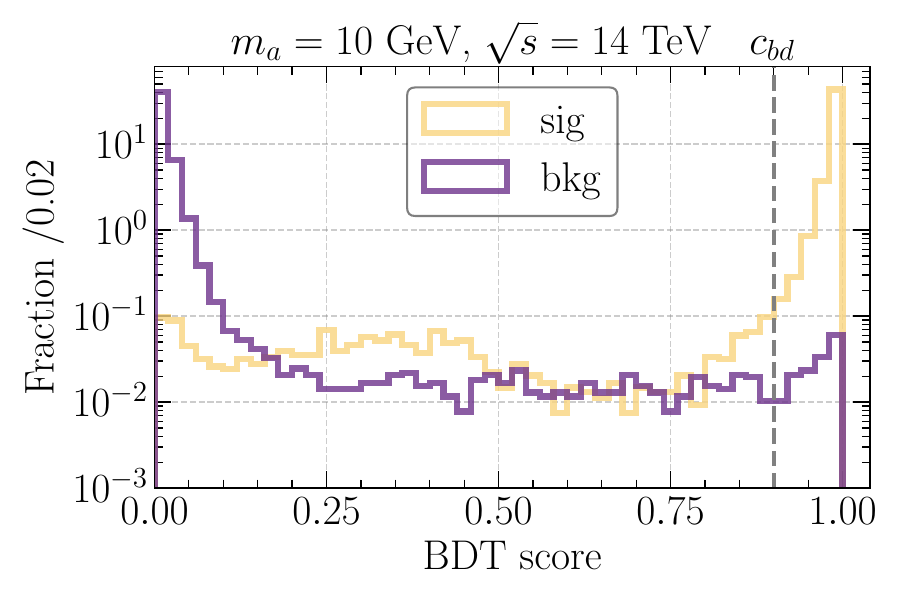}
      \minigraph{6.5cm}{-0.05in}{}{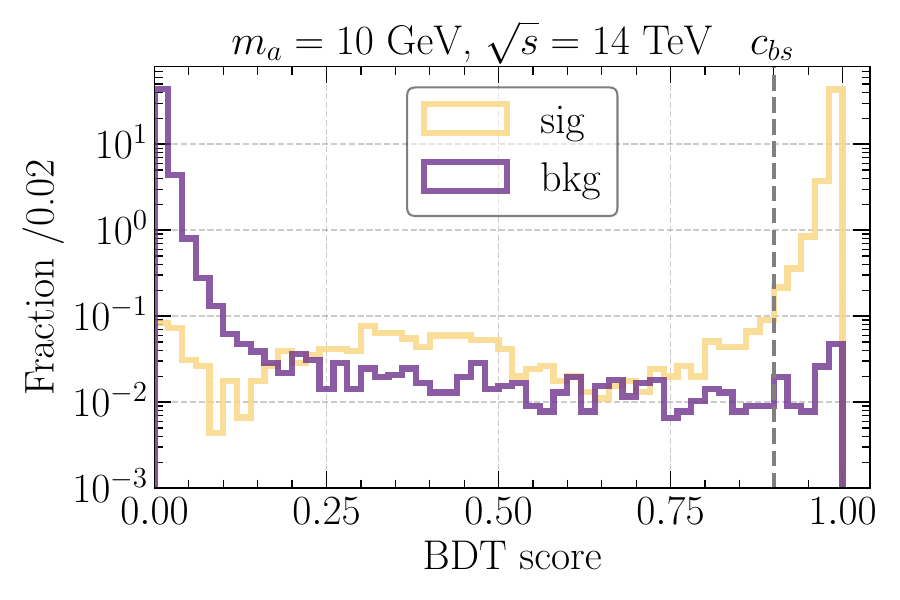}\\
      \minigraph{6.5cm}{-0.05in}{}{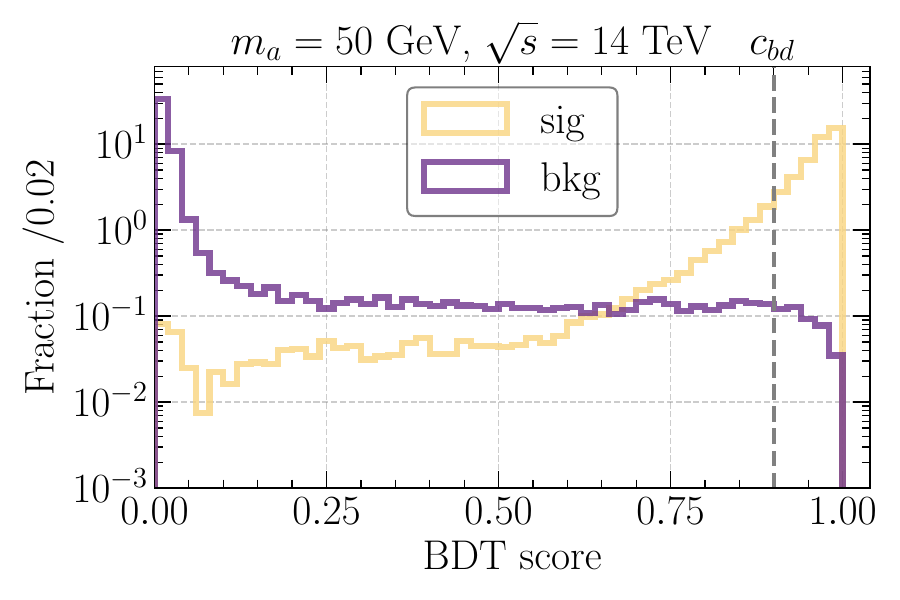}
      \minigraph{6.5cm}{-0.05in}{}{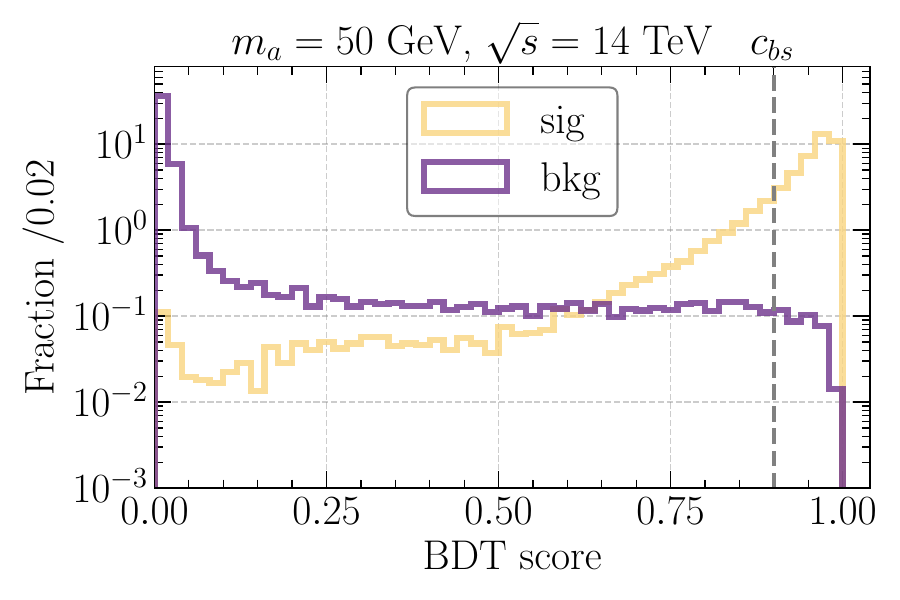}\\
      \minigraph{6.5cm}{-0.05in}{}{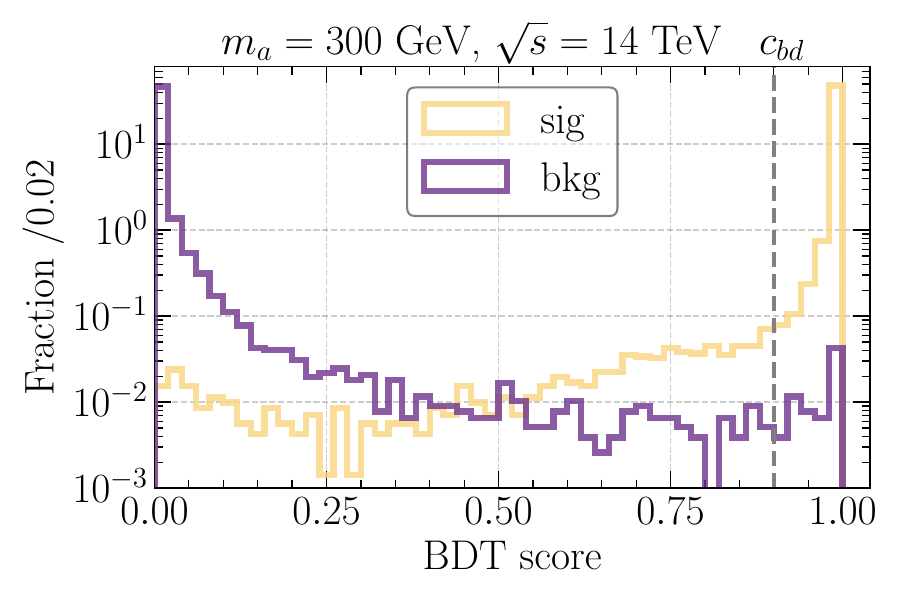}
      \minigraph{6.5cm}{-0.05in}{}{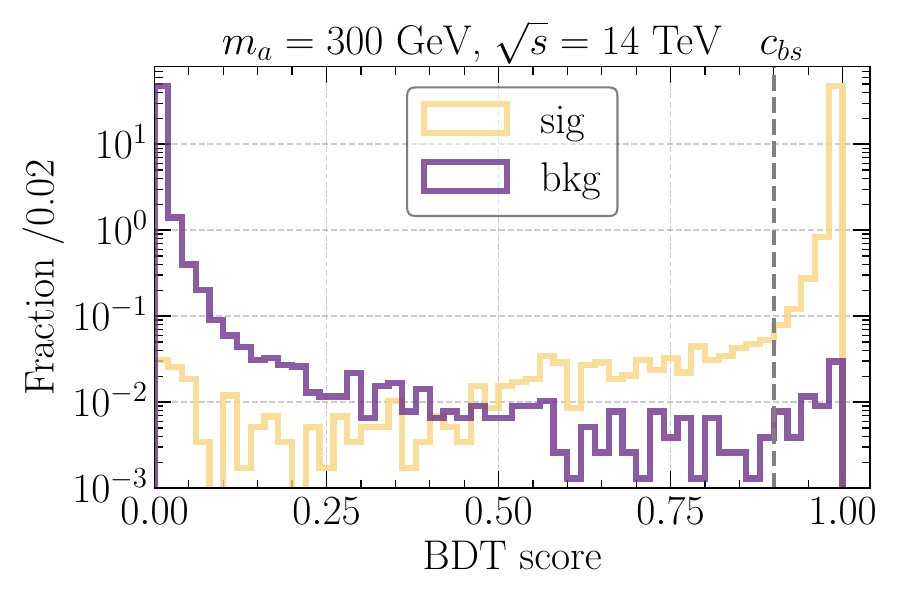}\\
      \minigraph{6.5cm}{-0.05in}{}{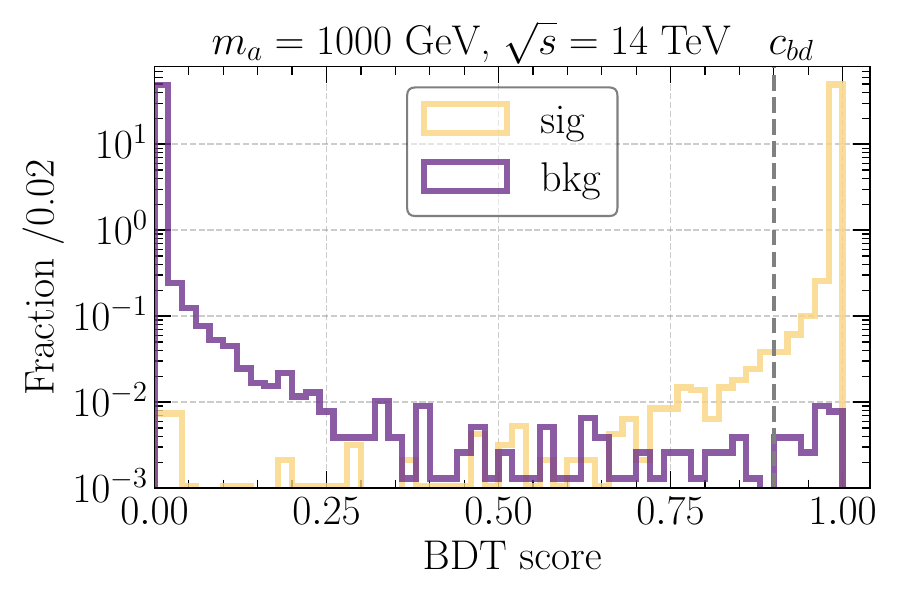}
      \minigraph{6.5cm}{-0.05in}{}{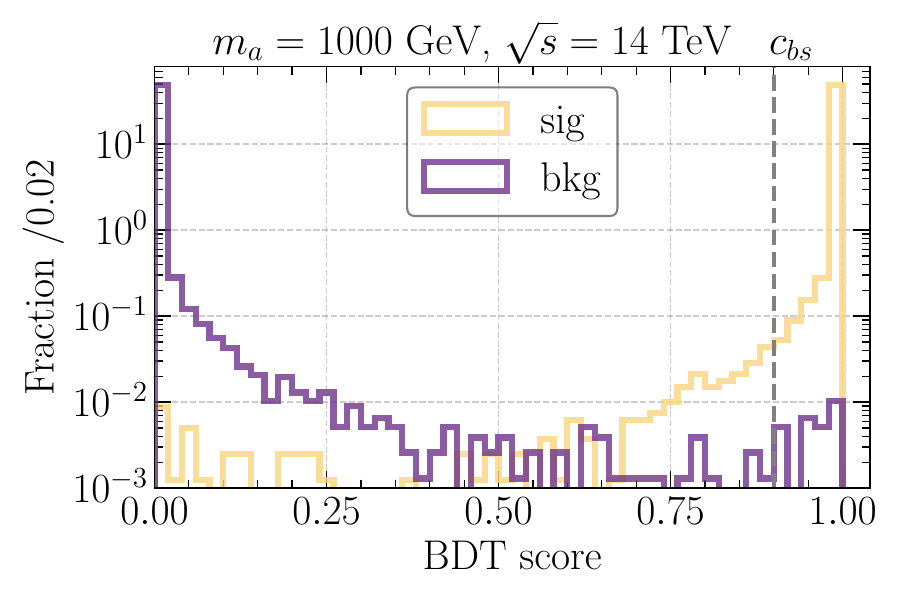}
\end{center}
\caption{ The BDT response score distribution of signal $bj\gamma\gamma$ (yellow) and total SM background (purple) with $m_a=10,~50,~300$ and 1000 GeV at HL-LHC with $\sqrt{s}=14$ TeV and $\mathcal{L}=3~{\rm ab}^{-1}$. The grey dash line is the BDT cut that maximizes the significance with fixed $|c_{bd}^{V(A)}|/f_a = 1~{\rm TeV}^{-1}$ (left four panels) or $|c_{bs}^{V(A)}|/f_a = 1~{\rm TeV}^{-1}$ (right four panels).}
\label{fig:aaBDT-score-HLLHC}
\end{figure}

\begin{figure}[th!]
\begin{center}
      \minigraph{6.5cm}{-0.05in}{}{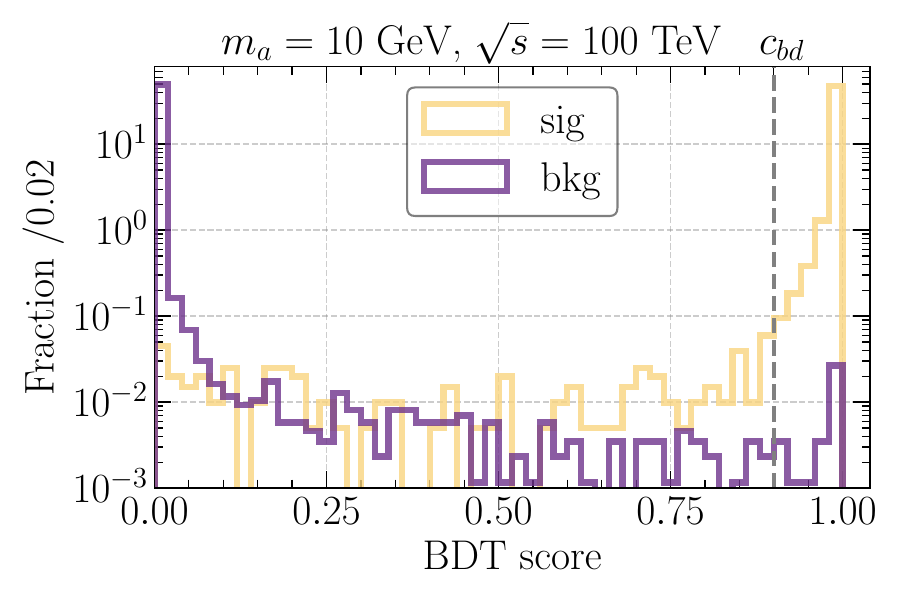}
      \minigraph{6.5cm}{-0.05in}{}{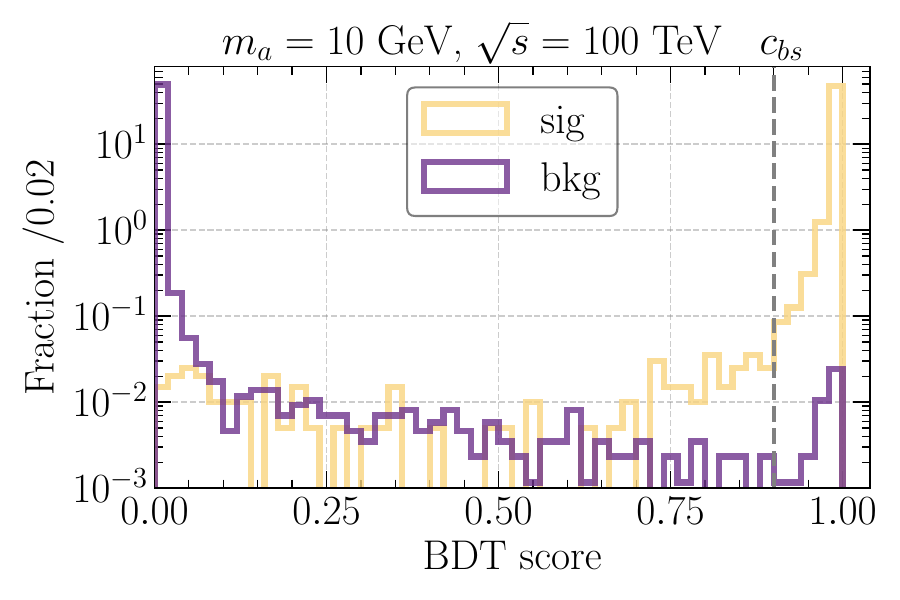}\\
      \minigraph{6.5cm}{-0.05in}{}{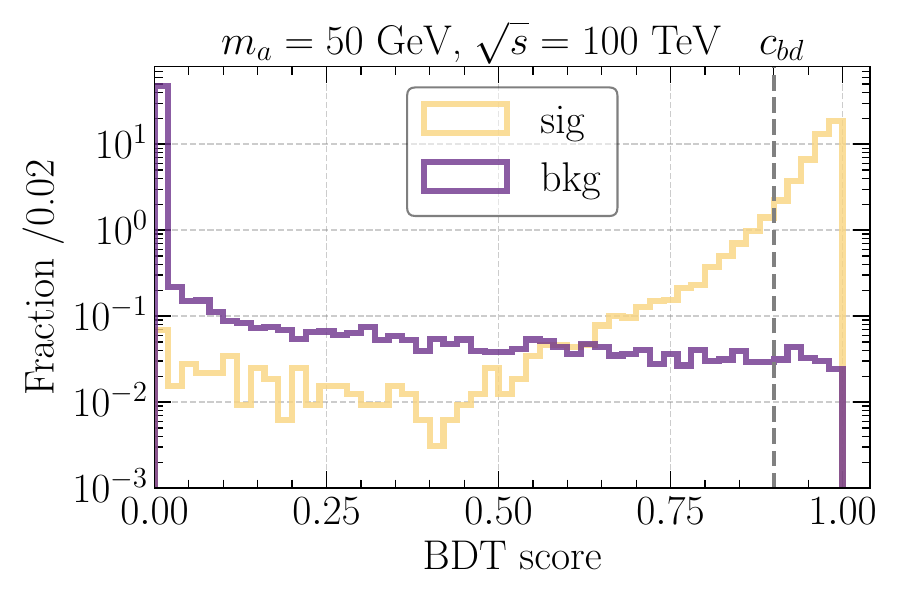}
      \minigraph{6.5cm}{-0.05in}{}{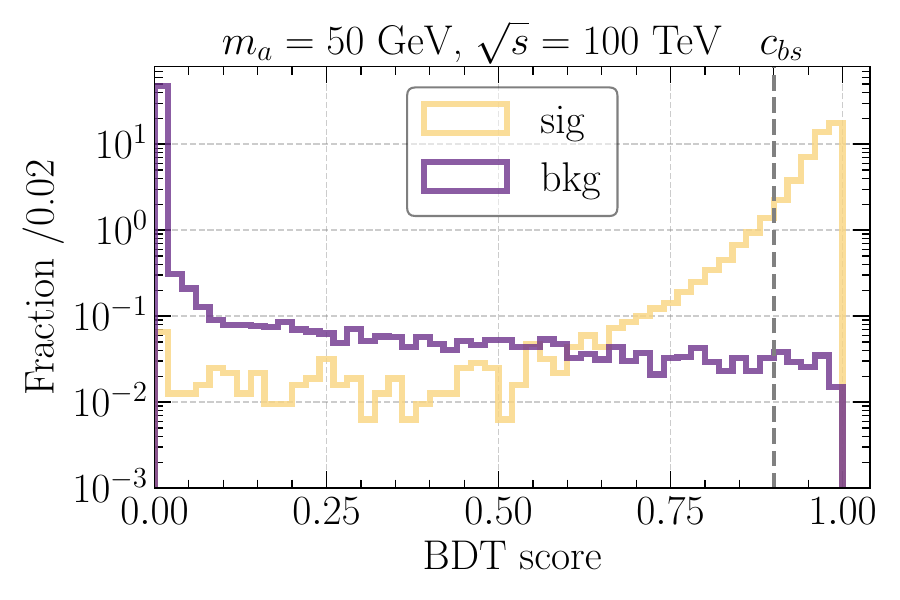}\\
      \minigraph{6.5cm}{-0.05in}{}{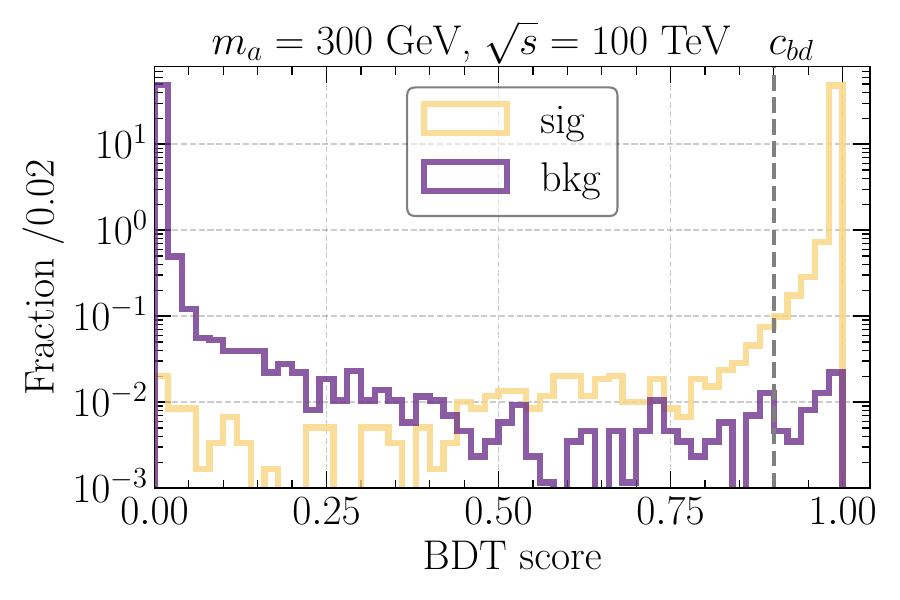}
      \minigraph{6.5cm}{-0.05in}{}{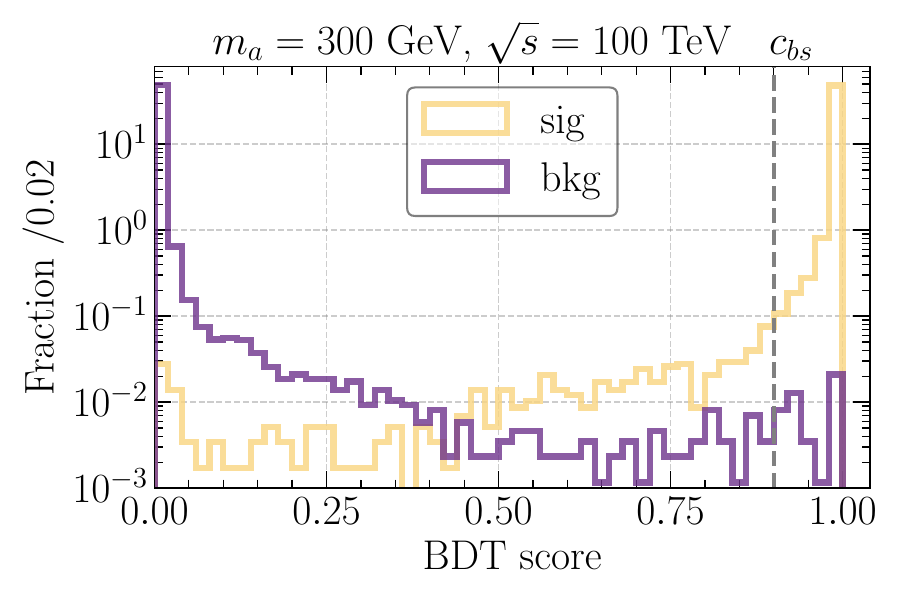}\\
      \minigraph{6.5cm}{-0.05in}{}{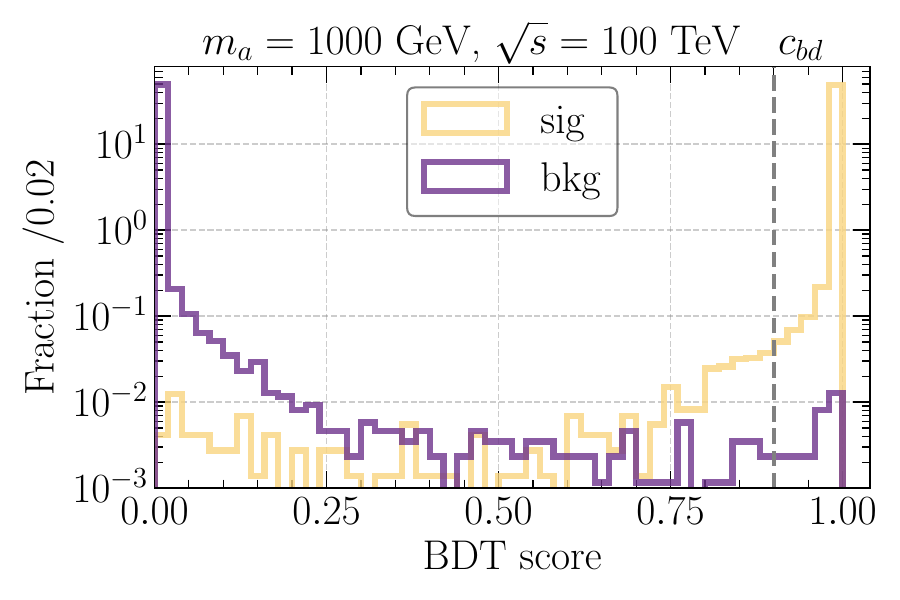}
      \minigraph{6.5cm}{-0.05in}{}{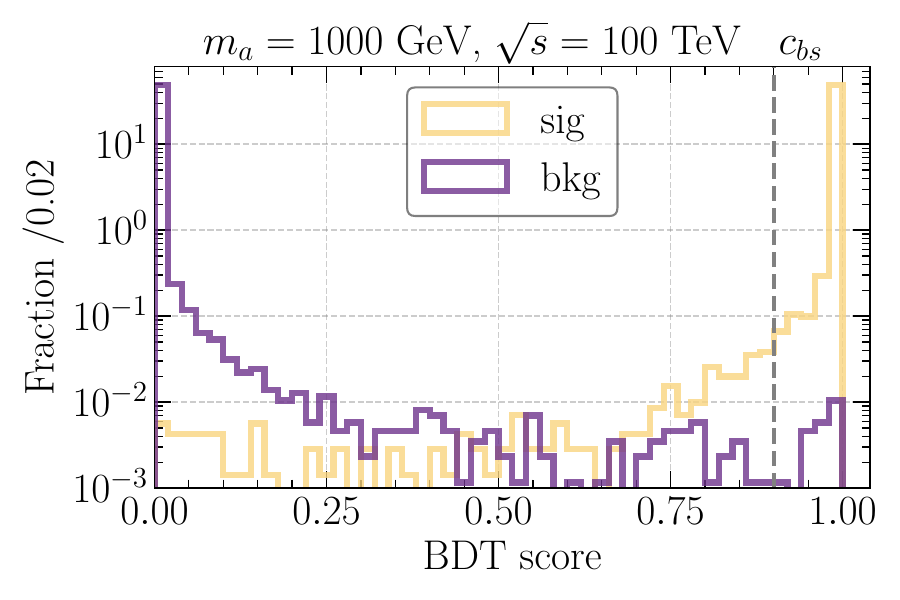}
\end{center}
\caption{ The BDT response score distribution of signal $bj\gamma\gamma$ (yellow) and total SM background (purple) with $m_a=10,~50,~300$ and 1000 GeV at FCC-hh with $\sqrt{s}=100$ TeV and $\mathcal{L}=30~{\rm ab}^{-1}$. The grey dash line is the BDT cut that maximizes the significance with fixed $|c_{bd}^{V(A)}|/f_a = 1~{\rm TeV}^{-1}$ (left four panels) or $|c_{bs}^{V(A)}|/f_a = 1~{\rm TeV}^{-1}$ (right four panels).}
\label{fig:aaBDT-score-FCC}
\end{figure}

For $pp\to jba$ with $a\to \gamma\gamma$, as the illustrations of HL-LHC ($\sqrt{s}=14$ TeV, $\mathcal{L}=3$ ab$^{-1}$) and FCC-hh ($\sqrt{s}=100$ TeV, $\mathcal{L}=30$ ab$^{-1}$), we also select four benchmarks ($m_a=10,~ 50, ~300$ and $1000$ GeV) to show the distribution of BDT-score in Fig.~\ref{fig:aaBDT-score-HLLHC} (HL-LHC) and Fig.~\ref{fig:aaBDT-score-FCC} (FCC-hh). The efficiencies of signal and backgrounds after applying the BDT-cut are shown in Table~\ref{tab:aaBDT-HLLHC} (HL-LHC) and Table~\ref{tab:aaBDT-FCC} (FCC-hh).

\begin{table}[htb!]
\centering
\begin{tabular}{c|c|c|c|c|c}
\hline
     $m_a$ & BDT cut & $\epsilon_{\rm sig.}$ & $\epsilon_{jj\gamma\gamma}$ & $\epsilon_{bb\gamma\gamma}$ &  $\mathcal{S}_{\rm max}$  \\
    \hline
    10   GeV& 0.90 & 9.66$\times10^{-1}$ & 3.89$\times10^{-3}$ & 2.78$\times10^{-3}$  &  9.30$\times10^{2}$\\
    50   GeV& 0.90 & 8.30$\times10^{-1}$ & 9.22$\times10^{-3}$ & 8.45$\times10^{-3}$  &  1.90$\times10^{2}$  \\
    300  GeV& 0.90 & 9.84$\times10^{-1}$ & 1.01$\times10^{-3}$ & 1.07$\times10^{-3}$  &  2.53$\times10^{1}$ \\
    1000 GeV& 0.90 & 9.95$\times10^{-1}$ & 1.01$\times10^{-3}$ & 5.99$\times10^{-4}$  &  5.37$\times10^{-1}$ \\
    \hline
    \hline
    10   GeV& 0.90 & 9.67$\times10^{-1}$ & 2.59$\times10^{-3}$ & 1.76$\times10^{-3}$ & 4.85$\times10^{2}$\\
    50   GeV& 0.90 & 7.97$\times10^{-1}$ & 8.07$\times10^{-3}$ & 7.10$\times10^{-3}$ & 6.46$\times10^{1}$  \\
    300  GeV& 0.90 & 9.85$\times10^{-1}$ & 1.15$\times10^{-3}$ & 1.20$\times10^{-3}$ & 5.78$\times10^{0}$ \\
    1000 GeV& 0.90 & 9.95$\times10^{-1}$ & 2.88$\times10^{-4}$ & 5.36$\times10^{-4}$ & 1.55$\times10^{-1}$ \\
    \hline
\end{tabular}
\caption{The BDT cut, cut efficiencies and achieved maximal significance in Eq.~(\ref{eqn-significance}) for the signal $bj\gamma\gamma$ and SM backgrounds at HL-LHC with $\sqrt{s}=14$ TeV and $\mathcal{L}=3~{\rm ab}^{-1}$. The benchmark masses are $m_a=10,~50,~300$ and 1000 GeV and the parameter is fixed as $|c_{bd}^{V(A)}|/f_a = 1~{\rm TeV}^{-1}$ (above the double line) or $|c_{bs}^{V(A)}|/f_a = 1~{\rm TeV}^{-1}$ (below the double line). The label ``$-$'' denotes the background at negligible level.}
\label{tab:aaBDT-HLLHC}
\end{table}

\begin{table}[htb!]
\centering
\begin{tabular}{c|c|c|c|c|c}
\hline
     $m_a$ & BDT cut & $\epsilon_{\rm sig.}$ & $\epsilon_{jj\gamma\gamma}$ & $\epsilon_{bb\gamma\gamma}$ &  $\mathcal{S}_{\rm max}$  \\
    \hline
    10   GeV& 0.90 & 9.89$\times10^{-1}$ & 9.01$\times10^{-4}$ & 5.95$\times10^{-4}$ & 8.94$\times10^{3}$\\
    50   GeV& 0.90 & 8.81$\times10^{-1}$ & 3.78$\times10^{-3}$ & 2.88$\times10^{-3}$ & 2.35$\times10^{3}$  \\
    300  GeV& 0.90 & 9.90$\times10^{-1}$ & 4.51$\times10^{-4}$ & 8.14$\times10^{-4}$ & 5.71$\times10^{2}$ \\
    1000 GeV& 0.90 & 9.95$\times10^{-1}$ & 2.70$\times10^{-4}$ & 5.01$\times10^{-4}$ & 4.24$\times10^{1}$ \\
    \hline
    \hline
    10   GeV& 0.90 & 9.90$\times10^{-1}$ & 9.01$\times10^{-4}$ & 5.01$\times10^{-4}$ & 6.70$\times10^{3}$\\
    50   GeV& 0.90 & 8.95$\times10^{-1}$ & 3.24$\times10^{-3}$ & 3.38$\times10^{-3}$ & 1.58$\times10^{3}$  \\
    300  GeV& 0.90 & 9.88$\times10^{-1}$ & 8.11$\times10^{-4}$ & 9.38$\times10^{-4}$ & 2.23$\times10^{2}$ \\
    1000 GeV& 0.90 & 9.93$\times10^{-1}$ & 4.51$\times10^{-4}$ & 5.63$\times10^{-4}$ & 1.20$\times10^{1}$ \\
    \hline
\end{tabular}
\caption{The BDT cut, cut efficiencies and achieved maximal significance in Eq.~(\ref{eqn-significance}) for the signal $bj\gamma\gamma$ and SM backgrounds at FCC-hh with $\sqrt{s}=100$ TeV and $\mathcal{L}=30~{\rm ab}^{-1}$. The benchmark masses are $m_a=10,~50,~300$ and 1000 GeV and the parameter is fixed as $|c_{bd}^{V(A)}|/f_a = 1~{\rm TeV}^{-1}$ (above the double line) or $|c_{bs}^{V(A)}|/f_a = 1~{\rm TeV}^{-1}$ (below the double line). The label ``$-$'' denotes the background at negligible level.}
\label{tab:aaBDT-FCC}
\end{table}

\clearpage
\bibliographystyle{JHEP}
\bibliography{refs}

\end{document}